\documentclass[12pt]{article}
\usepackage{amsmath,amsthm}
\usepackage{eucal}
\usepackage{amsfonts}
\gdef\C{{\Bbb C}}
\gdef\dZ{{\Bbb Z}}

\gdef\K{{\Bbb K}}
\gdef\R{{\Bbb R}}
\gdef\BH{{\Bbb H}}
\gdef\F{{\Bbb F}}
\newcommand{\End}{{\rm End}}
\newcommand{\spin}{{\bf Spin}}
\newcommand{\pin}{{\bf Pin}}
\newcommand{\Id}{{\rm Id}}

\newcommand{\Aut}{{\rm Aut}}

\newcommand{\M}{{\bf\sf M}}
\newcommand{\bi}{{\bf i}}
\newcommand{\bj}{{\bf j}}
\newcommand{\bk}{{\bf k}}
\newcommand{\bx}{{\bf x}}

\newcommand{\bZ}{{\bf Z}}
\newcommand{\Lip}{{\bf\Gamma}}
\newcommand{\cl}{C\kern -0.2em \ell}

\newcommand{\e}{\mbox{\bf e}}
\newcommand{\hs}{\hspace{0.2cm}}

\newcommand{\ar}{\renewcommand{\arraystretch}{1}}
\newcommand{\cA}{\mathcal{A}}

\newcommand{\cE}{\mathcal{E}}

\newcommand{\sA}{{\sf A}}
\newcommand{\sB}{{\sf B}}
\newcommand{\sE}{{\sf E}}
\newcommand{\sC}{{\sf C}}
\newcommand{\sI}{{\sf I}}
\newcommand{\sT}{{\sf T}}
\newcommand{\sW}{{\sf W}}
\newcommand{\p}{\prime}

\newcommand{\sAut}{{\sf Aut}}
\newtheorem{theorem}{Theorem}
\newtheorem{cor}{Corollary}
\begin{document}
\begin{center}
{\bf\Large Discrete Symmetries and Clifford Algebras}\\[0.5cm]
{\large V.V. Varlamov\footnote[2]{Department of Mathematics, Siberia State
University of Industry, Novokuznetsk 654007, Russia}}
\end{center}
\medskip
\begin{abstract}
An algebraic description of basic discrete symmetries (space reversal $P$,
time reversal $T$ and their combination $PT$) is studied. Discrete
subgroups of orthogonal groups of multidimensional spaces over the fields
of real and complex numbers are considered in terms of fundamental
automorphisms of Clifford algebras. In accordance with a division ring
structure, a complete classification of automorphism groups is established
for the Clifford algebras over the field of real numbers. The correspondence
between eight double coverings (D\c{a}browski groups) of the orthogonal group
and eight types of the real Clifford algebras is defined with the use of
isomorphisms between the automorphism groups and finite groups. Over the
field of complex numbers there is a correspondence between two
nonisomorphic double coverings of the complex orthogonal group and
two types of complex Clifford algebras. It is shown that these correspondences
associate with a well--known Atiyah--Bott--Shapiro periodicity.
Generalized Brauer--Wall groups are introduced on the extended sets of
the Clifford algebras. The structure of the inequality between the two
Clifford--Lipschitz groups with mutually opposite signatures is elucidated.
The physically important case of the two different double coverings of the
Lorentz groups is considered in details.
\end{abstract}
{\bf 1998 Physics and Astronomy Classification Scheme:} 02.10.Tq, 02.20.Df,
11.30.Er
\section{Introduction}
In 1909, Minkowski showed \cite{Min} that a causal structure of the world
is described by a 4--dimensional pseudo--Euclidean geometry. In accordance
with \cite{Min} the quadratic form $x^2+y^2+z^2-c^2t^2$ remains invariant
under the action of linear transformations of the four variables $x,y,z$ and 
$t$,
which form a general Lorentz group $G$. As known, the general Lorentz group
$G$ consists of an own Lorentz group $G_0$ and three reflections
(discrete transformations) $P,\,T,\,PT$, where $P$ and $T$ are space and
time reversal, and $PT$ is a so--called full reflection. The discrete
transformations $P,\,T$ and $PT$ added to an identical transformation
form a finite group. Thus, the general Lorentz group may be represented by
a semidirect product $G_0\odot\{1,P,T,PT\}$. Analogously, an orthogonal
group $O(p,q)$ of the real space $\R^{p,q}$ is represented by the semidirect
product of a connected component $O_0(p,q)$ and a discrete subgroup.

Further, a double covering of the orthogonal group $O(p,q)$ is a
Clifford--Lipschitz group $\pin(p,q)$ which is completely constructed within
a Clifford algebra $\cl_{p,q}$. In accordance with squares of elements of the
discrete subgroup ($a=P^2,\,b=T^2,\,c=(PT)^2$) there exist eight double
coverings (D\c{a}browski groups \cite{Dab88}) of the orthogonal group
defining by the signatures $(a,b,c)$, where $a,b,c\in\{-,+\}$. Such in brief is
a standard description scheme of the discrete transformations.

However, in this scheme there is one essential flaw. Namely, the
Clifford--Lipschitz group is an intrinsic notion of the algebra $\cl_{p,q}$
(a set of the all invertible elements of $\cl_{p,q}$), whereas the discrete
subgroup is introduced into the standard scheme in an external way, and the
choice of the signature $(a,b,c)$ of the discrete subgroup is not
determined by the signature of the space $\R^{p,q}$. Moreover, it is suggest
by default that for any signature $(p,q)$ of the vector space there exist
the all eight kinds of the discrete subgroups.

In the recent paper \cite{Var99}, to assimilate the discrete
transformations into an algebraic framework it has been shown that elements
of the discrete subgroup correspond to fundamental automorphisms of
the Clifford algebras. The set of the fundamental automorphisms added to
an identical automorphism forms a finite group, for which in virtue of
the Wedderburn--Artin theorem there exists a matrix representation.
The main subject of \cite{Var99} is the study of the homomorphism
$\C_{n+1}\rightarrow\C_n$ and its application in physics, where $\C_n$
is a Clifford algebra over the field of complex numbers $\F=\C$.

The main goal of the present paper is a more explicit and complete
formulation (in accordance with a division ring structure of the algebras
$\cl_{p,q}$) of the preliminary results obtained in \cite{Var99}.
The classification of automorphism groups of Clifford algebras over the
field of real numbers $\F=\R$ and a correspondence between eight
D\c{a}browski $\pin^{a,b,c}$--coverings of the group $O(p,q)$ and eight
types of $\cl_{p,q}$ are established in the section 3. It is shown that
the division ring structure of $\cl_{p,q}$ imposes hard restrictions on
existence and choice of the discrete subgroup, and the signature
$(a,b,c)$ depends upon the signature of the underlying space $\R^{p,q}$.
On the basis of obtained results, a nature of the inequality
$\pin(p,q)\not\simeq\pin(q,p)$ is elucidated in the section 4. As known,
the Lorentz groups $O(3,1)$ and $O(1,3)$ are isomorphic, whereas their
double coverings $\pin(3,1)$ and $\pin(1,3)$ are nonisomorphic.
With the help of Maple V package {\sc CLIFFORD} \cite{Abl96,Abl00},
a structure of the inequality
$\pin(3,1)\not\simeq\pin(1,3)$ is considered as an example that is,
all the possible
spinor representations of a Majorana algebra $\cl_{3,1}$ and a
spacetime algebra $\cl_{1,3}$, and corresponding automorphism groups, are
analysed in detail. In connection with this, it should be noted that
the general Lorentz group is a basis for (presently most profound in
both mathematical and physical viewpoints) Wigner's definition of
elementary particle as an irreducible representation of the
inhomogeneous Lorentz group \cite{Wig64}.

It is known that the Clifford algebras are modulo 8 periodic over the field
of real numbers and modulo 2 periodic over the field of complex
numbers (Atiyah--Bott-Shapiro periodicity \cite{AtBSh}). In virtue of this
periodicity, a structure of the Brauer--Wall group \cite{Wal64,BT88,Lou97}
is defined on the set of the Clifford algebras, where a group element is
$\cl$, and a group operation is a graded tensor product. The Brauer--Wall
group over the field $\F=\R$ is isomorphic to a cyclic group of eighth
order, and over the field $\F=\C$ to a cyclic group of second order.
Generalizations of the Brauer--Wall groups are considered in the section 5.
The Trautman diagrams of the generalized groups are defined as well.
\section{Preliminaries}In this section we will consider some basic facts about Clifford algebras
and Clifford--Lipschitz groups which we will widely use below.
Let $\F$ be a field of characteristic 0 $(\F=\R,\,\F=\C)$, where
$\R$ and $\C$ are the fields of real and complex numbers, respectively.
A Clifford algebra $\cl$ over a field $\F$ is an algebra with
$2^n$ basis elements: $\e_0$
(unit of the algebra) $\e_1,\e_2,\ldots,\e_n$ and products of the one--index
elements $\e_{i_1i_2\ldots i_k}=\e_{i_1}\e_{i_2}\ldots\e_{i_k}$.
Over the field $\F=\R$ the Clifford algebra denoted as $\cl_{p,q}$, where
the indices
$p$ and $q$ correspond to the indices of the quadratic form
\[
Q=x^2_1+\ldots+x^2_p-\ldots-x^2_{p+q}
\]
of a vector space $V$ associated with $\cl_{p,q}$. The multiplication law
of $\cl_{p,q}$ is defined by a following rule:
\begin{equation}\label{e1}
\e^2_i=\sigma(q-i)\e_0,\quad\e_i\e_j=-\e_j\e_i,
\end{equation}
where
\begin{equation}\label{e2}
\sigma(n)=\left\{\begin{array}{rl}
-1 & \mbox{if $n\leq 0$},\\
+1 & \mbox{if $n>0$}.
\end{array}\right.
\end{equation}
The square of a volume element $\omega=\e_{12\ldots n}$ ($n=p+q$) plays an
important role in the theory of Clifford algebras,
\begin{equation}\label{e3}
\omega^2=\left\{\begin{array}{rl}
-1 & \mbox{if $p-q\equiv 2,3,6,7\pmod{8}$},\\
+1 & \mbox{if $p-q\equiv 0,1,4,5\pmod{8}$}.
\end{array}\right.
\end{equation}  
A center $\bZ_{p,q}$ of the algebra $\cl_{p,q}$ consists of the unit $\e_0$ 
and the volume element $\omega$. The element $\omega=\e_{12\ldots n}$ 
belongs to a center when $n$ is odd. Indeed,
\begin{eqnarray}
\e_{12\ldots n}\e_i&=&(-1)^{n-i}\sigma(q-i)\e_{12\ldots i-1 i+1\ldots n},
\nonumber\\
\e_i\e_{12\ldots n}&=&(-1)^{i-1}\sigma(q-i)\e_{12\ldots i-1 i+1\ldots n},
\nonumber
\end{eqnarray}
therefore, $\omega\in\bZ_{p,q}$ if and only if $n-i\equiv i-1\pmod{2}$, 
that is, $n$ is odd. Further, using (\ref{e3}) we obtain
\begin{equation}\label{e4}
\bZ_{p,q}=\left\{\begin{array}{rl}
\phantom{1,}1 & \mbox{if $p-q\equiv 0,2,4,6\pmod{8}$},\\
1,\omega & \mbox{if $p-q\equiv 1,3,5,7\pmod{8}$}.
\end{array}\right.
\end{equation}

In Clifford algebra $\cl$ there exist four fundamental automorphisms.\\[0.2cm]
1) {\bf Identity}: An automorphism $\cA\rightarrow\cA$ and 
$\e_{i}\rightarrow\e_{i}$.\\
This automorphism is an identical automorphism of the algebra $\cl$. 
$\cA$ is an arbitrary element of $\cl$.\\[0.2cm]
2) {\bf Involution}: An automorphism $\cA\rightarrow\cA^\star$ and 
$\e_{i}\rightarrow-\e_{i}$.\\
In more details, for an arbitrary element $\cA\in\cl$ there exists a
decomposition
$
\cA=\cA^{\p}+\cA^{\p\p},
$
where $\cA^{\p}$ is an element consisting of homogeneous odd elements, and
$\cA^{\p\p}$ is an element consisting of homogeneous even elements,
respectively. Then the automorphism
$\cA\rightarrow\cA^{\star}$ is such that the element
$\cA^{\p\p}$ is not changed, and the element $\cA^{\p}$ changes sign:
$
\cA^{\star}=-\cA^{\p}+\cA^{\p\p}.
$
If $\cA$ is a homogeneous element, then
\begin{equation}\label{auto16}
\cA^{\star}=(-1)^{k}\cA,
\end{equation}
where $k$ is a degree of the element. It is easy to see that the
automorphism $\cA\rightarrow\cA^{\star}$ may be expressed via the volume
element $\omega=\e_{12\ldots p+q}$:
\begin{equation}\label{auto17}
\cA^{\star}=\omega\cA\omega^{-1},
\end{equation}
where
$\omega^{-1}=(-1)^{\frac{(p+q)(p+q-1)}{2}}\omega$. When $k$ is odd, 
for the basis elements 
$\e_{i_{1}i_{2}\ldots i_{k}}$ the sign changes, and when $k$ is even, the sign
is not changed.\\[0.2cm]
3) {\bf Reversion}: An antiautomorphism $\cA\rightarrow\widetilde{\cA}$ and
$\e_i\rightarrow\e_i$.\\
The antiautomorphism $\cA\rightarrow\widetilde{\cA}$ is a reversion of the
element $\cA$, that is the substitution of the each basis element
$\e_{i_{1}i_{2}\ldots i_{k}}\in\cA$ by the element
$\e_{i_{k}i_{k-1}\ldots i_{1}}$:
\[
\e_{i_{k}i_{k-1}\ldots i_{1}}=(-1)^{\frac{k(k-1)}{2}}
\e_{i_{1}i_{2}\ldots i_{k}}.
\]
Therefore, for any $\cA\in\cl_{p,q}$, we have
\begin{equation}\label{auto19}
\widetilde{\cA}=(-1)^{\frac{k(k-1)}{2}}\cA.
\end{equation}
4) {\bf Conjugation}: An antiautomorpism $\cA\rightarrow\widetilde{\cA^\star}$
and $\e_i\rightarrow-\e_i$.\\
This antiautomorphism is a composition of the antiautomorphism
$\cA\rightarrow\widetilde{\cA}$ with the automorphism
$\cA\rightarrow\cA^{\star}$. In the case of a homogeneous element from
the formulae (\ref{auto16}) and (\ref{auto19}), it follows
\begin{equation}\label{20}
\widetilde{\cA^{\star}}=(-1)^{\frac{k(k+1)}{2}}\cA.
\end{equation}

The Lipschitz group $\Lip_{p,q}$, also called the Clifford group, introduced
by Lipschitz in 1886 \cite{Lips}, may be defined as the subgroup of
invertible elements $s$ of the algebra $\cl_{p,q}$:
\[
\Lip_{p,q}=\left\{s\in\cl^+_{p,q}\cup\cl^-_{p,q}\;|\;\forall x\in\R^{p,q},\;
s\bx s^{-1}\in\R^{p,q}\right\}.
\]
The set $\Lip^+_{p,q}=\Lip_{p,q}\cap\cl^+_{p,q}$ is called {\it special
Lipschitz group} \cite{Che55}.

Let $N:\;\cl_{p,q}\rightarrow\cl_{p,q},\;N(\bx)=\bx\widetilde{\bx}$.
If $\bx\in\R^{p,q}$, then $N(\bx)=\bx(-\bx)=-\bx^2=-Q(\bx)$. Further, the
group $\Lip_{p,q}$ has a subgroup
\begin{equation}\label{Pin}
\pin(p,q)=\left\{s\in\Lip_{p,q}\;|\;N(s)=\pm 1\right\}.
\end{equation}
Analogously, {\it a spinor group} $\spin(p,q)$ is defined by the set
\begin{equation}\label{Spin}
\spin(p,q)=\left\{s\in\Lip^+_{p,q}\;|\;N(s)=\pm 1\right\}.
\end{equation}
It is obvious that $\spin(p,q)=\pin(p,q)\cap\cl^+_{p,q}$.
The group $\spin(p,q)$ contains a subgroup
\begin{equation}\label{Spin+}
\spin_+(p,q)=\left\{s\in\spin(p,q)\;|\;N(s)=1\right\}.
\end{equation}
It is easy to see that the groups $O(p,q),\,SO(p,q)$ and $SO_+(p,q)$ are
isomorphic, respectively, to the following quotient groups
\[
O(p,q)\simeq\pin(p,q)/\dZ_2,\quad
SO(p,q)\simeq\spin(p,q)/\dZ_2,\quad
SO_+(p,q)\simeq\spin_+(p,q)/\dZ_2,
\]
\begin{sloppypar}\noindent
where the kernel $\dZ_2=\{1,-1\}$. Thus, the groups $\pin(p,q)$, $\spin(p,q)$
and $\spin_+(p,q)$ are the double coverings of the groups $O(p,q),\,SO(p,q)$
and $SO_+(p,q)$, respectively.\end{sloppypar}

On the other hand, there exists a more detailed version of the $\pin$--group
(\ref{Pin}) proposed by D\c{a}browski in 1988 \cite{Dab88}. In general,
there are eight double coverings of the orthogonal group 
$O(p,q)$ \cite{Dab88,BD89}:
\[
\rho^{a,b,c}:\;\;\pin^{a,b,c}(p,q)\longrightarrow O(p,q),
\]
where $a,b,c\in\{+,-\}$. As known, the group $O(p,q)$ consists of four
connected components: identity connected component $O_0(p,q)$, and three
components corresponding to parity reversal $P$, time reversal
$T$, and the combination of these two $PT$, i.e., $O(p,q)=(O_0(p,q))\cup
P(Q_0(p,q))\cup T(O_0(p,q))\cup PT(O_0(p,q))$. Further, since the
four--element group (reflection group) $\{1,\,P,\,T,\,PT\}$ is isomorphic to
the finite group $\dZ_2\otimes\dZ_2$ 
(Gauss--Klein veergruppe \cite{Sal81a,Sal84}), then
$O(p,q)$ may be represented by a semidirect product $O(p,q)\simeq O_0(p,q)
\odot(\dZ_2\otimes\dZ_2)$. The signs of $a,b,c$ correspond to the signs of the
squares of the elements in $\pin^{a,b,c}(p,q)$ which cover space reflection
$P$, time reversal $T$ and a combination of these two
$PT$ ($a=-P^2,\,b=T^2,\,c=-(PT)^2$ in D\c{a}browski's notation \cite{Dab88} and
$a=P^2,\,b=T^2,\,c=(PT)^2$ in Chamblin's notation \cite{Ch94} which we will
use below).
An explicit form of the group $\pin^{a,b,c}(p,q)$ is given by the following
semidirect product
\begin{equation}\label{Pinabc}
\pin^{a,b,c}(p,q)\simeq\frac{(\spin_0(p,q)\odot C^{a,b,c})}{\dZ_2},
\end{equation}
where $C^{a,b,c}$ are the four double coverings of
$\dZ_2\otimes\dZ_2$. 
All the eight double coverings of the orthogonal group
$O(p,q)$ are given in the following table:
\begin{center}
{\renewcommand{\arraystretch}{1.4}
\begin{tabular}{|c|l|l|}\hline
$a$ $b$ $c$ & $C^{a,b,c}$ & Remark \\ \hline
$+$ $+$ $+$ & $\dZ_2\otimes\dZ_2\otimes\dZ_2$ & $PT=TP$\\
$+$ $-$ $-$ & $\dZ_2\otimes\dZ_4$ & $PT=TP$\\
$-$ $+$ $-$ & $\dZ_2\otimes\dZ_4$ & $PT=TP$\\
$-$ $-$ $+$ & $\dZ_2\otimes\dZ_4$ & $PT=TP$\\ \hline
$-$ $-$ $-$ & $Q_4$ & $PT=-TP$\\
$-$ $+$ $+$ & $D_4$ & $PT=-TP$\\
$+$ $-$ $+$ & $D_4$ & $PT=-TP$\\
$+$ $+$ $-$ & $D_4$ & $PT=-TP$\\ \hline
\end{tabular}
}
\end{center}
Here $\dZ_4$, $Q_4$, and $D_4$ are complex, quaternion, and
dihedral groups, respectively.
According to \cite{Dab88} the group $\pin^{a,b,c}(p,q)$ satisfying the
condition
$PT=-TP$ is called {\it Cliffordian}, and respectively {\it
non--Cliffordian} when $PT=TP$. 

One of the most fundamental theorems in the theory of associative algebras
is as follows
\begin{theorem}[{\rm Wedderburn--Artin}]
Any finite--dimensional associative simple algebra $\mathfrak{A}$ over the
field $\F$ is isomorphic to a full matrix algebra $\M_n(\K)$, where $n$ is
natural number defined unambiguously, and $\K$ a division ring defined
with an accuracy of isomorphism.
\end{theorem}
In accordance with this theorem, all properties of the initial algebra
$\mathfrak{A}$ are isomorphically transferred to the matrix algebra 
$\M_n(\K)$. Later on we will widely use this theorem. In its turn, for the
Clifford algebra $\cl_{p,q}$ over the field $\F=\R$ we have an isomorphism
$\cl_{p,q}\simeq\End_{\K}(I_{p,q})\simeq\M_{2^m}(\K)$, where $m=\frac{p+q}{2}$,
$I_{p,q}=\cl_{p,q}f$ is a minimal left ideal of $\cl_{p,q}$, and
$\K=f\cl_{p,q}f$ is a division ring of $\cl_{p,q}$. The primitive idempotent
of the algebra $\cl_{p,q}$ has a form
\[
f=\frac{1}{2}(1\pm\e_{\alpha_1})\frac{1}{2}(1\pm\e_{\alpha_2})\cdots\frac{1}{2}
(1\pm\e_{\alpha_k}),
\]
where $\e_{\alpha_1},\e_{\alpha_2},\ldots,\e_{\alpha_k}$ are commuting
elements with square 1 of the canonical basis of $\cl_{p,q}$ generating
a group of order $2^k$. The values of $k$ are defined by a formula
$k=q-r_{q-p}$, where $r_i$ are the Radon--Hurwitz numbers \cite{Rad22,Hur23},
values of which form a cycle of the period 8: $r_{i+8}=r_i+4$. The values of
all $r_i$ are
\begin{center}
\begin{tabular}{lcccccccc}
$i$  & 0 & 1 & 2 & 3 & 4 & 5 & 6 & 7\\ \hline
$r_i$& 0 & 1 & 2 & 2 & 3 & 3 & 3 & 3
\end{tabular}.
\end{center}
All the Clifford algebras $\cl_{p,q}$ over the field $\F=\R$ are divided
into eight different types with the following division ring structure:\\[0.3cm]
{\bf I}. Central simple algebras.
\begin{description}
\item[1)] Two types $p-q\equiv 0,2\pmod{8}$ with a division ring 
$\K\simeq\R$.
\item[2)] Two types $p-q\equiv 3,7\pmod{8}$ with a division ring
$\K\simeq\C$.
\item[3)] Two types $p-q\equiv 4,6\pmod{8}$ with a division ring
$\K\simeq\BH$.
\end{description}
{\bf II}. Semisimple algebras.
\begin{description}
\item[4)] The type $p-q\equiv 1\pmod{8}$ with a double division ring
$\K\simeq\R\oplus\R$.
\item[5)] The type $p-q\equiv 5\pmod{8}$ with a double quaternionic 
division ring $\K\simeq\BH\oplus\BH$.
\end{description}
Over the field $\F=\C$ there is an isomorphism $\C_n\simeq\M_{2^{n/2}}(\C)$
and there are two different types of complex Clifford algebras $\C_n$:
$n\equiv 0\pmod{2}$ and $n\equiv 1\pmod{2}$.

In virtue of the Wedderburn--Artin theorem, all fundamental automorphisms
of $\cl$ are transferred to the matrix algebra. Matrix representations of the
fundamental automorphisms of $\C_n$ was first obtained by Rashevskii in 1955
\cite{Rash}: 1) Involution: $\sA^\star=\sW\sA\sW^{-1}$, where $\sW$ is a
matrix of the automorphism $\star$ (matrix representation of the volume
element $\omega$); 2) Reversion: $\widetilde{\sA}=\sE\sA^{\sT}\sE^{-1}$, where
$\sE$ is a matrix of the antiautomorphism $\widetilde{\phantom{cc}}$
satisfying the conditions $\cE_i\sE-\sE\cE^{\sT}_i=0$ and 
$\sE^{\sT}=(-1)^{\frac{m(m-1)}{2}}\sE$, here $\cE_i=\gamma(\e_i)$ are matrix
representations of the units of the algebra $\cl$; 3) Conjugation:
$\widetilde{\sA^\star}=\sC\sA^{\sT}\sC^{-1}$, where $\sC=\sE\sW^{\sT}$ 
is a matrix of
the antiautomorphism $\widetilde{\star}$ satisfying the conditions
$\sC\cE^{\sT}+\cE_i\sC=0$ and
$\sC^{\sT}=(-1)^{\frac{m(m+1)}{2}}\sC$.

In the recent paper \cite{Var99}, it has been shown that space
reversal $P$, time reversal $T$ and combination $PT$ are correspond 
to the fundamental automorphisms 
$\cA\rightarrow\cA^\star$, $\cA\rightarrow\widetilde{\cA}$ and
$\cA\rightarrow\widetilde{\cA^{\star}}$, respectively. 
Moreover, there is an isomorphism
between the discrete subgroup $\{1,P,T,PT\}\simeq\dZ_2\otimes\dZ_2$
($P^2=T^2=(PT)^2=1,\;PT=TP$) of $O(p,q)$ and an automorphism group
$\Aut(\cl)=\{\Id,\star,\widetilde{\phantom{cc}},\widetilde{\star}\}$:
\[
{\renewcommand{\arraystretch}{1.4}
\begin{tabular}{|c||c|c|c|c|}\hline
        & $\Id$ & $\star$ & $\widetilde{\phantom{cc}}$ & $\widetilde{\star}$\\ \hline\hline
$\Id$   & $\Id$ & $\star$ & $\widetilde{\phantom{cc}}$ & $\widetilde{\star}$\\ \hline
$\star$ & $\star$ & $\Id$ & $\widetilde{\star}$ & $\widetilde{\phantom{cc}}$\\ \hline
$\widetilde{\phantom{cc}}$ & $\widetilde{\phantom{cc}}$ &$\widetilde{\star}$
& $\Id$ & $\star$ \\ \hline
$\widetilde{\star}$ & $\widetilde{\star}$ & $\widetilde{\phantom{cc}}$ &
$\star$ & $\Id$\\ \hline
\end{tabular}
\;\sim\;
\begin{tabular}{|c||c|c|c|c|}\hline
    & $1$ & $P$ & $T$ & $PT$\\ \hline\hline
$1$ & $1$ & $P$ & $T$ & $PT$\\ \hline
$P$ & $P$ & $1$ & $PT$& $T$\\ \hline
$T$ & $T$ & $PT$& $1$ & $P$\\ \hline
$PT$& $PT$& $T$ & $P$ & $1$\\ \hline
\end{tabular}
}
\]
Further, in the case $P^2=T^2=(PT)^2=\pm 1$ and $PT=-TP$, there is an
isomorphism between the group $\{1,P,T,PT\}$ and an automorphism group
$\sAut(\cl)=\{\sI,\sW,\sE,\sC\}$. So, for the Dirac algebra $\C_4$ in the
canonical $\gamma$--basis there exists a standard (Wigner) representation
$P=\gamma_0$ and $T=\gamma_1\gamma_3$ \cite{BLP89}, therefore,
$\{1,P,T,PT\}=\{1,\gamma_0,\gamma_1\gamma_3,\gamma_0\gamma_1\gamma_3\}$.
On the other hand, in the $\gamma$--basis, an automorphism group
of $\C_4$ has a form $\sAut(\C_4)=\{\sI,\sW,\sE,\sC\}=
\{\sI,\gamma_0\gamma_1\gamma_2\gamma_3,\gamma_1\gamma_3,\gamma_0\gamma_2\}$.
It has been shown \cite{Var99} that $\{1,P,T,PT\}=
\{1,\gamma_0,\gamma_1\gamma_3,\gamma_0\gamma_1\gamma_3\}\simeq\sAut(\C_4)
\simeq\dZ_4$, where $\dZ_4$ is a complex group with the signature
$(+,-,-)$. Generalizations of these results onto the algebras $\C_n$
are contained in the following two theorems:
\begin{theorem}[{\rm\cite{Var99}}]\label{taut}
Let $\sAut(\C_n)=\{\sI,\,\sW,\,\sE,\,\sC\}$ be a group of the fundamental
automorphisms of the algebra $\C_{n}$ $(n=2m)$, 
where $\sW=\cE_1\cE_2\cdots\cE_m\cE_{m+1}\cE_{m+2}\cdots\cE_{2m}$,
and $\sE=\cE_1\cE_2\cdots\cE_m$, $\sC=\cE_{m+1}\cE_{m+2}\cdots\cE_{2m}$ if
$m\equiv 1\pmod{2}$, and $\sE=\cE_{m+1}\cE_{m+2}\cdots\cE_{2m}$, $\sC=\cE_1E_2\cdots
\cE_m$ if $m\equiv 0\pmod{2}$. Let $\sAut_-(\C_n)$ and
$\sAut_+(\C_n)$ be the automorphism groups, in which all the elements
commute
$(m\equiv 0\pmod{2})$ and anticommute $(m\equiv 1\pmod{2})$, respectively. 
Then over the field $\F=\C$ there exist only two non--isomorphic groups:
$\sAut_-(\C_n)\simeq\dZ_2\otimes\dZ_2$ with the signature $(+,\,+,\,+)$ if
$n\equiv 0,4\pmod{8}$ and
$\sAut_+(\C_n)\simeq Q_4/\dZ_2$ with the signature
$(-,\,-,\,-)$ if
$n\equiv 2,6\pmod{8}$.
\end{theorem}
\begin{theorem}[{\rm\cite{Var99}}]\label{tgroup}
Let $\pin^{a,b,c}(n,\C)$ be a double covering of the complex orthogonal group
$O(n,\C)$ of the space $\C^{n}$ associated with the complex algebra
$\C_{n}$. A dimensionality of the algebra
$\C_{n}$ is even $(n=2m)$, squares of the symbols $a,b,c\in
\{-,+\}$ are correspond to squares of the elements of the finite group
$\sAut=\{\sI,\sW,\sE,\sC\}:\;a=\sW^2,\,b=\sE^2,\,c=\sC^2$, 
where $\sW,\,\sE$ and $\sC$
are the matrices of the fundamental automorphisms $\cA\rightarrow
\cA^\star,\,\cA\rightarrow\widetilde{\cA}$ and $\cA\rightarrow
\widetilde{\cA^\star}$ of $\C_{n}$, respectively.
Then over the field $\F=\C$, for the
algebra $\C_n$ there are two non--isomorphic double coverings of the group
$O(n,\C)$:\\
1) A non--Cliffordian group
\[
\pin^{+,+,+}(n,\C)\simeq\frac{(\spin_0(n,\C)\odot\dZ_2\otimes\dZ_2\otimes\dZ_2)}
{\dZ_2},
\]
if $n\equiv 0,4\pmod{8}$. \\
2) A Cliffordian group
\[
\pin^{-,-,-}(n,\C)\simeq\frac{(\spin_0(n,\C)\odot Q_4)}{\dZ_2},
\]
if $n\equiv 2,6\pmod{8}$.
\end{theorem}

\section{Discrete symmetries over the field $\F=\R$}
\begin{theorem}\label{tautr}
Let $\cl_{p,q}$ be a Clifford algebra over a field $\F=\R$ and let
$\sAut(\cl_{p,q})=\{\sI,\sW,\sE,\sC\}$ be a group of fundamental
automorphisms of the algebra $\cl_{p,q}$. Then for eight types of the 
algebras $\cl_{p,q}$ there exist, depending upon a division ring structure
of $\cl_{p,q}$, following isomorphisms between finite groups and groups
$\sAut(\cl_{p,q})$ with different signatures
$(a,b,c)$, where $a,b,c\in\{-,+\}$:\\[0.2cm]
1) $\K\simeq\R$, types $p-q\equiv 0,2\pmod{8}$.\\
If $\sE=\cE_{p+1}\cE_{p+2}\cdots\cE_{p+q}$ and $\sC=\cE_1\cE_2\cdots\cE_p$,
then Abelian groups $\sAut_-(\cl_{p,q})\simeq\dZ_2\otimes\dZ_2$
with the signature $(+,+,+)$ and $\sAut_-(\cl_{p,q})\simeq\dZ_4$ with the
signature
$(+,-,-)$ exist at $p,q\equiv 0\pmod{4}$ and $p,q\equiv 2\pmod{4}$, 
respectively,
for the type $p-q\equiv 0\pmod{8}$, and also Abelian groups
$\sAut_-(\cl_{p,q})\simeq\dZ_4$ with the signature $(-,-,+)$ and
$\sAut_-(\cl_{p,q})
\simeq\dZ_4$ with the signature $(-,+,-)$ exist 
at $p\equiv 0\pmod{4},\,
q\equiv 2\pmod{4}$ and $p\equiv 2\pmod{4},\,q\equiv 0\pmod{4}$ for the type
$p-q\equiv 2\pmod{8}$, respectively.\\
If $\sE=\cE_1\cE_2\cdots\cE_p$ and $\sC=\cE_{p+1}\cE_{p+2}\cdots\cE_{p+q}$,
then non--Abelian groups $\sAut_+(\cl_{p,q})\simeq D_4/\dZ_2$ with the
signature $(+,-,+)$ and $\sAut_+(\cl_{p,q})\simeq D_4/\dZ_2$ with the
signature
$(+,+,-)$ exist at $p,q\equiv 3\pmod{4}$ and $p,q\equiv 1\pmod{4}$, 
respectively,
for the type $p-q\equiv 0\pmod{8}$, and also non--Abelian groups
$\sAut_+(\cl_{p,q})\simeq Q_4/\dZ_2$ with $(-,-,-)$ and 
$\sAut_+(\cl_{p,q})\simeq
D_4/\dZ_2$ with $(-,+,+)$ exist at $p\equiv 3\pmod{4},\,q\equiv 1
\pmod{4}$ and $p\equiv 1\pmod{4},\,q\equiv 3\pmod{4}$ for the type
$p-q\equiv 2\pmod{8}$, respectively.\\[0.2cm]
2) $\K\simeq\BH$, types $p-q\equiv 4,6\pmod{8}$.\\
If $\sE=\cE_{j_1}\cE_{j_2}\cdots\cE_{j_k}$ is a product of $k$
skewsymmetric matrices (among which $l$ matrices have a square $+\sI$
and $t$ matrices have a square $-\sI$)
and $\sC=\cE_{i_1}\cE_{i_2}\cdots\cE_{i_{p+q-k}}$ is a product of $p+q-k$
symmetric matrices (among which $h$ matrices have a square $+\sI$ and
$g$ have a square $-\sI$),
then at $k\equiv 0\pmod{2}$ for the type $p-q\equiv 4\pmod{8}$ there exist
Abelian groups $\sAut_-(\cl_{p,q})\simeq\dZ_2\otimes\dZ_2$ with $(+,+,+)$
and $\sAut_-(\cl_{p,q})\simeq\dZ_4$ with $(+,-,-)$ if
$l-t,\,h-g\equiv 0,1,4,5\pmod{8}$ and
$l-t,\,h-g\equiv 2,3,6,7\pmod{8}$, respectively. And also at
$k\equiv 0\pmod{2}$ for the type $p-q\equiv 6\pmod{8}$ there exist
$\sAut_-(\cl_{p,q})\simeq\dZ_4$ with $(-,+,-)$ and 
$\sAut_-(\cl_{p,q})\simeq\dZ_4$
with $(-,-,+)$ if $l-t\equiv 0,1,4,5\pmod{8},\,
h-g\equiv 2,3,6,7\pmod{8}$ and $l-t\equiv 2,3,6,7\pmod{8},\,
h-g\equiv 0,1,4,5\pmod{8}$,respectively.\\
Inversely, if $\sE=\cE_{i_1}\cE_{i_2}\cdots\cE_{i_{p+q-k}}$ is a product of
$p+q-k$ symmetric matrices and 
$\sC=\cE_{j_1}\cE_{j_2}\cdots\cE_{j_k}$ is a product of $k$ skewsymmetric
matrices, then at $k\equiv 1\pmod{2}$
for the type $p-q\equiv 4\pmod{8}$ there exist non--Abelian groups
$\sAut_+(\cl_{p,q})
\simeq D_4/\dZ_2$ with $(+,-,+)$ and $\sAut_+(\cl_{p,q})\simeq D_4/\dZ_2$ with
$(+,+,-)$ if $h-g\equiv 2,3,6,7\pmod{8},\,l-t\equiv
0,1,4,5\pmod{8}$ and $h-g\equiv 0,1,4,5\pmod{8},\,l-t\equiv 2,3,6,7\pmod{8}$,
respectively.
And also at $k\equiv 1\pmod{2}$ for the type $p-q\equiv 6\pmod{8}$ there exist
$\sAut_+(\cl_{p,q})\simeq Q_4/\dZ_2$ with $(-,-,-)$ and $\sAut_+(\cl_{p,q})
\simeq D_4/\dZ_2$ with $(-,+,+)$ if $h-g,
\,l-t\equiv 2,3,6,7\pmod{8}$ and $h-g,\,l-t\equiv 0,1,4,5
\pmod{8}$, respectively.\\[0.2cm]
3) $\K\simeq\R\oplus\R,\,\K\simeq\BH\oplus\BH$, types $p-q\equiv 1,5\pmod{8}$.\\
For the algebras $\cl_{0,q}$ of the types $p-q\equiv 1,5\pmod{8}$ there exist
Abelian automorphism groups with the signatures
$(-,-,+)$, $(-,+,-)$ and non--Abelian automorphism groups with the signatures
$(-,-,-)$, $(-,+,+)$. Correspondingly, for the algebras $\cl_{p,0}$ of the
types $p-q\equiv 1,5\pmod{8}$ there exist Abelian groups with
$(+,+,+)$, $(+,-,-)$ and non--Abelian groups with $(+,-,+)$,
$(+,+,-)$. In general case for $\cl_{p,q}$, the types $p-q\equiv 1,5\pmod{8}$
admit all eight automorphism groups.\\[0.2cm]
4) $\K=\C$, types $p-q\equiv 3,7\pmod{8}$.\\
The types $p-q\equiv 3,7\pmod{8}$ admit the Abelian group $\sAut_-(\cl_{p,q})
\simeq\dZ_2\otimes\dZ_2$ with the signature $(+,+,+)$ if $p\equiv 0\pmod{2}$ and
$q\equiv 1\pmod{2}$, and also non--Abelian group 
$\sAut_+(\cl_{p,q})\simeq
Q_4/\dZ_2$ with the signature $(-,-,-)$ if $p\equiv 1\pmod{2}$ and
$q\equiv 0\pmod{2}$. 
\end{theorem} 
\begin{proof} Before we proceed to prove this theorem, let us consider in
more details a matrix (spinor) representations of the antiautomorphisms
$\cA\rightarrow\widetilde{\cA}$ and $\cA\rightarrow\widetilde{\cA^{\star}}$.
According to Wedderburn--Artin theorem
the antiautomorphism $\cA\rightarrow\widetilde{\cA}$ corresponds to
an antiautomorphism of the full matrix algebra $\M_{2^m}(\K)$:
$\sA\rightarrow\sA^{\sT}$, in virtue of the well--known relation 
$(\sA\sB)^{\sT}=
\sB^{\sT}\sA^{\sT}$, where $\sT$ is a symbol of transposition. On the other hand,
in the matrix representation of the elements $\cA\in\cl_{p,q}$, for the
antiautomorphism
$\cA\rightarrow\widetilde{\cA}$ we have $\sA\rightarrow\widetilde{\sA}$.
A composition of the two antiautomorphisms, $\sA^{\sT}\rightarrow\sA\rightarrow
\widetilde{\sA}$, gives an automorphism  $\sA^{\sT}\rightarrow\widetilde{\sA}$,
which is an internal automorphism of the algebra $\M_{2^m}(\K)$:
\begin{equation}\label{com}
\widetilde{\sA}=\sE\sA^{\sT}\sE^{-1},
\end{equation}
where $\sE$ is a matrix, by means of which the antiautomorphism $\cA\rightarrow
\widetilde{\cA}$ is expressed in the matrix representation of the
algebra $\cl_{p,q}$.
Under action of the antiautomorphism $\cA\rightarrow\widetilde{\cA}$
the units of $\cl_{p,q}$ remain unaltered, $\e_i\rightarrow\e_i$; therefore
in the matrix representation, we must demand $\cE_i\rightarrow\cE_i$,
where $\cE_i=\gamma(\e_i)$ also.
Therefore, for the definition of the matrix $\sE$
in accordance with (\ref{com}) we have
\begin{equation}\label{com1}
\cE_i\longrightarrow\cE_i=\sE\cE^{\sT}\sE^{-1}.
\end{equation}
Or, let $\{\cE_{\alpha_i}\}$ be a set consisting of symmetric matrices
($\cE^{\sT}_{\alpha_i}=\cE_{\alpha_i}$) and let $\{\cE_{\beta_j}\}$ be a set
consisting of skewsymmetric matrices ($\cE^{\sT}_{\beta_j}=-\cE_{\beta_j}$).
Then the transformation (\ref{com1}) may be rewritten in the following form:
\[
\cE_{\alpha_i}\longrightarrow\cE_{\alpha_i}=\sE\cE_{\alpha_i}\sE^{-1},\quad
\cE_{\beta_j}\longrightarrow\cE_{\beta_j}=-\sE\cE_{\beta_j}\sE^{-1}.
\]
Whence
\begin{equation}\label{commut}
\cE_{\alpha_i}\sE=\sE\cE_{\alpha_i},\quad \cE_{\beta_j}\sE=-\sE\cE_{\beta_j}.
\end{equation}
Thus, the matrix $\sE$ of the antiautomorphism $\cA\rightarrow\widetilde{\cA}$
commutes with a symmetric part of the spinbasis of the algebra $\cl_{p,q}$ and
anticommutes with a skewsymmetric part. An explicit form of the matrix $\sE$
in dependence on the type of the algebras $\cl_{p,q}$ will be found later, but
first let us define a general form of $\sE$, that is, let us show that for
the form of $\sE$ there are only two possibilities: 1) $\sE$ is a product of
symmetric matrices or 2) $\sE$ is a product of skewsymmetric matrices.
Let us prove this assertion another way: Let
$\sE=\cE_{\alpha_1}\cE_{\alpha_2}\cdots\cE_{\alpha_s}\cE_{\beta_1}\cE_{\beta_2}
\cdots\cE_{\beta_k}$ be a product of $s$ symmetric and $k$
skewsymmetric matrices. At this point $1< s+k\leq p+q$. The permutation
condition of the matrix $\sE$ with the symmetric basis matrices
$\cE_{\alpha_i}$ have a form
\begin{eqnarray}
\cE_{\alpha_i}\sE&=&
(-1)^{i-1}\sigma(\alpha_i)\cE_{\alpha_1}\cdots\cE_{\alpha_{i-1}}
\cE_{\alpha_{i+1}}\cdots\cE_{\alpha_s}\cE_{\beta_1}\cdots
\cE_{\beta_k},\nonumber\\
\sE\cE_{\alpha_i}&=&
(-1)^{k+s-i}\sigma(\alpha_i)\cE_{\alpha_1}\cdots\cE_{\alpha_{i-1}}
\cE_{\alpha_{i+1}}\cdots\cE_{\alpha_s}\cE_{\beta_1}\cdots
\cE_{\beta_k}.\label{commut1}
\end{eqnarray}
From here we obtain a comparison $k+s-i\equiv i-1\pmod{2}$, that is, at
$k+s\equiv 0\pmod{2}$ $\sE$ and $\cE_{\alpha_i}$ anticommute and at
$k+s\equiv 1\pmod{2}$ commute. Analogously, for the skewsymmetric part
we have
\begin{eqnarray}
\cE_{\beta_j}\sE&=&
(-1)^{s+j-1}\sigma(\beta_j)\cE_{\alpha_1}\cdots\cE_{\alpha_s}
\cE_{\beta_1}\cdots\cE_{\beta_{j-1}}\cE_{\beta_{j+1}}\cdots
\cE_{\beta_k},\nonumber\\
\sE\cE_{\beta_j}&=&
(-1)^{k-j}\sigma(\beta_i)\cE_{\alpha_1}\cdots\cE_{\alpha_s}
\cE_{\beta_1}\cdots\cE_{\beta_{j-1}}\cE_{\beta_{j+1}}\cdots
\cE_{\beta_k}.\label{commut2}
\end{eqnarray}
From the comparison $k-s\equiv 2j-1\pmod{2}$, it
follows that at $k-s\equiv 0\pmod{2}$,
$\sE$ and $\cE_{\beta_j}$ anticommute and at  $k-s\equiv 1\pmod{2}$
commute. Let $k+s=p+q$, then from (\ref{commut1}) we see that
at $p+q\equiv 0\pmod{2}$ $\sE$ and $\cE_{\alpha_i}$ anticommute, which is
inconsistent with (\ref{commut}). The case $p+q\equiv 1\pmod{2}$ is excluded,
since a dimensionality of $\cl_{p,q}$ is even (in the case of odd
dimensionality the algebra $\cl_{p+1,q}\,(\cl_{p,q+1})$ is isomorphic to
$\End_{\K\oplus\hat{\K}}(I_{p,q}\oplus\hat{I}_{p,q})\simeq\M_{2^m}(\K)\oplus
\M_{2^m}(\K)$, where $m=\frac{p+q}{2}$). Let suppose now that
$k+s<p+q$, that is, let us eliminate from the product $\sE$ one symmetric matrix,
then $k+s\equiv 1\pmod{2}$ and in virtue of (\ref{commut1}) the
matrices $\cE_{\alpha_i}$
that belong to $\sE$ commute with $\sE$, but the matrix that does not
belong to $\sE$ anticommutes with $\sE$. Thus, we came to a contradiction
with (\ref{commut}). It is obvious that elimination of two, three,
or more symmetric matrices from $\sE$ gives an analogous situation.
Now, Let us eliminate from
$\sE$ one skewsymmetric matrix, then $k+s\equiv 1\pmod{2}$ and
in virtue of (\ref{commut1}) $\sE$ and all $\cE_{\alpha_i}$ commute with each
other. Further, in virtue of (\ref{commut2}) the matrices $\cE_{\beta_j}$
that belong to the product $\sE$ commute with $\sE$, whereas the
the skewsymmetric matrix that does not belong to $\sE$ anticommute
with $\sE$. Therefore we came again to a contradiction with
(\ref{commut}). We come to an analogous situation if we eliminate two, three, or
more skewsymmetric matrices. Thus, the product
$\sE$ does not contain simultaneously symmetric and skewsymmetric matrices.
Hence it follows that the matrix of the antiautomorphism
$\cA\rightarrow\widetilde{\cA}$ is a product of only symmetric or only
skewsymmetric matrices.

Further, the matrix representations of the antiautomorphism
$\cA\rightarrow\widetilde{\cA^\star}$: 
$\widetilde{\sA^\star}=\sC\sA^{\sT}\sC^{-1}$ is defined in a similar manner.
First of all, since under action of the antiautomorphism $\widetilde{\star}$
we have $\e_i\rightarrow -\e_i$, in the matrix representation we must
demand $\cE_i\rightarrow -\cE_i$ also, or
\begin{equation}\label{com2}
\cE_i\longrightarrow -\cE_i=\sC\cE^{\sT}\sC^{-1}.
\end{equation}
Taking into account the symmetric $\{\cE_{\alpha_i}\}$ and the
skewsymmetric $\{\cE_{\beta_j}\}$ parts of the spinbasis we can write
the transformation (\ref{com2}) in the form
\[
\cE_{\alpha_i}\longrightarrow -\cE_{\alpha_i}=\sC\cE_{\alpha_i}\sC^{-1},\quad
\cE_{\beta_j}\longrightarrow\cE_{\beta_j}=\sC\cE_{\beta_j}\sC^{-1}.
\]
Hence it follows
\begin{equation}\label{commut3}
\sC\cE_{\alpha_i}=-\cE_{\alpha_i}\sC,\quad
\cE_{\beta_j}\sC=\sC\cE_{\beta_j}.
\end{equation}
Thus, in contrast with (\ref{commut}) the matrix $\sC$ of the antiautomorphism
$\widetilde{\star}$ anticommutes with the symmetric part of the spinbasis
of the algebra $\cl_{p,q}$ and commutes with the skewsymmetric part of the
same spinbasis. Further, in virtue of (\ref{auto17}) a matrix representation
of the automorphism $\star$ is defined as follows
\begin{equation}\label{W}
\sA^\star=\sW\sA\sW^{-1},
\end{equation}
where $\sW$ is a matrix representation of the volume element $\omega$.
The antiautomorphism $\cA\rightarrow\widetilde{\cA^\star}$, in turn, is
the composition of the antiautomorphism $\cA\rightarrow\widetilde{\cA}$ with
the automorphism $\cA\rightarrow\cA^\star$; therefore, from (\ref{com}) and
(\ref{W}) it follows (recall that the order of the composition of the
transformations (\ref{com}) and (\ref{W}) is not important, since
$\widetilde{\cA^\star}=(\widetilde{\cA})^\star=\widetilde{(\cA^\star)}$):
$\widetilde{\sA^\star}=\sW\sE\sA^{\sT}\sE^{-1}\sW^{-1}=\sE(\sW\sA\sW^{-1})^{\sT}
\sE^{-1}$, or
\begin{equation}\label{C}
\widetilde{\sA^\star}=(\sW\sE)\sA^{\sT}(\sW\sE)^{-1}=
(\sE\sW)\sA^{\sT}(\sE\sW)^{-1},
\end{equation}
since $\sW^{-1}=\sW^{\sT}$. Therefore, $\sC=\sE\sW$ or $\sC=\sW\sE$.
By this reason a general form of the matrix $\sC$ is similar to the form
of the matrix $\sE$, that is, $\sC$ is a product of symmetric or skewsymmetric
matrices only.

Let us consider in sequence definitions and permutation conditions of matrices
of the fundamental automorphisms (which are the elements of the groups
$\sAut(\cl_{p,q})$) for all eight types of the algebras
$\cl_{p,q}$, depending upon the division ring structure.\\[0.3cm]
1) The type $p-q\equiv 0\pmod{8}$, $\K\simeq\R$.\\
In this case according to Wedderburn--Artin theorem there is an isomorphism
$\cl_{p,q}\simeq\M_{2^m}(\R)$, where $m=\frac{p+q}{2}$. First, let consider
a case $p=q=m$. In the full matrix algebra
$\M_{2^m}(\R)$, in accordance with the signature of the algebra $\cl_{p,q}$ 
a choice of the symmetric and skewsymmetric matrices $\cE_i=\gamma(\e_i)$
is hardly fixed.
\begin{equation}\label{A1}
\cE^{\sT}_i=\left\{\begin{array}{rl}
\phantom{-}\cE_i, & \mbox{if $1\leq i\leq m$};\\
-\cE_i, & \mbox{if $m+1\leq i\leq 2m$},
\end{array}\right.
\end{equation}
That is, at this point the matrices of the first and second half of the basis
have a square $+\sI$ and $-\sI$, respectively.
Such a form of the basis (\ref{A1}) is explained by the following
reason: Over the field $\R$ there exist only symmetric matrices with a
square $+\sI$, and there exist no symmetric matrices with a square
$-\sI$. Inversely, skewsymmetric matrices over the field $\R$
only have the square $-\sI$.
Therefore, in this case the matrix of the antiautomorphism
$\cA\rightarrow\widetilde{\cA}$ is a product of $m$ symmetric matrices,
$\sE=\cE_1\cE_2\cdots\cE_m$, or is a product of $m$ skewsymmetric matrices,
$\sE=\cE_{m+1}\cE_{m+2}\cdots\cE_{2m}$. In accordance with (\ref{commut})
let us find permutation conditions of the matrix $\sE$ 
with the basis matrices $\cE_i$.
If $\sE=\cE_1\cE_2\cdots\cE_m$, and $\cE_i$ belong to the first half of the
basis (\ref{A1}), $1\leq i\leq m$, then
\begin{eqnarray}
\sE\cE_i&=&(-1)^{m-i}\cE_1\cE_2\cdots\cE_{i-1}\cE_{i+1}\cdots\cE_m,\nonumber\\
\cE_i\sE&=&(-1)^{i-1}\cE_1\cE_2\cdots\cE_{i-1}\cE_{i+1}\cdots\cE_m.\label{A2}
\end{eqnarray}
Therefore, we have a comparison $m-i\equiv i-1\pmod{2}$, whence
$m\equiv 2i-1\pmod{2}$. Thus, the matrix $\sE$ anticommutes
at $m\equiv 0\pmod{2}$ and commutes at $m\equiv 1\pmod{2}$
with the basis matrices $\cE_i$.
Further, let $\sE=\cE_1\cE_2\cdots\cE_m$, and  $\cE_i$ belong to the
second half of the basis $m+1\leq i\leq 2m$, then
\begin{equation}\label{A3}
\sE\cE_i=(-1)^m\cE_i\sE.
\end{equation}
Therefore at $m\equiv 0\pmod{2}$, $\sE$ commutes and at
$m\equiv 1\pmod{2}$ anticommutes with the matrices of the second half of the
basis.

Let now $\sE=\cE_{m+1}\cE_{m+2}\cdots\cE_{2m}$ be a product of
$m$ skewsymmetric matrices, then
\begin{equation}\label{A4}
\sE\cE_i=(-1)^m\cE_i\sE \quad (1\leq i\leq m)
\end{equation}
and
\begin{equation}\label{A4'}
\begin{array}{ccc}
\sE\cE_i&=&-(-1)^{m-i}\cE_{m+1}\cE_{m+2}\cdots\cE_{i-1}\cE_{i+1}\cdots\cE_{2m},\\
\cE_i\sE&=&-(-1)^{i-1}\cE_{m+1}\cE_{m+2}\cdots\cE_{i-1}\cE_{i+1}\cdots\cE_{2m},
\end{array}\quad (m+1\leq i\leq 2m)\end{equation}
that is, at $m\equiv 0\pmod{2}$ $\sE$ commutes with the matrices of the first
half of the basis (\ref{A1}) and anticommutes with the matrices of the second
half of (\ref{A1}).
At $m\equiv 1\pmod{2}$ $\sE$ anticommutes and commutes with the first and
the second half of the basis (\ref{A1}), respectively.

Let us find permutation conditions of the matrix $\sE$ with a matrix $\sW$
of the volume element (a matrix of the automorphism $\star$). Let
$\sE=\cE_1\cE_2\cdots\cE_m$, then
\begin{eqnarray}
\sE\sW&=&\cE_1\cE_2\cdots\cE_{m}\cE_1\cE_2\cdots\cE_{2m}
=(-1)^{\frac{m(m-1)}{2}}\cE_{m+1}\cE_{m+2}
\cdots\cE_{2m},\nonumber\\
\sW\sE&=&\cE_1\cE_2\cdots\cE_{2m}\cE_1\cE_2\cdots\cE_m=
(-1)^{\frac{m(3m-1)}{2}}\cE_{m+1}\cE_{m+2}\cdots\cE_{2m}.\label{A5}
\end{eqnarray}
Whence $\frac{m(3m-1)}{2}\equiv\frac{m(m-1)}{2}\pmod{2}$ and, therefore
at $m\equiv 0\pmod{2}$, $\sE$ and $\sW$ commute, and at $m\equiv 1\pmod{2}$
anticommute. It is easy to verify that analogous conditions take place if
$\sE=\cE_{m+1}\cE_{m+2}\cdots\cE_{2m}$ is the product of 
skewsymmetric matrices.

Since $\sC=\sE\sW$, then a matrix of the antiautomorphism
$\widetilde{\star}$ has a form $\sC=\cE_{m+1}\cE_{m+2}\cdots\cE_{2m}$ if
$\sE=\cE_1\cE_2\cdots\cE_m$ and correspondingly, $\sC=\cE_1\cE_2\cdots\cE_m$ if
$\sE=\cE_{m+1}\cE_{m+2}\cdots\cE_{2m}$. Therefore, permutation conditions
of the matrices $\sC$ and $\sW$ would be the same as that of
$\sE$ and $\sW$, that is, $\sC$ and $\sW$ commute if $m\equiv 0\pmod{2}$ and
anticommute if $m\equiv 1\pmod{2}$. It is easy to see that permutation
conditions of the matrix $\sC$ with the basis matrices $\cE_i$ are coincide
with (\ref{A2})--(\ref{A4'}).

Out of dependence on the choice of the matrices $\sE$ and $\sC$,
the permutation conditions between them in any of the two cases considered 
previously
are defined by the following relation
\begin{equation}\label{A6}
\sE\sC=(-1)^{m^2}\sC\sE,
\end{equation}
that is, the matrices $\sE$ and $\sC$ commute if $m\equiv 0\pmod{2}$ and
anticommute if $m\equiv 1\pmod{2}$.

Now, let us consider squares of the elements of the automorphism groups 
$\sAut(\cl_{p,q})$, $p-q\equiv 0\pmod{8}$, and $p=q=m$. For the matrices of
the automorphisms
$\widetilde{\phantom{cc}}$ and $\widetilde{\star}$ we have the following two
possibilities:\\
a) $\sE=\cE_1\cE_2\cdots\cE_m$, $\sC=\cE_{m+1}\cE_{m+2}\cdots\cE_{2m}$.
\begin{equation}\label{A7}
\sE^2=\left\{\begin{array}{rl}
+\sI, & \mbox{if $m\equiv 0,1\pmod{4}$},\\
-\sI, & \mbox{if $m\equiv 2,3\pmod{4}$};
\end{array}\right.\quad
\sC^2=\left\{\begin{array}{rl}
+\sI, & \mbox{if $m\equiv 0,3\pmod{4}$},\\
-\sI, & \mbox{if $m\equiv 1,2\pmod{4}$}.
\end{array}\right.
\end{equation}
b) $\sE=\cE_{m+1}\cE_{m+2}\cdots\cE_{2m}$, $\sC=\cE_1\cE_2\cdots\cE_m$.
\begin{equation}\label{A8}
\sE^2=\left\{\begin{array}{rl}
+\sI, & \mbox{if $m\equiv 0,3\pmod{4}$},\\
-\sI, & \mbox{if $m\equiv 1,2\pmod{4}$};
\end{array}\right.\quad
\sC^2=\left\{\begin{array}{rl}
+\sI, & \mbox{if $m\equiv 0,1\pmod{4}$},\\
-\sI, & \mbox{if $m\equiv 2,3\pmod{4}$}.
\end{array}\right.
\end{equation}
In virtue of (\ref{e3}), for the matrix of the automorphism $\star$ we have
always $\sW^2=+\sI$.

Now, we are in a position to define automorphism groups
for the type $p-q\equiv 0\pmod{8}$. First of all, let us consider Abelian groups.
In accordance with (\ref{A5}) and
(\ref{A6}), the automorphism group is Abelian if $m\equiv 0\pmod{2}$
($\sW,\sE$ and $\sC$ commute with each other). In virtue of (\ref{commut}) and
(\ref{A1}) the matrix $\sE$ should be commuted with the first (symmetric)
half and anticommuted with the second (skewsymmetric) half of the basis
(\ref{A1}). From (\ref{A2})--(\ref{A4'}) it is easy to see that this
condition is satisfied only if $\sE=\cE_{m+1}\cE_{m+2}\cdots\cE_{2m}$ and
$m\equiv 0\pmod{2}$. Correspondingly, in accordance with (\ref{commut3}) the
matrix $\sC$ should be anticommuted with the symmetric half of the basis
(\ref{A1}) and commuted with the skewsymmetric half of the same basis.
It is obvious that this condition is satisfied only if $\sC=\cE_1\cE_2\cdots
\cE_m$.
Therefore, when $m=p=q$ in accordance with
(\ref{A8}), there exists an Abelian group $\sAut_-(\cl_{p,q})\simeq
\dZ_2\otimes\dZ_2$ with the signature $(+,+,+)$ if $p,q\equiv 0\pmod{4}$,
and $\sAut_-(\cl_{p,q})\simeq\dZ_4$ 
with the signature $(+,-,-)$
if $p,q\equiv 2\pmod{4}$. Further, in accordance with (\ref{A5}) and (\ref{A6}),
the automorphism group is non--Abelian if $m\equiv 1\pmod{2}$.
In this case, from (\ref{A2})--(\ref{A4'}) it follows that the matrix $\sE$
commutes with the symmetric half and anticommutes with the skewsymmetric half
of the basis (\ref{A1}) if and only if $\sE=\cE_1\cE_2\cdots\cE_m$
is a product of $m$ symmetric matrices, $m\equiv 1\pmod{8}$. In its turn,
the matrix $\sC$ anticommutes with the symmetric half and commutes with the
skewsymmetric half of the basis (\ref{A1}) if and only if $\sC=\cE_{m+1}
\cE_{m+2}\cdots\cE_{2m}$.
Therefore,
in accordance with (\ref{A7}), there exist non--Abelian groups
$\sAut_+(\cl_{p,q})\simeq D_4/\dZ_2$
with the signature $(+,-,+)$ if $p,q\equiv 3\pmod{4}$, and 
$\sAut_+(\cl_{p,q})\simeq D_4/\dZ_2$ with the signature $(+,+,-)$ if
$p,q\equiv 1\pmod{4}$.

In addition to the previously considered case $p=q$, the type
$p-q\equiv 0\pmod{8}$ also admits two particular cases in relation with the
algebras 
$\cl_{p,0}$ and $\cl_{0,q}$. In these cases, a spinbasis is defined as follows
\[
\begin{array}{ccc}
\cE^{\sT}_i&=&\phantom{-}\cE_i\quad\mbox{for the algebras $\cl_{8t,0}$},\\
\cE^{\sT}_i&=&-\cE_i\quad\mbox{for the algebras $\cl_{0,8t}$}.
\end{array}\quad t=1,2,\ldots
\]
that is, a spinbasis of the algebra $\cl_{8t,0}$ consists of only symmetric
matrices, and that of $\cl_{0,8t}$ consists of only
skewsymmetric matrices. According to (\ref{commut}), for the algebra
$\cl_{p,0}$ the matrix $\sE$ should commute with all
$\cE_i$. It is obvious that we cannot take the matrix
$\sE$ of the form $\cE_1\cE_2\cdots\cE_s$, where $1<s<p$, since at
$s\equiv 0\pmod{2}$ $\sE$ and $\cE_i$ anticommute, which contradicts with
(\ref{commut}), and at $s\equiv 1\pmod{2}$ $\sE$ and $\cE_i$ that belong to
$\sE$ commute with each other, whereas $\cE_i$ that do not belong
to $\sE$
anticommute with $\sE$, which again conradicts with (\ref{commut}).
The case $s=p$ is also excluded, since $p$ is even. Therefore, only one
possibility remains, that is, the matrix $\sE$ is proportional to the
unit matrix, $\sE\sim\sI$. At this point, from (\ref{C}) it follows that
$\sC\sim\cE_1\cE_2\cdots\cE_p$ and we see that the conditions (\ref{commut3})
are satisfied.
Thus, the matrices
$\sE\sim\sI$, $\sC=\sE\sW$ and $\sW$ of the fundamental automorphisms
$\cA\rightarrow\widetilde{\cA},\,\cA\rightarrow\widetilde{\cA^\star}$ and
$\cA\rightarrow\cA^\star$ of the algebra $\cl_{p,0}$ ($p\equiv 0\pmod{8}$)
from an Abelian group $\sAut_-(\cl_{p,0})\simeq\dZ_2\otimes
\dZ_2$. Further, for the algebras $\cl_{0,8t}$, in accordance with (\ref{commut})
the matrix $\sE$ should anticommute with all $\cE_i$. It is easy to
see that we cannot take the matrix $\sE$ of the form $\cE_1\cE_2\cdots\cE_k$, 
where
$1<k<q$, since at $k\equiv 0\pmod{2}$ the matrix $\sE$ and the matrices $\cE_i$
that belong to $\sE$ anticommute with each other, whereas
$\cE_i$ that do not belong to $\sE$ commute with $\sE$, which
contradicts with
(\ref{commut}). Inversely, if $k\equiv 1\pmod{2}$, $\sE$ and $\cE_i$
that belong to $\sE$ commute, but $\sE$ and $\cE_i$ that do not 
belong to $\sE$ anticommute,
which also contradicts with (\ref{commut}). It is obvious that in this case 
$\sE\sim\sI$ is excluded; therefore, $\sE\sim\cE_1\cE_2\cdots\cE_q$.
In this case, according to (\ref{commut3}) the matrix $\sC$ is proportional
to the unit matrix.
Thus, the matrices $\sE\sim\cE_1\cE_2\cdots\cE_q$, $\sC=\sE\sW\sim\sI$ and
$\sW$ of the automorphisms $\cA\rightarrow\widetilde{\cA},\,
\cA\rightarrow\widetilde{\cA^\star}$ and $\cA\rightarrow\cA^\star$ of the
algebra $\cl_{0,q}$ ($q\equiv 0\pmod{8}$) from the group
$\sAut_-(\cl_{0,q})\simeq\dZ_2\otimes\dZ_2$.\\[0.3cm]
2) The type $p-q\equiv 2\pmod{8}$, $\K\simeq\R$.\\
In virtue of the isomorphism $\cl_{p,q}\simeq\M_{2^{\frac{p+q}{2}}}(\R)$ for
the type $p-q\equiv 2\pmod{8}$ in accordance with the signature of the
algebra $\cl_{p,q}$, we have the following basis
\begin{equation}\label{B1}
\cE^{\sT}_i=\left\{\begin{array}{rl}
\phantom{-}\cE_i, & \mbox{if $1\leq i\leq p$},\\
-\cE_i, & \mbox{if $p+1\leq i\leq p+q$}.
\end{array}\right.
\end{equation}
Therefore, in this case the matrix of the antiautomorphism 
$\widetilde{\phantom{cc}}$ is a product of
$p$ symmetric matrices, $\sE=\cE_1\cE_2\cdots\cE_p$
or is a product of $q$ skewsymmetric matrices, $\sE=\cE_{p+1}\cE_{p+2}
\cdots\cE_{p+q}$. Let us find permutation conditions of the matrix $\sE$ with the
basis matrices $\cE_i$. Let $\sE=\cE_1\cE_2\cdots\cE_p$, then
\begin{equation}\label{B2}
\begin{array}{ccc}
\sE\cE_i&=&(-1)^{p-i}\cE_1\cE_2\cdots\cE_{i-1}\cE_{i+1}\cdots\cE_p,\\
\cE_i\sE&=&(-1)^{i-1}\cE_1\cE_2\cdots\cE_{i-1}\cE_{i+1}\cdots\cE_p
\end{array}\quad(1\leq i\leq p)
\end{equation}
and
\begin{equation}\label{B3}
\sE\cE_i=(-1)^p\cE_i\sE, \quad (p+1\leq i\leq p+q)
\end{equation}
that is, at $p\equiv 0\pmod{2}$ the matrix $\sE$  anticommutes with the 
symmetric and commutes with the skewsymmetric part of the basis
(\ref{B1}). Correspondingly, at $p\equiv 1\pmod{2}$ $\sE$ commutes with the
symmetric and anticommutes with the skewsymmetric part of the basis (\ref{B1}).

Analogously, let $\sE=\cE_{p+1}\cE_{p+2}\cdots\cE_{p+q}$, then
\begin{equation}\label{B4}
\sE\cE_i=(-1)^q\cE_i\sE\quad (1\leq i\leq p)
\end{equation}
and
\begin{equation}\label{B5}
\begin{array}{ccc}
\sE\cE_i&=&-(-1)^{q-i}\cE_{p+1}\cE_{p+2}\cdots\cE_{i-1}\cE_{i+1}\cdots
\cE_{p+q};\\
\cE_i\sE&=&-(-1)^{i-1}\cE_{p+1}\cE_{p+2}\cdots\cE_{i-1}\cE_{i+1}\cdots
\cE_{p+q},
\end{array}\quad(p+1\leq i\leq p+q)
\end{equation}
that is, at $q\equiv 0\pmod{2}$ the matrix $\sE$ commutes with the symmetric
and anticommutes with the skewsymmetric part of the basis (\ref{B1}).
Correspondingly, at $q\equiv 1\pmod{2}$ $\sE$ anticommutes with the symmetric
and commutes with the skewsymmetric part of (\ref{B1}).

Further, permutation conditions of the matrices $\sE=\cE_1\cE_2
\cdots\cE_p$ and $\sW$ are defined by the following relations:
\begin{eqnarray}
\sE\sW&=&\cE_1\cE_2\cdots\cE_p\cE_1\cE_2\cdots\cE_{p+q}=(-1)^{\frac{p(p-1)}{2}}
\cE_{p+1}\cE_{p+2}\cdots\cE_{p+q},\nonumber\\
\sW\sE&=&\cE_1\cE_2\cdots\cE_{p+q}\cE_1\cE_2\cdots\cE_p=
(-1)^{\frac{p(p-1)}{2}+pq}\cE_{p+1}\cE_{p+2}\cdots\cE_{p+q}.\label{B6}
\end{eqnarray}
From a comparison $\frac{p(p-1)}{2}+pq\equiv\frac{p(p-1)}{2}\pmod{2}$ it follows
that the matrices $\sE$ and $\sW$ commute with each other if
$pq\equiv 0\pmod{2}$ and anticommute if
$pq\equiv 1\pmod{2}$. If we take $\sE=\cE_{p+1}\cE_{p+2}\cdots
\cE_{p+q}$, then the relations
\begin{eqnarray}
\sE\sW&=&\cE_{p+1}\cE_{p+2}\cdots\cE_{p+q}\cE_1\cE_2\cdots\cE_{p+q}=
(-1)^{\frac{q(q+1)}{2}+pq}\cE_1\cE_2\cdots\cE_p,\nonumber\\
\sW\sE&=&\cE_1\cE_2\cdots\cE_{p+q}\cE_{p+1}\cE_{p+2}\cdots\cE_{p+q}=
(-1)^{\frac{q(q+1)}{2}}\cE_1\cE_2\cdots\cE_p\label{B7}
\end{eqnarray}
give analogous permutation conditions for $\sE$ and $\sW$
($pq\equiv 0,1\pmod{2}$). It is obvious that permutation conditions of
$\sC$ (the matrix of the antiautomorphism $\widetilde{\star}$) with the
basis matrices $\cE_i$ and
with $\sW$ 
are analogous to the conditions (\ref{B2})--(\ref{B5}) and
(\ref{B6})--(\ref{B7}), respectively.

Out of dependence on the choice of the matrices $\sE$ and $\sC$,
permutation conditions between them are defined by a relation
\begin{equation}\label{B8}
\sE\sC=(-1)^{pq}\sC\sE,
\end{equation}
that is, $\sE$ and $\sC$ commute if $pq\equiv 0\pmod{2}$ and anticommute
if $pq\equiv 1\pmod{2}$.

For the squares of the automorphisms $\widetilde{\phantom{cc}}$ and
$\widetilde{\star}$ we have following two possibilities:\\[0.2cm]
a) $\sE=\cE_1\cE_2\cdots\cE_p$, $\sC=\cE_{p+1}\cE_{p+2}\cdots\cE_{p+q}$.
\begin{equation}\label{B9}
\sE^2=\left\{\begin{array}{rl}
+\sI, & \mbox{if $p\equiv 0,1\pmod{4}$};\\
-\sI, & \mbox{if $p\equiv 2,3\pmod{4}$},
\end{array}\right.\quad
\sC^2=\left\{\begin{array}{rl}
+\sI, & \mbox{if $q\equiv 0,3\pmod{4}$};\\
-\sI, & \mbox{if $q\equiv 1,2\pmod{4}$}.
\end{array}\right.
\end{equation}
b) $\sE=\cE_{p+1}\cE_{p+2}\cdots\cE_{p+q}$, $\sC=\cE_1\cE_2\cdots\cE_p$.
\begin{equation}\label{B10}
\sE^2=\left\{\begin{array}{rl}
+\sI, & \mbox{if $q\equiv 0,3\pmod{4}$};\\
-\sI, & \mbox{if $q\equiv 1,2\pmod{4}$},
\end{array}\right.\quad
\sC^2=\left\{\begin{array}{rl}
+\sI, & \mbox{if $p\equiv 0,1\pmod{4}$};\\
-\sI, & \mbox{if $p\equiv 2,3\pmod{4}$}.
\end{array}\right.
\end{equation}
For the type $p-q\equiv 2\pmod{8}$ in virtue of (\ref{e3}) a square of the
matrix $\sW$ is always equal to $-\sI$.

Now, let us consider automorphism groups for the type
$p-q\equiv 2\pmod{8}$. In accordance with (\ref{B6})--(\ref{B8}), the
automorphism group $\sAut(\cl_{p,q})$ is Abelian if $pq\equiv 0\pmod{2}$.
Further, in virtue of (\ref{commut}) and (\ref{B1}) the matrix of the
antiautomorphism
$\widetilde{\phantom{cc}}$ should commute with the symmetric part of the
basis (\ref{B1}) and anticommute with the skewsymmetric part of the same
basis. From (\ref{B2})--(\ref{B5}), it is easy to see that
this condition is satisfied at $pq\equiv 0\pmod{2}$ if and only if
$\sE=\cE_{p+1}\cE_{p+2}\cdots
\cE_{p+q}$ is a product of $q$ skewsymmetric matrices (recall that for the
type $p-q\equiv 2\pmod{8}$, the numbers $p$ and $q$ are both even or both
odd). Correspondingly, in accordance with (\ref{commut3}), the matrix $\sC$
should anticommute with the symmetric part of the basis (\ref{B1}) and
commute with skewsymmetric part of the same basis. It is obvious that this
requirement is satisfied if and only if $\sC=\cE_1\cE_2\cdots\cE_p$ is a
product of $p$ symmetric matrices.
Thus, in accordance with (\ref{B10}), there exist Abelian groups
$\sAut_-(\cl_{p,q})\simeq\dZ_4$ with the signature $(-,-,+)$ if
$p\equiv 0\pmod{4}$ and $q\equiv 2\pmod{4}$, and 
with the signature $(-,+,-)$ if
$p\equiv 2\pmod{4}$ and $q\equiv 0\pmod{4}$. Further, according to
(\ref{B6})--(\ref{B8}), the automorphism group is non--Abelian if
$pq\equiv 1\pmod{2}$. In this case, from (\ref{B2})--(\ref{B5}) it follows that
the matrix of the antiautomorphism $\widetilde{\phantom{cc}}$ commutes with
the symmetric part of the basis (\ref{B1}) and anticommutes with the
skewsymmetric part if and only if $\sE=\cE_1\cE_2\cdots\cE_p$ is a product
of $p$ symmetric matrices. In its turn, the matrix $\sC$ anticommutes with
the symmetric part of the basis (\ref{B1}) and commutes with the skewsymmetric
part of the same basis if and only if $\sC=\cE_{p+1}\cE_{p+2}\cdots\cE_{p+q}$.
Therefore in accordance with
(\ref{B9}), there exists a non--Abelian group 
$\sAut_+(\cl_{p,q})\simeq Q_4/\dZ_2$ with the signature
$(-,-,-)$ if $p\equiv 3\pmod{4}$ and $q\equiv 1\pmod{4}$ and 
$\sAut_+(\cl_{p,q})\simeq D_4/\dZ_2$ with the
signature
$(-,+,+)$ if $p\equiv 1\pmod{4}$ and $q\equiv 3\pmod{4}$.\\[0.3cm]
3) The type $p-q\equiv 6\pmod{8}$, $\K\simeq\BH$.\\
First of all, over the ring $\K\simeq\BH$ there exists no fixed basis of
the form (\ref{A1}) or (\ref{B1}) for the matrices $\cE_i$. In general,
a number of the skewsymmetric matrices does not coincide with a number of
matrices with the negative square ($\cE^2_j=-\sI$) as it takes place for the
types $p-q\equiv 0,2\pmod{8}$. Thus, the matrix 
$\sE$ is a product of skewsymmetric matrices $\cE_j$, among which there are
matrices with positive and negative squares, or $\sE$ is a product of
symmetric matrices
$\cE_i$, among which also there are matrices with '$+$' and '$-$' squares.
Let $k$ be a number of the skewsymmetric matrices $\cE_j$ of a spinbasis of
the algebra
$\cl_{p,q}$, $0\leq k\leq p+q$. Among the matrices $\cE_j$, $l$ have 
'$+$'-square
and $t$ matrices have '$-$'-square. Let $0<k<p+q$ and let
$\sE=\cE_{j_1}\cE_{j_2}\cdots\cE_{j_k}$ be a matrix of the antiautomorphism
$\cA\rightarrow\widetilde{\cA}$, then permutation conditions of the matrix
$\sE$ with the matrices $\cE_{i_r}$ of the symmetric part
($0<r\leq p+q-k$) and with the matrices $\cE_{j_u}$ of the skewsymmetric part
($0<u\leq k$) have the respective form
\begin{equation}\label{C1}
\sE\cE_{i_r}=(-1)^k\cE_{i_r}\sE\quad (0<r\leq p+q-k),
\end{equation}
\begin{equation}\label{C2}
\begin{array}{ccc}
\sE\cE_{j_u}&=&(-1)^{k-u}\sigma(j_u)\cE_{j_1}\cE_{j_2}\cdots\cE_{j_{u-1}}
\cE_{j_{u+1}}\cdots\cE_{j_k},\\
\cE_{j_u}\sE&=&(-1)^{u-1}\sigma(j_u)\cE_{j_1}\cE_{j_2}\cdots\cE_{j_{u-1}}
\cE_{j_{u+1}}\cdots\cE_{j_k},
\end{array}\quad (0<u\leq k)
\end{equation}
that is, at $k\equiv 0\pmod{2}$ the matrix $\sE$ commutes with the symmetric part
and anticommutes with the skewsymmetric part of the spinbasis. Correspondingly,
at $k\equiv 1\pmod{2}$, $\sE$ anticommutes with the symmetric and commutes
with the skewsymmetric part. Further, let $\sE=\cE_{i_1}\cE_{i_2}
\cdots\cE_{i_{p+q-k}}$ be a product of the symmetric matrices, then
\begin{equation}\label{C3}
\begin{array}{ccc}
\sE\cE_{i_r}&=&(-1)^{p+q-k}\sigma(i_r)\cE_{i_1}\cE_{i_2}\cdots\cE_{i_{r-1}}
\cE_{i_{r+1}}\cdots\cE_{i_{p+q-k}},\\
\cE_{i_r}\sE&=&(-1)^{r-1}\sigma(i_r)\cE_{i_1}\cE_{i_2}\cdots\cE_{i_{r-1}}
\cE_{i_{r+1}}\cdots\cE_{i_{p+q-k}},
\end{array}\quad (0<r\leq p+q-k)
\end{equation}
\begin{equation}\label{C4}
\sE\cE_{j_u}=(-1)^{p+q-k}\cE_{j_u}\sE,\quad (0<u\leq k)
\end{equation}
that is, at $p+q-k\equiv 0\pmod{2}$ the matrix $\sE$ anticommutes with the
symmetric part and commutes with the skewsymmetric part of the spinbasis.
Correspondingly, at $p+q-k\equiv 1\pmod{2}$ $\sE$ commutes with the symmetric
part and anticommutes with the skewsymmetric part. It is easy to see that
permutations conditions of the matrix $\sC$ with the basis matrices $\cE_i$
are coincide with (\ref{C1})--(\ref{C4}).

For the permutation conditions of the matrices $\sW=\cE_{i_1}\cE_{i_2}\cdots
\cE_{i_{p+q-k}}\cE_{j_1}\cE_{j_2}\cdots\cE_{j_k}$, $\sE=\cE_{j_1}\cE_{j_2}
\cdots\cE_{j_k}$, and $\sC=\cE_{i_1}\cE_{i_2}\cdots\cE_{i_{p+q-k}}$ we have
\begin{eqnarray}
\sE\sW&=&(-1)^{\frac{k(k-1)}{2}+t+k(p+q-k)}\cE_{i_1}\cE_{i_2}\cdots
\cE_{i_{p+q-k}},\nonumber\\
\sW\sE&=&(-1)^{\frac{k(k-1)}{2}+t}\cE_{i_1}\cE_{i_2}\cdots\cE_{i_{p+q-k}}.
\label{C5}
\end{eqnarray}
\begin{equation}\label{C6}
\sE\sC=(-1)^{k(p+q-k}\sC\sE.
\end{equation}
Hence it follows that the matrices $\sW,\,\sE$, and $\sC$ commute at
$k(p+q-k)\equiv 0\pmod{2}$ and anticommute at $k(p+q-k)\equiv 1\pmod{2}$.
It is easy to verify that permutation conditions for the matrices
$\sE=\cE_{i_1}\cE_{i_2}\cdots\cE_{i_{p+q-k}}$, $\sC=\cE_{j_1}\cE_{j_2}\cdots
\cE_{j_k}$ would be the same.

In accordance with (\ref{commut}), (\ref{commut3}),
(\ref{C1})--(\ref{C4}), and also with
(\ref{C5})--(\ref{C6}), the Abelian automorphism groups for the type
$p-q\equiv 6\pmod{8}$ exist only if $\sE=\cE_{j_1}\cE_{j_2}\cdots
\cE_{j_k}$ and $\sC=\cE_{i_1}\cE_{i_2}\cdots\cE_{i_{p+q-k}}$, 
$k\equiv 0\pmod{2}$. Let
$l$ and $t$ be the quantities of the matrices in the
product $\cE_{j_1}\cE_{j_2}\cdots
\cE_{j_k}$, which have '$+$' and '$-$'-squares, respectively,
and also let $h$ and $g$ be the quantities of the matrices with the same
meaning in the product $\cE_{i_1}\cE_{i_2}
\cdots\cE_{i_{p+q-k}}$. Then, the group $\sAut_-(\cl_{p,q})\simeq\dZ_4$
with the signature $(-,+,-)$ exists if $l-t\equiv 0,1,4,5\pmod{8}$ and
$h-g\equiv 2,3,6,7\pmod{8}$ (recall that for the type $p-q\equiv 6\pmod{8}$
we have 
$\sW^2=-\sI$), and also, the group $\sAut_-(\cl_{p,q})\simeq\dZ_4$ with the
signature
$(-,-,+)$ exists if $l-t\equiv 2,3,6,7\pmod{8}$ and $h-g\equiv 0,1,4,5
\pmod{8}$. Further, from (\ref{commut}), (\ref{commut3}),
and (\ref{C1})--(\ref{C6}), it follows
that the non--Abelian automorphism groups exist only if
$\sE=\cE_{i_1}\cE_{i_2}\cdots
\cE_{i_{p+q-k}}$ and $\sC=\cE_{j_1}\cE_{j_2}\cdots\cE_{j_k}$,
$k\equiv 1\pmod{2}$. At this point the group
$\sAut_+(\cl_{p,q})\simeq Q_4/\dZ_2$ with the signature $(-,-,-)$
exists if $h-g\equiv 2,3,6,7\pmod{8}$ and $l-t\equiv 2,3,6,7\pmod{8}$.
Correspondingly, the group $\sAut_+(\cl_{p,q})\simeq D_4/\dZ_2$ with the
signature
$(-,+,+)$ exists if $h-g\equiv 0,1,4,5\pmod{8}$ and $l-t\equiv 0,1,4,5
\pmod{8}$. In absence of the skewsymmetric matrices, $k=0$, the spinbasis
of $\cl_{p,q}$ contains only symmetric matrices. In this case, from
(\ref{commut}), it follows that the matrix of the antiautomorphism $\cA
\rightarrow\widetilde{\cA}$ should commute with all the basis matrices.
It is obvious that this condition is satisfied if and only if $\sE$ is
proportional to the unit matrix. At this point, from (\ref{C}), it follows
that $\sC\sim\cE_1\cE_2\cdots\cE_{p+q}$ and we see that condition
(\ref{commut3}) is satisfied.
Thus, we have the Abelian group
$\sAut_-(\cl_{p,q})\simeq\dZ_4$ with the signature $(-,+,-)$. In other
degenerate case $k=p+q$, the spinbasis of $\cl_{p,q}$ contains only
skewsymmetric matrices; therefore, the matrix $\sE$ should  
anticommute with all the basis matrices. This condition is satisfied
if and only if $\sE\sim\cE_1\cE_2\cdots\cE_{p+q}$. In its turn, the matrix
$\sC$ commutes with all the basis matrices if and only if $\sC\sim\sI$.
It is easy to see
that in this case we have the group $\sAut_-(\cl_{p,q})\simeq\dZ_4$
with the signature $(-,-,+)$.\\[0.3cm]
4) The type $p-q\equiv 4\pmod{8}$, $\K\simeq\BH$.\\
It is obvious that a proof for this type is analogous to the case
$p-q\equiv 6\pmod{8}$, where also $\K\simeq\BH$. For the
type $p-q\equiv 4\pmod{8}$
we have $\sW^2=+\sI$. As well as for the type $p-q\equiv 6\pmod{8}$,
the Abelian groups exist only if $\sE=\cE_{j_1}\cE_{j_2}
\cdots\cE_{j_k}$ and $\sC=\cE_{i_1}\cE_{i_2}\cdots\cE_{i_{p+q-k}}$,
$k\equiv 0\pmod{2}$. At this point the group
$\sAut_-(\cl_{p,q})\simeq\dZ_2\otimes\dZ_2$ with $(+,+,+)$ exists if
$l-t,h-g\equiv 0,1,4,5\pmod{8}$, and also the group $\sAut_-(\cl_{p,q})
\simeq\dZ_4$ with $(+,-,-)$ exists if $l-t,h-g\equiv 2,3,6,7\pmod{8}$.
Correspondingly, the non--Abelian group exist only if $\sE$ is a product
of $k$ skewsymmetric matrices and $\sC$ is a product of $p+q-k$ symmetric
matrices,
$k\equiv 1\pmod{2}$.
The group $\sAut_+(\cl_{p,q})\simeq D_4/\dZ_2$ with $(+,-,+)$
exists if $h-g\equiv 2,3,6,7\pmod{8}$, $l-t\equiv 0,1,4,5\pmod{8}$,
and the group $\sAut_+(\cl_{p,q})\simeq D_4/\dZ_2$ with $(+,+,-)$ exists
if $h-g\equiv 0,1,4,5\pmod{8}$, $l-t\equiv 2,3,6,7\pmod{8}$.
For the type $p-q\equiv 4\pmod{8}$ both the degenerate cases $k=0$ and
$k=p+q$ give rise to the group 
$\sAut_-(\cl_{p,q})\simeq\dZ_2\otimes\dZ_2$.\\[0.3cm]
5) The type $p-q\equiv 1\pmod{8}$, $\K\simeq\R\oplus\R$.\\
In this case a dimensionality  $p+q$ is odd and the algebra $\cl_{p,q}$ is
semi--simple. Over the ring $\K\simeq\R\oplus\R$ the algebras of this type
decompose into a direct sum of two subalgebras with even dimensionality.
At this point there exist two types of decomposition \cite{Rash,Port69}:
\begin{eqnarray}
\cl_{p,q}&\simeq&\cl_{p,q-1}\oplus\cl_{p,q-1},\label{D1}\\
\cl_{p,q}&\simeq&\cl_{q,p-1}\oplus\cl_{q,p-1},\label{D1'}
\end{eqnarray}
where each algebra $\cl_{p,q-1}$ ($\cl_{q,p-1}$) is obtained by means
of either of the two central idempotents 
$\frac{1}{2}(1\pm\e_1\e_2\ldots\e_{p+q})$
and isomorphisms
\begin{eqnarray}
\cl^+_{p,q}&\simeq&\cl_{p,q-1},\label{D2'}\\
\cl^+_{p,q}&\simeq&\cl_{q,p-1}.\label{D2}
\end{eqnarray}
In general, the structure of the ring $\K\simeq\R\oplus\R$ in virtue of
the decompositions
(\ref{D1})--(\ref{D1'}) and isomorphisms  (\ref{D2'})--(\ref{D2}) 
admits all eight kinds
of the automorphism groups, since the subalgebras in the direct sums
(\ref{D1})--(\ref{D1'}) have the type $p-q\equiv 2\pmod{8}$ or the type
$p-q\equiv 0\pmod{8}$. More precisely, for the algebras $\cl_{0,q}$ of the type
$p-q\equiv 1\pmod{8}$, the subalgebras in the direct sum (\ref{D1}) have the
type $p-q\equiv 2\pmod{8}$ and only this type; therefore, in accordance with
previously obtained conditions for the type $p-q\equiv 2\pmod{8}$, we have
four and only four kinds of the automorphism groups with the signatures
$(-,-,+),\,(-,+,-)$ and $(-,-,-),\,(-,+,+)$.
Further, for the algebra $\cl_{p,0}$ ($p-q\equiv 1\pmod{8}$) the subalgebras
in the direct sum (\ref{D1'}) have the type $p-q\equiv 0\pmod{8}$; therefore,
in this case there exist four and only four kinds of the automorphism groups
with the signatures $(+,+,+),\,(+,-,-)$ and $(+,-,+),\,(+,+,-)$. In general
case, $\cl_{p,q}$, the type $p-q\equiv 1\pmod{8}$ admits all eight kinds of the
automorphism groups.\\[0.3cm]
6) The type $p-q\equiv 5\pmod{8}$, $\K\simeq\BH\oplus\BH$.\\
In this case the algebra $\cl_{p,q}$ is also semi--simple and, therefore,
we have decompositions of the form (\ref{D1})--(\ref{D1'}). By analogy with
the type $p-q\equiv 1\pmod{8}$, a structure of the double quaternionic ring
$\K\simeq\BH\oplus\BH$ in virtue of the decompositions (\ref{D1})--(\ref{D1'}) 
and isomorphisms (\ref{D2'})--(\ref{D2}) 
is also admits, in a general case, all eight kinds
of the automorphism groups, since the subalgebras in the direct sums
(\ref{D1})--(\ref{D1'}) have the type $p-q\equiv 6\pmod{8}$ or the type
$p-q\equiv 4\pmod{8}$. More precisely, for the algebras $\cl_{0,q}$ of the type
$p-q\equiv 5\pmod{8}$, the subalgebras in the direct sum (\ref{D1}) have the
type $p-q\equiv 6\pmod{8}$ and only this type; therefore, in accordance with
previously obtained results for the quaternionic rings we have four and only
four kinds of the automorphism groups with the signatures $(-,+,-),\,(-,-,+)$ 
and $(-,-,-),\,(-,+,+)$. Analogously, for the algebras $\cl_{p,0}$ 
($p-q\equiv 5\pmod{8}$), the subalgebras in the direct sum (\ref{D1'}) have the
type $p-q\equiv 4\pmod{8}$; therefore, in this case there exist four and only
four kinds of the automorphism groups with the signatures
$(+,+,+),\,(+,-,-)$ and $(+,-,+),\,(+,+,-)$. In general case, $\cl_{p,q}$,
the type $p-q\equiv 5\pmod{8}$ admits all eight kinds of the automorphism
groups.\\[0.3cm]
7) The type $p-q\equiv 3\pmod{8}$, $\K\simeq\C$.\\
For this type a center $\bZ$ of the algebra $\cl_{p,q}$ consists of the unit
and the volume element $\omega=\e_1\e_2\ldots\e_{p+q}$, since $p+q$ is odd
and the element $\omega$ commutes with all the basis elements of the algebra
$\cl_{p,q}$. Moreover,
$\omega^2=-1$, hence it follows that $\bZ\simeq\R\oplus i\R$. Thus, for the
algebras $\cl_{p,q}$ of the type $p-q\equiv 3\pmod{8}$, there exists an
isomorphism
\begin{equation}\label{E1}
\cl_{p,q}\simeq\C_{n-1},
\end{equation}
where $n=p+q$. It is easy to see that the algebra $\C_{n-1}=\C_{2m}$ in 
(\ref{E1})
is a complex algebra with even dimensionality, where $m$ is either even or
odd. More precisely, the number $m$ is even if $p\equiv 0\pmod{2}$ and
$q\equiv 1\pmod{2}$, and odd if $p\equiv 1\pmod{2}$ and
$q\equiv 0\pmod{2}$. In accordance with Theorem \ref{taut} 
at $m\equiv 0\pmod{2}$
the algebra $\C_{2m}$ admits the Abelian group $\sAut_-(\C_{2m})\simeq
\dZ_2\otimes\dZ_2$ with $(+,+,+)$, and at $m\equiv 1\pmod{2}$
the non--Abelian group $\sAut_+(\C_{2m})\simeq Q_4/\dZ_2$ with $(-,-,-)$.
Hence it follows the statement of the theorem for this type.\\[0.3cm]
8) The type $p-q\equiv 7\pmod{8}$, $\K\simeq\C$.\\
It is obvious that for this type the isomorphism (\ref{E1}) also takes
place.
Therefore, the type $p-q\equiv 7\pmod{8}$ admits the group $\sAut_-(\cl_{p,q})
\simeq\dZ_2\otimes\dZ_2$ if $p\equiv 0\pmod{2}$ and $q\equiv 1\pmod{2}$,
and also the group $\sAut_+(\cl_{p,q})\simeq Q_4/\dZ_2$ if $p\equiv 1\pmod{2}$
and $q\equiv 0\pmod{2}$.
\end{proof}\begin{cor}
The matrices $\sE$ and $\sC$ of the antiautomorphisms 
$\cA\rightarrow\widetilde{\cA}$ and $\cA\rightarrow\widetilde{\cA^\star}$
over the field $\F=\R$ satisfy the following conditions
\begin{equation}\label{condt}
\sE^{\sT}=(-1)^{\frac{m(m-1)}{2}}\sE,\quad
\sC^{\sT}=(-1)^{\frac{m(m+1)}{2}}\sC,
\end{equation}
that is, $\sE$ is symmetric if $m\equiv 0,1\pmod{4}$ and skewsymmetric if
$m\equiv 2,3\pmod{4}$. Correspondingly, $\sC$ is symmetric if 
$m\equiv 0,3\pmod{4}$ and skewsymmetric if $m\equiv 1,2\pmod{4}$.
\end{cor}
\begin{proof}
Let us consider first the types with the ring $\K\simeq\R$. As follows
from Theorem \ref{tautr}, the type $p-q\equiv 0\pmod{8}$ admits the
Abelian automorphism groups $(\sE\sC=\sC\sE)$ if $\sE$ is the product of
$q$ skewsymmetric matrices ($q\equiv 0,2\pmod{4}$) and $\sC$ is the product
of $p$ symmetric matrices ($p\equiv 0,2\pmod{4}$). Therefore,
\begin{multline}
\sE^{\sT}=(\cE_{m+1}\cE_{m+2}\cdots\cE_{2m})^{\sT}=\cE^{\sT}_{2m}\cdots
\cE^{\sT}_{m+2}\cE^{\sT}_{m+1}=\\
(-\cE_{2m})\cdots(-\cE_{m+2})(-\cE_{m+1})=\cE_{2m}\cdots\cE_{m+2}\cE_{m+1}=
(-1)^{\frac{q(q-1)}{2}}\sE,\label{T1}
\end{multline}
\begin{equation}
\sC^{\sT}=(\cE_1\cE_2\cdots\cE_m)^{\sT}=\cE^{\sT}_m\cdots\cE^{\sT}_2
\cE^{\sT}_1=\cE_m\cdots\cE_2\cE_1=(-1)^{\frac{p(p-1)}{2}}\sC.\label{T2}
\end{equation}
Further, the type $p-q\equiv 0\pmod{8}$ admits the non--Abelian automorphism
groups ($\sE\sC=-\sC\sE$) if $\sE$ is the product of $p$ symmetric matrices
($p\equiv 1,3\pmod{4}$) and $\sC$ is the product of $q$ skewsymmetric
matrices ($q\equiv 1,3\pmod{4}$). In this case, we have
\begin{equation}
\sE^{\sT}=(\cE_1\cE_2\cdots\cE_m)^{\sT}=\cE^{\sT}_m\cdots\cE^{\sT}_2
\cE^{\sT}_1=\cE_m\cdots\cE_2\cE_1=(-1)^{\frac{p(p-1)}{2}}\sE,\label{T3}
\end{equation}
\begin{multline}
\sC^{\sT}=(\cE_{m+1}\cE_{m+2}\cdots\cE_{2})^{\sT}=\cE^{\sT}_{2m}\cdots
\cE^{\sT}_{m+2}\cE^{\sT}_{m+1}=\\
(-\cE_{2m})\cdots(-\cE_{m+2})(-\cE_{m+1})=-\cE_{2m}\cdots\cE_{m+2}\cE_{m+1}=
-(-1)^{\frac{q(q-1)}{2}}\sC.
\label{T4}
\end{multline}
In the degenerate case $\cl_{p,0}$, $p\equiv 0\pmod{8}$, we have $\sE\sim\sI$
and $\sC\sim\cE_1\cE_2\cdots\cE_p$. Therefore, $\sE$ is always symmetric
and $\sC^{\sT}=(-1)^{\frac{p(p-1)}{2}}\sC$. In other degenerate case
$\cl_{0,q}$, $q\equiv 0\pmod{8}$, we have $\sE\sim\cE_1\cE_2\cdots\cE_q$ and
$\sC\sim\sI$; therefore, $\sE^{\sT}=(-1)^{\frac{q(q-1)}{2}}\sE$ and $\sC$
is always symmetric.

Since for the type $p-q\equiv 0\pmod{8}$ we have $p=q=m$, or $m=p$ and
$m=q$ for the degenerate cases (both degenerate cases correspond to the
Abelian group $\dZ_2\otimes\dZ_2$), it is easy to see that the
formulas (\ref{T1}) and (\ref{T3}) coincide with the first formula of
(\ref{condt}). For the matrix $\sC$, we can unite the formulas (\ref{T2})
and (\ref{T4}) into the formula which coincides with the second formula of
(\ref{condt}). Indeed, the factor $(-1)^{\frac{m(m+1)}{2}}$ does not change
sign in $\sC^{\sT}=(-1)^{\frac{m(m+1)}{2}}\sC$ when $m$ is even and
changes sign when $m$ is odd, which is equivalent to both formulas
(\ref{T2}) and (\ref{T4}).

Further, the following real type $p-q\equiv 2\pmod{8}$ admits the Abelian
automorphism groups if $\sE=\cE_{p+1}\cE_{p+2}\cdots\cE_{p+q}$ and
$\sC=\cE_1\cE_2\cdots\cE_p$, where $p$ and $q\equiv 0,2\pmod{4}$. Therefore,
\begin{multline}
\sE^{\sT}=(\cE_{p+1}\cE_{p+2}\cdots\cE_{p+q})^{\sT}=\cE^{\sT}_{p+q}\cdots
\cE^{\sT}_{p+2}\cE^{\sT}_{p+1}=\\
(-\cE_{p+q})\cdots(-\cE_{p+2})(-\cE_{p+1})=\cE_{p+q}\cdots\cE_{p+2}\cE_{p+1}
=(-1)^{\frac{q(q-1)}{2}}\sE,\label{T5}
\end{multline}
\begin{equation}
\sC^{\sT}=(\cE_1\cE_2\cdots\cE_p)^{\sT}=\cE^{\sT}_p\cdots\cE^{\sT}_2
\cE^{\sT}_1=\cE_p\cdots\cE_2\cE_1=(-1)^{\frac{p(p-1)}{2}}\sC.\label{T6}
\end{equation}
Correspondingly, the type $p-q\equiv 2\pmod{8}$ admits the non--Abelian
automorphism groups if $\sE=\cE_1\cE_2\cdots\cE_p$ and $\sC=\cE_{p+1}\cE_{p+2}
\cdots\cE_{p+q}$, where $p$ and $q\equiv 1,3\pmod{4}$. In this case, we have
\begin{equation}
\sE^{\sT}=(\cE_1\cE_2\cdots\cE_p)^{\sT}=\cE^{\sT}_p\cdots\cE^{\sT}_2
\cE^{\sT}_1=\cE_p\cdots\cE_2\cE_1=(-1)^{\frac{p(p-1)}{2}}\sE,\label{T7}
\end{equation}
\begin{multline}
\sC^{\sT}=(\cE_{p+1}\cE_{p+2}\cdots\cE_{p+q})^{\sT}=\cE^{\sT}_{p+q}\cdots
\cE^{\sT}_{p+2}\cE^{\sT}_{p+1}=\\
(-\cE_{p+q})\cdots(-\cE_{p+2})(-\cE_{p+1})=-\cE_{p+q}\cdots\cE_{p+2}
\cE_{p+1}=-(-1)^{\frac{q(q-1)}{2}}\sC.\label{T8}
\end{multline}
It is easy to see that formulas (\ref{T5})--(\ref{T8}) are similar to the
formulas (\ref{T1})--(\ref{T4}) and, therefore, the conditions (\ref{condt})
hold for the type $p-q\equiv 2\pmod{8}$.

Analogously, the quaternionic types $p-q\equiv 4,6\pmod{8}$ admit the
Abelian automorphism groups if $\sE=\cE_{j_1}\cE_{j_2}\cdots\cE_{j_k}$ and
$\sC=\cE_{i_1}\cE_{i_2}\cdots\cE_{i_{p+q-k}}$, where $k$ and $p+q-k$ are
even (Theorem \ref{tautr}). Transposition of these matrices gives
\begin{multline}
\sE^{\sT}=(\cE_{j_1}\cE_{j_2}\cdots\cE_{j_k})^{\sT}=\cE^{\sT}_{j_k}\cdots
\cE^{\sT}_{j_2}\cE^{\sT}_{j_1}=\\
(-\cE_{j_k})\cdots(-\cE_{j_2})(-\cE_{j_1})=\cE_{j_k}\cdots\cE_{j_2}
\cE_{j_1}=(-1)^{\frac{k(k-1)}{2}}\sE,\label{T9}
\end{multline}
\begin{multline}
\sC^{\sT}=(\cE_{i_1}\cE_{i_2}\cdots\cE_{i_{p+q-k}})^{\sT}=\\
\cE^{\sT}_{i_{p+q-k}}\cdots\cE^{\sT}_{i_2}\cE^{\sT}_{i_1}=\cE_{i_{p+q-k}}\cdots
\cE_{i_2}\cE_{i_1}=(-1)^{\frac{(p+q-k)(p+q-k-1)}{2}}\sC.\label{T10}
\end{multline}
The non--Abelian automorphism groups take place for the types $p-q\equiv
4,6\pmod{8}$ if $\sE=\cE_{i_1}\cE_{i_2}\cdots\cE_{i_{p+q-k}}$ and
$\sC=\cE_{j_1}\cE_{j_2}\cdots\cE_{j_k}$, where $k$ and $p+q-k$ are odd.
In this case we have
\begin{multline}
\sE^{\sT}=(\cE_{i_1}\cE_{i_2}\cdots\cE_{i_{p+q-k}})^{\sT}=\\
\cE^{\sT}_{i_{p+q-k}}\cdots\cE^{\sT}_{i_2}\cE^{\sT}_{i_1}=\cE_{p+q-k}\cdots
\cE_{i_2}\cE_{i_1}=(-1)^{\frac{(p+q-k)(p+q-k-1)}{2}}\sE,\label{T11}
\end{multline}
\begin{multline}
\sC^{\sT}=(\cE_{j_1}\cE_{j_2}\cdots\cE_{j_k})^{\sT}=\cE^{\sT}_{j_k}\cdots
\cE^{\sT}_{j_2}\cE^{\sT}_{j_1}=\\
(-\cE_{j_k})\cdots(-\cE_{j_2})(-\cE_{j_1})=-\cE_{j_k}\cdots\cE_{j_1}
\cE_{j_1}=-(-1)^{\frac{k(k-1)}{2}}\sC.\label{T12}
\end{multline}
As it takes place for these two types considered here, we again come to the
same situation. Therefore, the conditions (\ref{condt}) hold for the
quaternionic types $p-q\equiv 4,6\pmod{8}$.

In virtue of the isomorphism (\ref{E1}) and the Theorem \ref{tautr}, the
matrices $\sE$ and $\sC$ for the types $p-q\equiv 3,7\pmod{8}$ with the
ring $\K\simeq\C$ have the following form: $\sE=\cE_1\cE_2\cdots\cE_m$,
$\sC=\cE_{m+1}\cE_{m+2}\cdots\cE_{2m}$ if $m\equiv 1\pmod{2}$ ($\sE\sC=-
\sC\sE$) and $\sE=\cE_{m+1}\cE_{m+2}\cdots\cE_{2m}$, $\sC=\cE_1\cE_2\cdots
\cE_m$ if $m\equiv 0\pmod{2}$ ($\sE\sC=\sC\sE$). It is obvious that for
these types the conditions (\ref{condt}) hold.

Finally, for the semi--simple types $p-q\equiv 1,5\pmod{8}$ in virtue of the
decompositions (\ref{D1})--(\ref{D1'}) we have the formulas 
(\ref{T1})--(\ref{T4}) or (\ref{T5})--(\ref{T8}) in case of the ring
$\K\simeq\R\oplus\R$ ($p-q\equiv 1\pmod{8}$) and the formulas
(\ref{T9})--(\ref{T12}) in case of the ring $\K\simeq\BH\oplus\BH$
($p-q\equiv 5\pmod{8}$).
\end{proof}

An algebraic structure of the discrete transformations is defined by the
isomorphism $\{\Id,\star,\widetilde{\phantom{cc}},\widetilde{\star}\}\simeq
\{1,P,T,PT\}$ \cite{Var99}. Using (\ref{Pin}) or (\ref{Pinabc}), we can apply
this structure to the double coverings of the orthogonal group $O(p,q)$.
Obviously, in case of the types $p-q\equiv 0,2,4,6\pmod{8}$, it is
established directly. Further, in virtue of the isomorphism (\ref{E1}) for
the types $p-q\equiv 3,7\pmod{8}$, we have
\[
\pin(p,q)\simeq\pin(n-1,\C),
\]
where $n=p+q$. Analogously, for the semi--simple types $p-q\equiv 1,5\pmod{8}$
in virtue of the decompositions (\ref{D1})--(\ref{D1'}) the algebra $\cl_{p,q}$
is isomorphic to a direct sum of two mutually annihilating simple ideals
$\frac{1}{2}(1\pm\omega)\cl_{p,q}$: $\cl_{p,q}\simeq\frac{1}{2}(1+\omega)
\cl_{p,q}\oplus\frac{1}{2}(1-\omega)\cl_{p,q}$, where 
$\omega=\e_{12\cdots p+q}$. At this point, each ideal is isomorphic to
$\cl_{p,q-1}$ or $\cl_{q,p-1}$. Therefore, for the Clifford--Lipschitz
groups of these types we have the following isomorphisms
\begin{eqnarray}
\pin(p,q)&\simeq&\pin(p,q-1)\bigcup\e_{12\ldots p+q}\pin(p,q-1),\nonumber\\
\pin(p,q)&\simeq&\pin(q,p-1)\bigcup\e_{12\ldots p+q}\pin(q,p-1).\nonumber
\end{eqnarray} 
\begin{theorem}\label{tgroupr}
Let $\pin^{a,b,c}(p,q)$ be a double covering of the orthogonal group
$O(p,q)$ of the real space $\R^{p,q}$ associated with the algebra
$\cl_{p,q}$.
The squares of symbols $a,b,c\in
\{-,+\}$ correspond to the squares of the elements of a finite group
$\sAut(\cl_{p,q})=\{\sI,\sW,\sE,\sC\}:\;a=\sW^2,\,b=\sE^2,\,c=\sC^2$, 
where $\sW,\sE$ and $\sC$
are the matrices of the fundamental automorphisms $\cA\rightarrow
\cA^\star,\,\cA\rightarrow\widetilde{\cA}$ and $\cA\rightarrow
\widetilde{\cA^\star}$ of the algebra $\cl_{p,q}$, respectively.
Then over the field $\F=\R$ 
in dependence on a division ring structure of the algebra $\cl_{p,q}$,
there exist eight double coverings of the orthogonal group $O(p,q)$:\\[0.2cm]
1) A non--Cliffordian group
\[
\pin^{+,+,+}(p,q)\simeq\frac{(\spin_0(p,q)\odot\dZ_2\otimes\dZ_2\otimes\dZ_2)}
{\dZ_2},
\]
exists if $\K\simeq\R$ and the numbers $p$ and $q$ form the type 
$p-q\equiv 0\pmod{8}$ and $p,q\equiv 0\pmod{4}$, and also if
$p-q\equiv 4\pmod{8}$ and $\K\simeq\BH$. The algebras $\cl_{p,q}$ with the
rings $\K\simeq\R\oplus\R,\,\K\simeq\BH\oplus\BH$ ($p-q\equiv 1,5\pmod{8}$)
admit the group $\pin^{+,+,+}(p,q)$ if in the direct sums there are
addendums of the type
$p-q\equiv 0\pmod{8}$ or $p-q\equiv 4\pmod{8}$. The types $p-q\equiv 3,7
\pmod{8}$, $\K\simeq\C$ admit a non--Cliffordian group $\pin^{+,+,+}(p+q-1,
\C)$ if $p\equiv 0\pmod{2}$ and $q\equiv 1\pmod{2}$. Further, 
non--Cliffordian groups
\[
\pin^{a,b,c}(p,q)\simeq\frac{(\spin_0(p,q)\odot(\dZ_2\otimes\dZ_4)}{\dZ_2},
\]
with $(a,b,c)=(+,-,-)$ exist if $p-q\equiv 0\pmod{8}$, 
$p,q\equiv 2\pmod{4}$ and $\K\simeq\R$, and also if
$p-q\equiv 4\pmod{8}$ and $\K\simeq\BH$. Non--Cliffordian
groups with the signatures
$(a,b,c)=(-,+,-)$ and $(a,b,c)=(-,-,+)$ exist over the ring
$\K\simeq\R$ ($p-q\equiv 2\pmod{8}$) if $p\equiv
2\pmod{4},\,q\equiv 0\pmod{4}$ and $p\equiv 0\pmod{4},\,q\equiv 2\pmod{4}$,
respectively,
and also these groups exist over the ring $\K\simeq\BH$ if
$p-q\equiv 6\pmod{8}$. 
The algebras $\cl_{p,q}$ with the rings
$\K\simeq\R\oplus\R,\,\K\simeq\BH\oplus\BH$ ($p-q\equiv 1,5\pmod{8}$)
admit the group $\pin^{+,-,-}(p,q)$ if in the direct sums there are addendums
of the type $p-q\equiv 0\pmod{8}$ or $p-q\equiv 4\pmod{8}$, and also admit the
groups $\pin^{-,+,-}(p,q)$ and $\pin^{-,-,+}(p,q)$ if in the direct sums
there are addendums of the type $p-q\equiv 2\pmod{8}$ 
or $p-q\equiv 6\pmod{8}$.\\[0.2cm]
2) A Cliffordian group
\[
\pin^{-,-,-}(p,q)\simeq\frac{(\spin_0(p,q)\odot Q_4)}{\dZ_2},
\]
exists if $\K\simeq\R$ ($p-q\equiv 2\pmod{8}$) and $p\equiv 3\pmod{4},\,
q\equiv 1\pmod{4}$, and also if $p-q\equiv 6\pmod{8}$ and $\K\simeq\BH$.
The algebras $\cl_{p,q}$ with the rings 
$\K\simeq\R\oplus\R,\,\K\simeq\BH\oplus\BH$ ($p-q\equiv 1,5\pmod{8}$)
admit the group $\pin^{-,-,-}(p,q)$ if in the direct sums there are 
addendums of the type
$p-q\equiv 2\pmod{8}$ or $p-q\equiv 6\pmod{8}$. The types $p-q\equiv 3,7
\pmod{8}$, $\K\simeq\C$ admit a Cliffordian group $\pin^{-,-,-}(p+q-1,\C)$,
if $p\equiv 1\pmod{2}$ and $q\equiv 0\pmod{2}$. Further, Cliffordian groups
\[
\pin^{a,b,c}(p,q)\simeq\frac{(\spin_0(p,q)\odot D_4)}{\dZ_2},
\]
with $(a,b,c)=(-,+,+)$ exist if $\K\simeq\R$ ($p-q\equiv 2\pmod{8}$)
and $p\equiv 1\pmod{4},\,q\equiv 3\pmod{4}$,
and also if $p-q\equiv 6\pmod{8}$ and $\K\simeq\BH$. Cliffordian groups with
the signatures
$(a,b,c)=(+,-,+)$ and $(a,b,c)=(+,+,-)$ exist over the ring
$\K\simeq\R$ ($p-q\equiv 0\pmod{8}$) if
$p,q\equiv 3\pmod{4}$ and $p,q\equiv 1\pmod{4}$, respectively,
and also these groups
exist over the ring $\K\simeq\BH$ if $p-q\equiv 4\pmod{8}$.
The algebras $\cl_{p,q}$ with the rings 
$\K\simeq\R\oplus\R,\,\K\simeq\BH\oplus\BH$ ($p-q\equiv 1,5\pmod{8}$)
admit the group $\pin^{-,+,+}(p,q)$ if in the direct sums there are addendums
of the type $p-q\equiv 2\pmod{8}$ or $p-q\equiv 6\pmod{8}$, and also admit the
groups $\pin^{+,-,+}(p,q)$ and $\pin^{+,+,-}(p,q)$ if in the direct sums there
are addendums of the type $p-q\equiv 0\pmod{8}$ or $p-q\equiv 4\pmod{8}$.
\end{theorem}\section{The structure of $\pin(p,q)\not\simeq\pin(q,p)$}
It is easy to see that the definitions (\ref{Pin}) and (\ref{Pinabc}) are
equivalent. Moreover, Salingaros showed \cite{Sal81a,Sal82,Sal84} that there
are isomorphisms $\dZ_2\otimes\dZ_2\simeq\cl_{1,0}$ and $\dZ_4\simeq\cl_{0,1}$.
Further, since $\cl^+_{p,q}\simeq\cl^+_{q,p}$, in accordance with the
definition (\ref{Spin}), it follows that $\spin(p,q)\simeq\spin(q,p)$.
On the other hand, since in a general case $\cl_{p,q}\not\simeq\cl_{q,p}$,
from the definition (\ref{Pin}) it follows that $\pin(p,q)\not\simeq\pin(q,p)$
(or $\pin^{a,b,c}(p,q)\not\simeq\pin^{a,b,c}(q,p)$). In connection with this,
some authors \cite{CDD82,DM88,KT89,CGT95,CGT98,FrTr99} 
used notations $\pin^+\simeq\pin(p,q)$ and
$\pin^-\simeq\pin(q,p)$. In Theorems \ref{tautr} and \ref{tgroupr}, we
have establish a relation between the signatures $(p,q)=(\underbrace{+,+,
\ldots,+}_{p\,\text{times}},\underbrace{-,-,\ldots,-}_{q\,\text{times}})$
of the spaces $\R^{p,q}$ and the signatures $(a,b,c)$ of the automorphism
groups of $\cl_{p,q}$ and corresponding D\c{a}browski groups. This relation
allows to completely define the structure of the inequality $\pin(p,q)\not
\simeq\pin(q,p)$ ($\pin^+\not\simeq\pin^-$). Indeed, from (\ref{Pinabc}) and
(\ref{Spin+}), it follows that $\spin_0(p,q)\simeq\spin_0(q,p)$, therefore,
a nature of the inequality $\pin(p,q)\not\simeq\pin(q,p)$ wholly lies in
the double covering $C^{a,b,c}$ of the discrete subgroup. For example,
in accordance with Theorem \ref{tgroupr} for the type $p-q\equiv 2\pmod{8}$
with the division ring $\K\simeq\R$ there exist the groups $\pin^{a,b,c}(p,q)
\simeq\pin^+$, where double coverings of the discrete subgroup have the form:
1) $C^{-,-,-}\simeq Q_4$ if $p\equiv 3\pmod{4}$ and $q\equiv 1\pmod{4}$;
2) $C^{-,+,+}\simeq D_4$ if $p\equiv 1\pmod{4}$ and $q\equiv 3\pmod{4}$;
3) $C^{-,-,+}\simeq\dZ_2\otimes\dZ_4$ if $p\equiv 0\pmod{4}$ and $q\equiv 2
\pmod{4}$; 4) $C^{-,+,-}\simeq\dZ_2\otimes\dZ_4$ if $p\equiv 2\pmod{4}$ and
$q\equiv 0\pmod{4}$. Whereas the groups with opposite signature, $\pin^-
\simeq\pin^{a,b,c}(q,p)$, have the type $q-p\equiv 6\pmod{8}$ with the
ring $\K\simeq\BH$. In virtue of the more wide ring $\K\simeq\BH$, there
exists a far greater choice of the discrete subgroups for each concrete
kind of $\pin^{a,b,c}(q,p)$. Thus,
\[
\begin{array}{rcl}
\pin^{a,b,c}(p,q)&\not\simeq&\pin^{a,b,c}(q,p)\\
p-q\equiv 2\pmod{8}&&q-p\equiv 6\pmod{8}.
\end{array}
\]
Further, the type $p-q\equiv 1\pmod{8}$ with the ring $\K\simeq\R\oplus\R$
in virtue of Theorems \ref{tautr} and \ref{tgroupr} admits the group
$\pin^{a,b,c}(p,q)\simeq\pin^+$, where the double covering $C^{a,b,c}$
adopts all the eight possible values. Whereas the opposite type
$q-p\equiv 7\pmod{8}$ with the ring $\K\simeq\C$ admits the group
$\pin^{a,b,c}(q,p)\simeq\pin^-$, where for the double covering $C^{a,b,c}$ of
the discrete subgroup there are only two possibilities: 1) $C^{+,+,+}\simeq
\dZ_2\otimes\dZ_2\otimes\dZ_2$ if $p\equiv 0\pmod{2}$ and $q\equiv 1\pmod{2}$;
2) $C^{-,-,-}\simeq Q_4$ if $p\equiv 1\pmod{2}$ and $q\equiv 0\pmod{2}$.
The analogous situation takes place for the two mutually opposite types
$p-q\equiv 3\pmod{8}$ with $\K\simeq\C$ and $q-p\equiv 5\pmod{8}$ with
$\K\simeq\BH\oplus\BH$. Therefore,
\[
\begin{array}{rcl}
\pin^{a,b,c}(p,q)&\not\simeq&\pin^{a,b,c}(q,p)\\
p-q\equiv 1\pmod{8}&&q-p\equiv 7\pmod{8};\\
\pin^{a,b,c}(p,q)&\not\simeq&\pin^{a,b,c}(q,p)\\
p-q\equiv 3\pmod{8}&&q-p\equiv 5\pmod{8}.
\end{array}
\]
It is easy to see that an opposite type to the type $p-q\equiv 0\pmod{8}$
with the ring $\K\simeq\R$ is the same type $q-p\equiv 0\pmod{8}$. Therefore,
in virtue of Theorems \ref{tautr} and \ref{tgroupr} double coverings
$C^{a,b,c}$ for the groups $\pin^{a,b,c}(p,q)\simeq\pin^+$ and $\pin^{a,b,c}
(q,p)\simeq\pin^-$ coincide. The same is the situation for the type
$p-q\equiv 4\pmod{8}$ with $\K\simeq\BH$, which has the opposite type
$q-p\equiv 4\pmod{8}$. Thus,
\[
\begin{array}{rcl}
\pin^{a,b,c}(p,q)&\simeq&\pin^{a,b,c}(q,p)\\
p-q\equiv 0\pmod{8}&&q-p\equiv 0\pmod{8};\\
\pin^{a,b,c}(p,q)&\simeq&\pin^{a,b,c}(q,p)\\
p-q\equiv 4\pmod{8}&&q-p\equiv 4\pmod{8}.
\end{array}
\]
We will call the types $p-q\equiv 0\pmod{8}$ and $p-q\equiv 4\pmod{8}$,
which coincide with their opposite types, {\it neutral types}.\\[0.3cm]
{\bf Example}. Let us consider a structure of the inequality $\pin(3,1)\not
\simeq\pin(1,3)$. The groups $\pin(3,1)$ and $\pin(1,3)$ are two different
double coverings of the general Lorentz group. These groups play an
important role in physics \cite{CWM88,DW90,DG90,DWGK}. 
As follows from (\ref{Pin})
the group $\pin(3,1)$ is completely defined in the framework of the
Majorana algebra $\cl_{3,1}$, which has the type $p-q\equiv 2\pmod{8}$ and
the division ring $\K\simeq\R$. As noted previously, the structure of the
inequality $\pin(p,q)\not\simeq\pin(q,p)$ is defined by the double covering
$C^{a,b,c}$. From Theorems \ref{tautr} and \ref{tgroupr}, it follows that
the algebra $\cl_{3,1}\simeq\M_4(\R)$ admits one and only 
one group $\pin^{-,-,-}(3,1)$, where a double covering of the discrete
subgroup has a form $C^{-,-,-}\simeq Q_4$. Indeed, let us consider a matrix
representation of the units of $\cl_{3,1}$, using the Maple V and the
{\sc CLIFFORD} package developed by Rafa\l Ab\l amowicz 
\cite{Abl96,Abl98,Abl00}. Let $f=\frac{1}{4}(1+\e_1)(1+\e_{34})$ be a
primitive idempotent of the algebra $\cl_{3,1}$ (prestored idempotent for
$\cl_{3,1}$ in {\sc CLIFFORD}), then a following {\sc CLIFFORD} command
sequence gives:\\[0.3cm]
{\tt
>\hs restart:with(Cliff4):with(double):\\[0.1cm]
>\hs dim := 4:\\[0.1cm]
>\hs eval(makealiases(dim)):\\[0.1cm]
>\hs B := linalg(diag(1,1,1,-1)):\\[0.1cm]
>\hs clibasis := cbasis(dim):\\[0.1cm]
>\hs data := clidata(B):\\[0.1cm]
>\hs f := data[4]:\\[0.1cm]
>\hs left\_sbasis := minimalideal(clibasis,f,'left'):\\[0.1cm]
>\hs Kbasis := Kfield(left\_sbasis,f):\\[0.1cm]
>\hs SBgens := left\_sbasis[2]:FBgens := Kbasis[2]:\\[0.1cm]
>\hs K\_basis := spinorKbasis(SBgens,f,FBgens,'left'):\\[0.1cm]
>\hs for i from 1 to 4 do}
\begin{gather}
{\tt E[i] := spinorKrepr(e.i.,K\_basis[1],FBgens,'left') od;}\nonumber\\
\ar
E_1:=\begin{bmatrix}
Id & 0 & 0 & 0\\
0 &-Id & 0 & 0\\
0 & 0 &-Id & 0\\
0 & 0 & 0 & Id
\end{bmatrix},\quad
E_2:=\begin{bmatrix}
0 & Id & 0 & 0\\
Id& 0 & 0 & 0\\
0 & 0 & 0 & Id\\
0 & 0 & Id& 0
\end{bmatrix},\nonumber\\
E_3:=\ar\begin{bmatrix}
0 & 0 & Id & 0\\
0 & 0 & 0 &-Id\\
Id & 0 & 0 & 0\\
0 &-Id & 0 & 0
\end{bmatrix},\quad
E_4:=\begin{bmatrix}
0 & 0 & -Id & 0\\
0 & 0 & 0 & Id\\
Id & 0 & 0 & 0\\
0 & -Id & 0 & 0
\end{bmatrix}.\label{matr}
\end{gather}
It is easy to see that the matrices (\ref{matr}) build up a basis of the
form (\ref{B1}). Since the condition $pq\equiv 1\pmod{2}$ is satisfied for
the algebra $\cl_{3,1}$, the automorphism group $\sAut(\cl_{3,1})$ is
non--Abelian. In accordance with (\ref{commut}), the matrix $\sE$ should 
commute with a symmetric part of the basis (\ref{matr}) and anticommute
with a skewsymmetric part of (\ref{matr}). In this case, as follows from
(\ref{B2})--(\ref{B5}) and (\ref{matr}), the matrix $\sE$ is a product of
$p=3$ symmetric matrices, that is,
\[
\sE=\cE_1\cE_2\cE_3=\ar\begin{pmatrix}
0 & 0 & 0 & -1\\
0 & 0 & -1& 0\\
0 & 1 & 0 & 0\\
1 & 0 & 0 & 0
\end{pmatrix}.
\]
Further, matrices of the automorphisms $\star$ and $\widetilde{\star}$ for
the basis (\ref{matr}) have a form
\[
\sW=\cE_1\cE_2\cE_3\cE_4=\ar\begin{pmatrix}
0 & 1 & 0 & 0\\
-1& 0 & 0 & 0\\
0 & 0 & 0 & 1\\
0 & 0 &-1 & 0
\end{pmatrix},\quad\sC=\sE\sW=\begin{pmatrix}
0 & 0 & 1 & 0\\
0 & 0 & 0 & -1\\
-1 & 0 & 0 & 0\\
0 & 1 & 0 & 0
\end{pmatrix}.
\]
Thus, a group of the fundamental automorphisms of the algebra $\cl_{3,1}$
in the matrix representation is defined by a finite group $\{\sI,\sW,\sE,\sC\}
\sim\{I,\cE_{1234},\cE_{123},\cE_4\}$. The multiplication table of this
group has a form
\begin{equation}\label{tab}
{\renewcommand{\arraystretch}{1.4}
\begin{tabular}{|c||c|c|c|c|}\hline
             & $I$ & $\cE_{1234}$ & $\cE_{123}$ & $\cE_4$ \\ \hline\hline
$I$          & $I$ & $\cE_{1234}$ & $\cE_{123}$ & $\cE_4$ \\ \hline
$\cE_{1234}$ & $\cE_{1234}$ & $-I$ & $\cE_4$ & $-\cE_{123}$ \\ \hline
$\cE_{123}$ & $\cE_{123}$ & $-\cE_4$ & $-I$ & $\cE_{1234}$ \\ \hline
$\cE_4$ & $\cE_4$ & $\cE_{123}$ & $-\cE_{1234}$ & $-I$ \\ \hline
\end{tabular}\;\sim\;
\begin{tabular}{|c||c|c|c|c|}\hline
      & $\sI$ & $\sW$ & $\sE$ & $\sC$ \\ \hline\hline
$\sI$ & $\sI$ & $\sW$ & $\sE$ & $\sC$ \\ \hline
$\sW$ & $\sW$ & $-\sI$& $\sC$ & $-\sE$ \\ \hline
$\sE$ & $\sE$ & $-\sC$& $-\sI$& $\sW$ \\ \hline
$\sC$ & $\sC$ & $\sE$ & $-\sW$& $-\sI$ \\ \hline 
\end{tabular}
}.
\end{equation}
From the table, it follows that $\sAut_+(\cl_{3,1})\simeq\{\sI,\sW,\sE,\sC\}
\simeq Q_4/\dZ_2$ and, therefore, the algebra $\cl_{3,1}$ admits a
Cliffordian group $\pin^{-,-,-}(3,1)$ (Theorem \ref{tgroupr}). It is easy
to verify that the double covering $C^{-,-,-}\simeq Q_4$ is an invariant
fact for the algebra $\cl_{3,1}$, that is, $C^{-,-,-}$ does not depend on the
choice of the matrix representation. Indeed, for each  two commuting
elements of the algebra $\cl_{3,1}$ there exist four different primitive
idempotents that generate four different matrix representations of $\cl_{3,1}$.
The invariability of the previously mentioned fact is easily verified with the
help of a procedure {\tt commutingelements} of the {\sc CLIFFORD} package,
which allows to consider in sequence all the possible primitive idempotents
of the algebra $\cl_{3,1}$ and their corresponding matrix representations.

Now, let us consider discrete subgroups of the double covering $\pin(1,3)$.
The group $\pin(1,3)$, in turn, is completely constructed within the
spacetime algebra $\cl_{1,3}$ that has the opposite (in relation to the
Majorana algebra $\cl_{3,1}$) type $p-q\equiv 6\pmod{8}$ with the division
ring $\K\simeq\BH$. According to Wedderburn--Artin theorem, in this case
there is an isomorphism $\cl_{1,3}\simeq\M_2(\BH)$. The following
{\sc CLIFFORD} command sequence allows to find matrix representations of the
units of the algebra $\cl_{1,3}$ for a prestored primitive idempotent
$f=\frac{1}{2}(1+\e_{14})$:
\begin{eqnarray}
\text{\tt >}&&\text{\tt restart:with(Cliff4):with(double):}\\
\text{\tt >}&&\text{\tt dim := 4: eval(makealiases(dim)):}\\
\text{\tt >}&&\text{\tt B := linalg(diag(1,-1,-1,-1)):}\\
\text{\tt >}&&\text{\tt clibasis := cbasis(dim):}\\
\text{\tt >}&&\text{\tt data := clidata(B): f := data[4]:}\\
\text{\tt >}&&\text{\tt left\_sbasis := minimalideal(clibasis,f,'left'):}
\label{m6}\\
\text{\tt >}&&\text{\tt Kbasis := Kfield(left\_sbasis,f):}\\
\text{\tt >}&&\text{\tt SBgens := left\_sbasis[2]: FBgens := Kbasis[2]:}\\
\text{\tt >}&&\text{\tt K\_basis := spinorKbasis(SBgens,f,FBgens,'left'):}\\
\text{\tt >}&&\text{\tt for i from 1 to 4 do}
\end{eqnarray}
\begin{gather}
\text{\tt E[i] := spinorKrepr(e.i.,K\_basis[1],FBgens,'left') od;}\label{m10}\\
E_1:=\ar\begin{bmatrix}
0 & Id\\
Id & 0
\end{bmatrix},\quad
E_2:=\begin{bmatrix}
e2 & 0\\
0 & -e2
\end{bmatrix},\quad
E_3:=\begin{bmatrix}
e3 & 0\\
0 & -e3
\end{bmatrix},\quad
E_4:=\begin{bmatrix}
0 & -Id\\
Id & 0
\end{bmatrix}.\label{matr2}
\end{gather}
At this point, the division ring $\K\simeq\BH$ is generated by a set
$\{1,\e_2,\e_3,\e_{23}\}\simeq\{1,\bi,\bj,\bk\}$, where $\bi,\bj,\bk$ are
well-known quaternion units. The basis (\ref{matr2}) contains three
symmetric matrices and one skewsymmetric matrix. Therefore, in accordance
with (\ref{commut}) and (\ref{C1})--(\ref{C6}) the matrix of the
antiautomorphism $\cA\rightarrow\widetilde{\cA}$ is a product of symmetric
matrices of the basis (\ref{matr2}). Thus,
\begin{equation}\label{set}
\sW=\cE_1\cE_2\cE_3\cE_4=\ar\begin{pmatrix}
\bk & 0\\
0 & -\bk
\end{pmatrix},\quad
\sE=\cE_1\cE_2\cE_3=\begin{pmatrix}
0 & \bk\\
\bk & 0
\end{pmatrix},\quad
\sC=\sE\sW=\begin{pmatrix}
0 & 1\\
-1 & 0
\end{pmatrix}.
\end{equation}
It is easy to verify that a set of the matrices (\ref{set}) added by the
unit matrix forms the non--Abelian group $\sAut_+(\cl_{1,3})\simeq
Q_4/\dZ_2$ with a multiplication table of the form (\ref{tab}). Therefore,
the spacetime algebra $\cl_{1,3}$ admits the Cliffordian group
$\pin^{-,-,-}(1,3)$, where a double covering of the discrete subgroup has
the form $C^{-,-,-}\simeq Q_4$. However, as follows from Theorem
\ref{tautr}, in virtue of the more wide ring $\K\simeq\BH$ the group
$\pin^{-,-,-}(1,3)$ does not the only possible for the algebra
$\cl_{1,3}\simeq\M_2(\BH)$. Indeed, looking over all the possible commuting
elements of the algebra $\cl_{1,3}$ we find with the help of the procedure
{\tt commutingelements} that\\
{\tt >\hs L1 := commutingelements(clibasis);}
\begin{equation}\label{m11}
L1:=[e1]
\end{equation}
{\tt >\hs L2 := commutingelements(remove(member,clibasis,L1));}
\begin{equation}\label{m12}
L2:=[e12]
\end{equation}
{\tt >\hs L3 := commutingelements(remove(member,clibasis,[op(L1),op(L2)]));}
\begin{equation}\label{m13}
L3:=[e13]
\end{equation}
{\tt >\hs L4 := commutingelements(remove(member,clibasis,[op(L1),op(L2),\\
\phantom{>\hs}op(L3)]));}
\begin{equation}\label{m14}
L4:=[e14]
\end{equation}
{\tt >\hs L5 := commutingelements(remove(member,clibasis,[op(L1),op(L2),\\
\phantom{>\hs}op(L3),op(L4)]));}
\begin{equation}\label{m15}
L5:=[e234]
\end{equation}
{\tt >\hs f := cmulQ((1/2)$\ast$(Id + e2we3we4);}
\begin{equation}\label{m16}
f:=\frac{1}{2}Id+\frac{1}{2}e234
\end{equation}
{\tt >\hs type(f,primitiveidemp);}
\begin{equation}\label{m17}
true
\end{equation}
It is easy to verify that primitive idempotents $\frac{1}{2}(1\pm\e_1),\,
\frac{1}{2}(1\pm\e_{12}),\,\frac{1}{2}(1\pm\e_{13})$ and $\frac{1}{2}(1\pm
\e_{14})$ constructed by means of the commuting elements $\e_1,\,\e_{12},\,
\e_{13}$ and $\e_{14}$ generate matrix representations that give rise to
the group $\sAut_+(\cl_{1,3})\simeq Q_4/\dZ_2$. However, the situation 
changes for the element $\e_{234}$ and corresponding primitive idempotent
$\frac{1}{2}(1+\e_{234})$ ($\frac{1}{2}(1-\e_{234})$). Indeed, executing
the commands (\ref{m16}) and (\ref{m17}) and subsequently the commands
(\ref{m6})--(\ref{m10}), we find that\\
{\tt >\hs for i from 1 to 4 do}
\begin{gather}
\text{\tt E[i] := spinorKrepr(e.i.,Kbasis[1],FBgens,'left') od;}\nonumber\\
E_1:=\ar\begin{bmatrix}
0 & Id\\
Id & 0
\end{bmatrix},\quad
E_2:=\begin{bmatrix}
e2 & 0\\
0 & -e2
\end{bmatrix},\quad
E_3:=\begin{bmatrix}
e34 & 0\\
0 & -e34
\end{bmatrix},\quad
E_4:=\begin{bmatrix}
e4 & 0\\
0 & -e4
\end{bmatrix},\label{matr3}
\end{gather}
where the division ring $\K\simeq\BH$ is generated by a set $\{1,\e_2,\e_4,
\e_{24}\}\simeq\{1,\bi,\bj,\bk\}$. The basis (\ref{matr3}) consists of 
symmetric matrices only. Therefore, in accordance with (\ref{commut}), the
matrix $\sE$ should commute with all the matrices of the basis
(\ref{matr3}). It is obvious that this condition is satisfied only if $\sE$
is proportional to the unit matrix (recall that any element of the
automorphism group may be multiplied by an arbitrary factor $\eta\in\F$, in
this case $\F=\R$). Further, a set of the matrices $\sW=\cE_1\cE_2\cE_3\cE_4$,
$\sE\sim\sI$, $\sC=\sE\sW$ added by the unit matrix forms a finite group
with a following multiplication table
\[
{\renewcommand{\arraystretch}{1.4}
\begin{tabular}{|c||c|c|c|c|}\hline
      & $\sI$ & $\sW$ & $\sE$ & $\sC$\\ \hline\hline
$\sI$ & $\sI$ & $\sW$ & $\sE$ & $\sC$\\ \hline
$\sW$ & $\sW$ & $-\sI$& $\sC$ & $-\sE$\\ \hline
$\sE$ & $\sE$ & $\sC$ & $\sI$ & $\sW$\\ \hline 
$\sC$ & $\sC$ & $-\sE$& $\sW$ & $-\sI$\\ \hline 
\end{tabular}
}.
\]
As follows from the table, we have in this case the Abelian group
$\sAut_-(\cl_{1,3})\simeq\dZ_4$ with the signature $(-,+,-)$. Thus, the
spacetime algebra $\cl_{1,3}$ admits the group $\pin^{-,+,-}(1,3)$, where
a double covering of the discrete subgroup has the form 
$C^{-,+,-}\simeq\dZ_2\otimes\dZ_4$.

The fulfilled analysis explicitly shows a difference between the two
double coverings $\pin(3,1)$ and $\pin(1,3)$ of the Lorentz group. Since
double coverings of the connected components of both groups $\pin(3,1)$
and $\pin(1,3)$ are isomorphic, $\spin_0(3,1)\simeq\spin_0(1,3)$, the
nature of the difference between them consists in a concrete form and a
number of the double covering $C^{a,b,c}$ of the discrete subgroups. So,
for the Majorana algebra $\cl_{3,1}$, all the existing primitive idempotents
$\frac{1}{4}(1\pm\e_1)(1\pm\e_{34}),\,\frac{1}{4}(1\pm\e_1)(1\pm\e_{24}),\,
\frac{1}{4}(1\pm\e_2)(1\pm\e_{14}),\,\frac{1}{4}(1\pm\e_3)(1\pm\e_{134}),\,
\frac{1}{4}(1\pm\e_{34})(1\pm\e_{234})$ generate 20 matrix representations,
each of which gives rise to the double covering $C^{-,-,-}\simeq Q_4$.
On the other hand, for the spacetime algebra $\cl_{1,3}$ primitive
idempotents $\frac{1}{2}(1\pm\e_{14}),\,\frac{1}{2}(1\pm\e_1),\,
\frac{1}{2}(1\pm\e_{12}),\,\frac{1}{2}(1\pm\e_{13})$ generate 8 matrix
representations with $C^{-,-,-}\simeq Q_4$, whereas remaining two primitive
idempotents $\frac{1}{2}(1\pm\e_{234})$ generate matrix representations
with $C^{-,+,-}\simeq\dZ_2\otimes\dZ_4$.\\[0.3cm]
{\bf Remark}. Physicists commonly use a transition from some given
signature to its opposite (signature change) by means of a replacement
$\cE_i\rightarrow i\cE_i$ (so--called Wick rotation). However, such a
transition is unsatisfactory from a mathematical viewpoint. For example,
we can use the replacement $\cE_i\rightarrow i\cE_i$ for a transition
from the spacetime algebra $\cl_{1,3}\simeq\M_2(\BH)$ to the Majorana
algebra $\cl_{3,1}\simeq\M_4(\R)$ since $i\in\M_2(\BH)$, whereas an inverse
transition $\cl_{3,1}\rightarrow\cl_{1,3}$ can not be performed by the
replacement $\cE_i\rightarrow i\cE_i$, since $i\not\in\M_4(\R)$. The
mathematically correct alternative to the Wick rotation is a
tilt--transformation introduced by Lounesto \cite{Lou93}. The
tilt--transformation is expressed by a map $ab\rightarrow a_+b_++b_+a_-+
b_-a_+-b_-a_-$, where $a_\pm,b_\pm\in\cl^\pm_{p,q}$. The further developing
of the tilt--transformation and its application for a formulation of
physical theories in the spaces with different signatures has been
considered in the recent paper \cite{MPV00}.
\section{Discrete transformations and Brauer--Wall groups}
The algebra $\cl$ is naturally $\dZ_2$--graded. Let
$\cl^+$ (correspondingly $\cl^-$) be a set consisting of all even
(correspondingly odd) 
elements of the algebra $\cl$. The set $\cl^+$ is a subalgebra of
$\cl$. It is obvious that
$\cl=\cl^+\oplus\cl^-$, and also $\cl^+\cl^+
\subset\cl^+,\,\cl^+\cl^-\subset\cl^-,\,
\cl^-\cl^+\subset\cl^-,\,\cl^-\cl^-\subset
\cl^+$. A degree $\deg a$ of the even (correspondingly odd) 
element $a\in\cl$ is equal
to 0 (correspondingly 1). 
Let $\mathfrak{A}$ and $\mathfrak{B}$ be the two associative
$\dZ_2$--graded algebras over the field $\F$; then a multiplication of
homogeneous elements
$\mathfrak{a}^\prime\in\mathfrak{A}$ and $\mathfrak{b}\in\mathfrak{B}$ in a
graded tensor product
$\mathfrak{A}\hat{\otimes}\mathfrak{B}$ is defined as follows: 
$(\mathfrak{a}\otimes \mathfrak{b})(\mathfrak{a}^\prime
\otimes \mathfrak{b}^\prime)=(-1)^{\deg\mathfrak{b}\deg\mathfrak{a}^\prime}
\mathfrak{a}\mathfrak{a}^\prime\otimes\mathfrak{b}\mathfrak{b}^\prime$.
The graded tensor product of the two graded central simple algebras is also
graded central simple 
\cite[Theorem 2]{Wal64}. The Clifford algebra $\cl_{p,q}$ is central simple
if $p-q\not\equiv 1,5\pmod{8}$. It is known that for the Clifford algebra
with odd dimensionality, the isomorphisms are as follows:
$\cl^+_{p,q+1}\simeq\cl_{p,q}$ and $\cl^+_{p+1,q}\simeq\cl_{q,p}$ 
\cite{Rash,Port69}. Thus, $\cl^+_{p,q+1}$ and $\cl^+_{p+1,q}$
are central simple algebras. Further, in accordance with Chevalley Theorem
\cite{Che55} for the graded tensor product there is an isomorphism
$\cl_{p,q}\hat{\otimes}\cl_{p^{\p},q^{\p}}\simeq
\cl_{p+p^{\p},q+q^{\p}}$. Two algebras $\cl_{p,q}$ and $\cl_{p^{\p},q^{\p}}$
are said to be of the same class if $p+q^{\p}\equiv p^{\p}+q\pmod{8}$.
The graded central simple Clifford algebras over the field $\F=\R$
form eight similarity classes, which, as it is easy to see, coincide
with the eight types of the algebras $\cl_{p,q}$.
The set of these 8 types (classes) forms a Brauer--Wall group $BW_{\R}$
\cite{Wal64} that is isomorphic to a cyclic group $\dZ_8$. Thus, the
algebra $\cl_{p,q}$ is an element of the Brauer--Wall group, and a group
operation is the graded tensor product $\hat{\otimes}$.
A cyclic structure of the group $BW_{\R}\simeq\dZ_8$ may be represented on
the Trautman diagram (spinorial clock) \cite{BTr87,BT88} (Fig. 1) by means
of a transition $\cl^+_{p,q}\stackrel{h}{\longrightarrow}\cl_{p,q}$ 
(the round on the diagram is realized by an hour--hand). At this point, the type
of the algebra is defined on the diagram by an equality
$q-p=h+8r$, where $h\in\{0,\ldots,7\}$, $r\in\dZ$.

\[
\unitlength=0.5mm
\begin{picture}(100.00,110.00)

\put(97,67){$\C$}\put(105,64){$p-q\equiv 7\!\!\!\!\pmod{8}$}
\put(80,80){1}
\put(75,93.3){$\cdot$}
\put(75.5,93){$\cdot$}
\put(76,92.7){$\cdot$}
\put(76.5,92.4){$\cdot$}
\put(77,92.08){$\cdot$}
\put(77.5,91.76){$\cdot$}
\put(78,91.42){$\cdot$}
\put(78.5,91.08){$\cdot$}
\put(79,90.73){$\cdot$}
\put(79.5,90.37){$\cdot$}
\put(80,90.0){$\cdot$}
\put(80.5,89.62){$\cdot$}
\put(81,89.23){$\cdot$}
\put(81.5,88.83){$\cdot$}
\put(82,88.42){$\cdot$}
\put(82.5,87.99){$\cdot$}
\put(83,87.56){$\cdot$}
\put(83.5,87.12){$\cdot$}
\put(84,86.66){$\cdot$}
\put(84.5,86.19){$\cdot$}
\put(85,85.70){$\cdot$}
\put(85.5,85.21){$\cdot$}
\put(86,84.69){$\cdot$}
\put(86.5,84.17){$\cdot$}
\put(87,83.63){$\cdot$}
\put(87.5,83.07){$\cdot$}
\put(88,82.49){$\cdot$}
\put(88.5,81.9){$\cdot$}
\put(89,81.29){$\cdot$}
\put(89.5,80.65){$\cdot$}
\put(90,80){$\cdot$}
\put(90.5,79.32){$\cdot$}
\put(91,78.62){$\cdot$}
\put(91.5,77.89){$\cdot$}
\put(92,77.13){$\cdot$}
\put(92.5,76.34){$\cdot$}
\put(93,75.51){$\cdot$}
\put(93.5,74.65){$\cdot$}
\put(94,73.74){$\cdot$}
\put(94.5,72.79){$\cdot$}
\put(96.5,73.74){\vector(1,-2){1}}
\put(80,20){3}
\put(97,31){$\BH$}\put(105,28){$p-q\equiv 6\!\!\!\!\pmod{8}$}
\put(75,6.7){$\cdot$}
\put(75.5,7){$\cdot$}
\put(76,7.29){$\cdot$}
\put(76.5,7.6){$\cdot$}
\put(77,7.91){$\cdot$}
\put(77.5,8.24){$\cdot$}
\put(78,8.57){$\cdot$}
\put(78.5,8.91){$\cdot$}
\put(79,9.27){$\cdot$}
\put(79.5,9.63){$\cdot$}
\put(80,10){$\cdot$}
\put(80.5,10.38){$\cdot$}
\put(81,10.77){$\cdot$}
\put(81.5,11.17){$\cdot$}
\put(82,11.58){$\cdot$}
\put(82.5,12.00){$\cdot$}
\put(83,12.44){$\cdot$}
\put(83.5,12.88){$\cdot$}
\put(84,13.34){$\cdot$}
\put(84.5,13.8){$\cdot$}
\put(85,14.29){$\cdot$}
\put(85.5,14.79){$\cdot$}
\put(86,15.3){$\cdot$}
\put(86.5,15.82){$\cdot$}
\put(87,16.37){$\cdot$}
\put(87.5,16.92){$\cdot$}
\put(88,17.5){$\cdot$}
\put(88.5,18.09){$\cdot$}
\put(89,18.71){$\cdot$}
\put(89.5,19.34){$\cdot$}
\put(90,20){$\cdot$}
\put(90.5,20.68){$\cdot$}
\put(91,21.38){$\cdot$}
\put(91.5,22.11){$\cdot$}
\put(92,22.87){$\cdot$}
\put(92.5,23.66){$\cdot$}
\put(93,24.48){$\cdot$}
\put(93.5,25.34){$\cdot$}
\put(94,26.25){$\cdot$}
\put(94.5,27.20){$\cdot$}
\put(95,28.20){$\cdot$}
\put(20,80){7}
\put(25,93.3){$\cdot$}
\put(24.5,93){$\cdot$}
\put(24,92.7){$\cdot$}
\put(23.5,92.49){$\cdot$}
\put(23,92.08){$\cdot$}
\put(22.5,91.75){$\cdot$}
\put(22,91.42){$\cdot$}
\put(21.5,91.08){$\cdot$}
\put(21,90.73){$\cdot$}
\put(20.5,90.37){$\cdot$}
\put(20,90){$\cdot$}
\put(19.5,89.62){$\cdot$}
\put(19,89.23){$\cdot$}
\put(18.5,88.83){$\cdot$}
\put(18,88.42){$\cdot$}
\put(17.5,87.99){$\cdot$}
\put(17,87.56){$\cdot$}
\put(16.5,87.12){$\cdot$}
\put(16,86.66){$\cdot$}
\put(15.5,86.19){$\cdot$}
\put(15,85.70){$\cdot$}
\put(14.5,85.21){$\cdot$}
\put(14,84.69){$\cdot$}
\put(13.5,84.17){$\cdot$}
\put(13,83.63){$\cdot$}
\put(12.5,83.07){$\cdot$}
\put(12,82.49){$\cdot$}
\put(11.5,81.9){$\cdot$}
\put(11,81.29){$\cdot$}
\put(10.5,80.65){$\cdot$}
\put(10,80){$\cdot$}
\put(9.5,79.32){$\cdot$}
\put(9,78.62){$\cdot$}
\put(8.5,77.89){$\cdot$}
\put(8,77.13){$\cdot$}
\put(7.5,76.34){$\cdot$}
\put(7,75.51){$\cdot$}
\put(6.5,74.65){$\cdot$}
\put(6,73.79){$\cdot$}
\put(5.5,72.79){$\cdot$}
\put(5,71.79){$\cdot$}
\put(20,20){5}
\put(25,6.7){$\cdot$}
\put(24.5,7){$\cdot$}
\put(24,7.29){$\cdot$}
\put(23.5,7.6){$\cdot$}
\put(23,7.91){$\cdot$}
\put(22.5,8.24){$\cdot$}
\put(22,8.57){$\cdot$}
\put(21.5,8.91){$\cdot$}
\put(21,9.27){$\cdot$}
\put(20.5,9.63){$\cdot$}
\put(20,10){$\cdot$}
\put(19.5,10.38){$\cdot$}
\put(19,10.77){$\cdot$}
\put(18.5,11.17){$\cdot$}
\put(18,11.58){$\cdot$}
\put(17.5,12){$\cdot$}
\put(17,12.44){$\cdot$}
\put(16.5,12.88){$\cdot$}
\put(16,13.34){$\cdot$}
\put(15.5,13.8){$\cdot$}
\put(15,14.29){$\cdot$}
\put(14.5,14.79){$\cdot$}
\put(14,15.3){$\cdot$}
\put(13.5,15.82){$\cdot$}
\put(13,16.37){$\cdot$}
\put(12.5,16.92){$\cdot$}
\put(12,17.5){$\cdot$}
\put(11.5,18.09){$\cdot$}
\put(11,18.71){$\cdot$}
\put(10.5,19.34){$\cdot$}
\put(10,20){$\cdot$}
\put(9.5,20.68){$\cdot$}
\put(9,21.38){$\cdot$}
\put(8.5,22.11){$\cdot$}
\put(8,22.87){$\cdot$}
\put(7.5,23.66){$\cdot$}
\put(7,24.48){$\cdot$}
\put(6.5,25.34){$\cdot$}
\put(6,26.25){$\cdot$}
\put(5.5,27.20){$\cdot$}
\put(5,28.20){$\cdot$}
\put(13,97){$\R\oplus\R$}\put(-55,105){$p-q\equiv 1\!\!\!\!\pmod{8}$}
\put(50,93){0}
\put(50,100){$\cdot$}
\put(49.5,99.99){$\cdot$}
\put(49,99.98){$\cdot$}
\put(48.5,99.97){$\cdot$}
\put(48,99.96){$\cdot$}
\put(47.5,99.94){$\cdot$}
\put(47,99.91){$\cdot$}
\put(46.5,99.86){$\cdot$}
\put(46,99.84){$\cdot$}
\put(45.5,99.8){$\cdot$}
\put(45,99.75){$\cdot$}
\put(44.5,99.7){$\cdot$}
\put(44,99.64){$\cdot$}
\put(43.5,99.57){$\cdot$}
\put(43,99.51){$\cdot$}
\put(42.5,99.43){$\cdot$}
\put(42,99.35){$\cdot$}
\put(41.5,99.27){$\cdot$}
\put(41,99.18){$\cdot$}
\put(40.5,99.09){$\cdot$}
\put(40,98.99){$\cdot$}
\put(39.5,98.88){$\cdot$}
\put(39,98.77){$\cdot$}
\put(38.5,98.66){$\cdot$}
\put(38,98.54){$\cdot$}
\put(37.5,98.41){$\cdot$}
\put(37,98.28){$\cdot$}
\put(50.5,99.99){$\cdot$}
\put(51,99.98){$\cdot$}
\put(51.5,99.97){$\cdot$}
\put(52,99.96){$\cdot$}
\put(52.5,99.94){$\cdot$}
\put(53,99.91){$\cdot$}
\put(53.5,99.86){$\cdot$}
\put(54,99.84){$\cdot$}
\put(54.5,99.8){$\cdot$}
\put(55,99.75){$\cdot$}
\put(55.5,99.7){$\cdot$}
\put(56,99.64){$\cdot$}
\put(56.5,99.57){$\cdot$}
\put(57,99.51){$\cdot$}
\put(57.5,99.43){$\cdot$}
\put(58,99.35){$\cdot$}
\put(58.5,99.27){$\cdot$}
\put(59,99.18){$\cdot$}
\put(59.5,99.09){$\cdot$}
\put(60,98.99){$\cdot$}
\put(60.5,98.88){$\cdot$}
\put(61,98.77){$\cdot$}
\put(61.5,98.66){$\cdot$}
\put(62,98.54){$\cdot$}
\put(62.5,98.41){$\cdot$}
\put(63,98.28){$\cdot$}
\put(68,97){$\R$}\put(73,105){$p-q\equiv 0\!\!\!\!\pmod{8}$}
\put(50,7){4}
\put(68,2){$\BH\oplus\BH$}\put(90,-4){$p-q\equiv 5\!\!\!\!\pmod{8}$}
\put(50,0){$\cdot$}
\put(50.5,0){$\cdot$}
\put(51,0.01){$\cdot$}
\put(51.5,0.02){$\cdot$}
\put(52,0.04){$\cdot$}
\put(52.5,0.06){$\cdot$}
\put(53,0.09){$\cdot$}
\put(53.5,0.12){$\cdot$}
\put(54,0.16){$\cdot$}
\put(54.5,0.2){$\cdot$}
\put(55,0.25){$\cdot$}
\put(55.5,0.3){$\cdot$}
\put(56,0.36){$\cdot$}
\put(56.5,0.42){$\cdot$}
\put(57,0.49){$\cdot$}
\put(57.5,0.56){$\cdot$}
\put(58,0.64){$\cdot$}
\put(58.5,0.73){$\cdot$}
\put(59,0.82){$\cdot$}
\put(59.5,0.91){$\cdot$}
\put(60,1.01){$\cdot$}
\put(60.5,1.11){$\cdot$}
\put(61,1.22){$\cdot$}
\put(61.5,1.34){$\cdot$}
\put(62,1.46){$\cdot$}
\put(62.5,1.59){$\cdot$}
\put(63,1.72){$\cdot$}
\put(49.5,0){$\cdot$}
\put(49,0.01){$\cdot$}
\put(48.5,0.02){$\cdot$}
\put(48,0.04){$\cdot$}
\put(47.5,0.06){$\cdot$}
\put(47,0.09){$\cdot$}
\put(46.5,0.12){$\cdot$}
\put(46,0.16){$\cdot$}
\put(45.5,0.2){$\cdot$}
\put(45,0.25){$\cdot$}
\put(44.5,0.3){$\cdot$}
\put(44,0.36){$\cdot$}
\put(43.5,0.42){$\cdot$}
\put(43,0.49){$\cdot$}
\put(42.5,0.56){$\cdot$}
\put(42,0.64){$\cdot$}
\put(41.5,0.73){$\cdot$}
\put(41,0.82){$\cdot$}
\put(40.5,0.91){$\cdot$}
\put(40,1.01){$\cdot$}
\put(39.5,1.11){$\cdot$}
\put(39,1.22){$\cdot$}
\put(38.5,1.34){$\cdot$}
\put(38,1.46){$\cdot$}
\put(37.5,1.59){$\cdot$}
\put(37,1.72){$\cdot$}
\put(28,3){$\BH$}\put(-40,-4){$p-q\equiv 4\!\!\!\!\pmod{8}$}
\put(93,50){2}
\put(98.28,63){$\cdot$}
\put(98.41,62.5){$\cdot$}
\put(98.54,62){$\cdot$}
\put(98.66,61.5){$\cdot$}
\put(98.77,61){$\cdot$}
\put(98.88,60.5){$\cdot$}
\put(98.99,60){$\cdot$}
\put(99.09,59.5){$\cdot$}
\put(99.18,59){$\cdot$}
\put(99.27,58.5){$\cdot$}
\put(99.35,58){$\cdot$}
\put(99.43,57.5){$\cdot$}
\put(99.51,57){$\cdot$}
\put(99.57,56.5){$\cdot$}
\put(99.64,56){$\cdot$}
\put(99.7,55.5){$\cdot$}
\put(99.75,55){$\cdot$}
\put(99.8,54.5){$\cdot$}
\put(99.84,54){$\cdot$}
\put(99.86,53.5){$\cdot$}
\put(99.91,53){$\cdot$}
\put(99.94,52.5){$\cdot$}
\put(99.96,52){$\cdot$}
\put(99.97,51.5){$\cdot$}
\put(99.98,51){$\cdot$}
\put(99.99,50.5){$\cdot$}
\put(100,50){$\cdot$}
\put(98.28,37){$\cdot$}
\put(98.41,37.5){$\cdot$}
\put(98.54,38){$\cdot$}
\put(98.66,38.5){$\cdot$}
\put(98.77,39){$\cdot$}
\put(98.88,39.5){$\cdot$}
\put(98.99,40){$\cdot$}
\put(99.09,40.5){$\cdot$}
\put(99.18,41){$\cdot$}
\put(99.27,41.5){$\cdot$}
\put(99.35,42){$\cdot$}
\put(99.43,42.5){$\cdot$}
\put(99.51,43){$\cdot$}
\put(99.57,43.5){$\cdot$}
\put(99.64,44){$\cdot$}
\put(99.7,44.5){$\cdot$}
\put(99.75,45){$\cdot$}
\put(99.8,45.5){$\cdot$}
\put(99.84,46){$\cdot$}
\put(99.86,46.5){$\cdot$}
\put(99.91,47){$\cdot$}
\put(99.94,47.5){$\cdot$}
\put(99.96,48){$\cdot$}
\put(99.97,48.5){$\cdot$}
\put(99.98,49){$\cdot$}
\put(99.99,49.5){$\cdot$}
\put(7,50){6}
\put(1,32){$\C$}\put(-65,29){$p-q\equiv 3\!\!\!\!\pmod{8}$}
\put(1.72,63){$\cdot$}
\put(1.59,62.5){$\cdot$}
\put(1.46,62){$\cdot$}
\put(1.34,61.5){$\cdot$}
\put(1.22,61){$\cdot$}
\put(1.11,60.5){$\cdot$}
\put(1.01,60){$\cdot$}
\put(0.99,59.5){$\cdot$}
\put(0.82,59){$\cdot$}
\put(0.73,58.5){$\cdot$}
\put(0.64,58){$\cdot$}
\put(0.56,57.5){$\cdot$}
\put(0.49,57){$\cdot$}
\put(0.42,56.5){$\cdot$}
\put(0.36,56){$\cdot$}
\put(0.3,55.5){$\cdot$}
\put(0.25,55){$\cdot$}
\put(0.2,54.5){$\cdot$}
\put(0.16,54){$\cdot$}
\put(0.12,53.5){$\cdot$}
\put(0.09,53){$\cdot$}
\put(0.06,52.5){$\cdot$}
\put(0.04,52){$\cdot$}
\put(0.02,51.5){$\cdot$}
\put(0.01,51){$\cdot$}
\put(0,50.5){$\cdot$}
\put(0,50){$\cdot$}
\put(1.72,37){$\cdot$}
\put(1.59,37.5){$\cdot$}
\put(1.46,38){$\cdot$}
\put(1.34,38.5){$\cdot$}
\put(1.22,39){$\cdot$}
\put(1.11,39.5){$\cdot$}
\put(1.01,40){$\cdot$}
\put(0.99,40.5){$\cdot$}
\put(0.82,41){$\cdot$}
\put(0.73,41.5){$\cdot$}
\put(0.64,42){$\cdot$}
\put(0.56,42.5){$\cdot$}
\put(0.49,43){$\cdot$}
\put(0.42,43.5){$\cdot$}
\put(0.36,44){$\cdot$}
\put(0.3,44.5){$\cdot$}
\put(0.25,45){$\cdot$}
\put(0.2,45.5){$\cdot$}
\put(0.16,46){$\cdot$}
\put(0.12,46.5){$\cdot$}
\put(0.09,47){$\cdot$}
\put(0.06,47.5){$\cdot$}
\put(0.04,48){$\cdot$}
\put(0.02,48.5){$\cdot$}
\put(0.01,49){$\cdot$}
\put(0,49.5){$\cdot$}
\put(0.5,67){$\R$}\put(-65,75){$p-q\equiv 2\!\!\!\!\pmod{8}$}
\end{picture}
\]
\vspace{2ex}
\begin{center}
\begin{minipage}{25pc}{\small
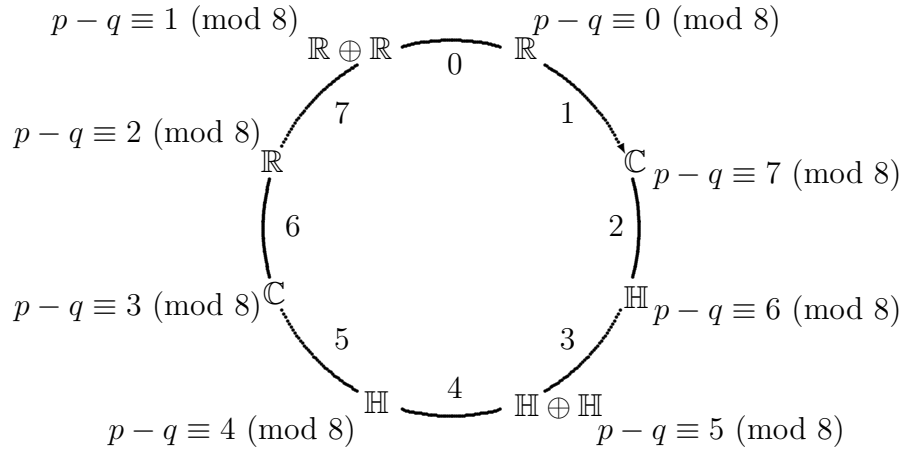
{\bf Fig.1} The Trautman diagram for the Brauer--Wall group
$BW_{\R}\simeq\dZ_8$}
\end{minipage}
\end{center}
\medskip
It is obvious that a group structure over $\cl_{p,q}$, defined by
$BW_{\R}\simeq\dZ_8$, immediately relates with the Atiyah--Bott--Shapiro
periodicity \cite{AtBSh}. In accordance with \cite{AtBSh}, the Clifford
algebra over the field $\F=\R$ is modulo 8 periodic:
$\cl_{p+8,q}\simeq\cl_{p,q}\otimes\cl_{8,0}\,(\cl_{p,q+8}\simeq\cl_{p,q}
\otimes\cl_{0,8})$.

Coming back to Theorem \ref{tautr} we see that for each type of 
algebra $\cl_{p,q}$ there exists some set of the automorphism groups.
If we take into account this relation, then the cyclic structure of a
generalized group $BW^{a,b,c}_{\R}$ would look as follows (Fig. 2).
First of all, the semi--simple algebras $\cl_{p,q}$ with the rings
$\K\simeq\R\oplus\R$ and $\K\simeq\BH\oplus\BH$ ($p-q\equiv 1,5\pmod{8}$)
form an axis of the eighth order, which defines the cyclic group $\dZ_8$.
Further, the neutral types $p-q\equiv 0\pmod{8}$ ($\K\simeq\R$) and
$p-q\equiv 4\pmod{8}$ ($\K\simeq\BH$), which in common admit the
automorphism groups with the signatures $(+,b,c)$, form an axis of the
fourth order corresponding to the cyclic group $\dZ_4$. Analogously, the
two mutually opposite types $p-q\equiv 2\pmod{8}$ ($\K\simeq\R$) and
$p-q\equiv 6\pmod{8}$ ($\K\simeq\BH$), which in common admit the
automorphism groups with the signatures $(-,b,c)$, also form an axis
of the fourth order. Finally, the types $p-q\equiv 3,7\pmod{8}$ ($\K\simeq\C$)
with the $(+,+,+)$ and $(-,-,-)$ automorphism groups form an axis of the
second order. Therefore, $BW^{a,b,c}_{\R}\simeq\dZ_2\otimes(\dZ_4)^2\otimes
\dZ_8$, where $(\dZ_4)^2=\dZ_4\otimes\dZ_4$.

\[
\unitlength=0.5mm
\begin{picture}(100.00,200.00)(0,-50)

\put(97,67){$\C$}\put(105,64){$p-q\equiv 7\!\!\!\!\pmod{8}$}
\put(107,82){$(+,+,+)$}
\put(107,75){$(-,-,-)$}
\put(80,80){1}
\put(75,93.3){$\cdot$}
\put(75.5,93){$\cdot$}
\put(76,92.7){$\cdot$}
\put(76.5,92.4){$\cdot$}
\put(77,92.08){$\cdot$}
\put(77.5,91.76){$\cdot$}
\put(78,91.42){$\cdot$}
\put(78.5,91.08){$\cdot$}
\put(79,90.73){$\cdot$}
\put(79.5,90.37){$\cdot$}
\put(80,90.0){$\cdot$}
\put(80.5,89.62){$\cdot$}
\put(81,89.23){$\cdot$}
\put(81.5,88.83){$\cdot$}
\put(82,88.42){$\cdot$}
\put(82.5,87.99){$\cdot$}
\put(83,87.56){$\cdot$}
\put(83.5,87.12){$\cdot$}
\put(84,86.66){$\cdot$}
\put(84.5,86.19){$\cdot$}
\put(85,85.70){$\cdot$}
\put(85.5,85.21){$\cdot$}
\put(86,84.69){$\cdot$}
\put(86.5,84.17){$\cdot$}
\put(87,83.63){$\cdot$}
\put(87.5,83.07){$\cdot$}
\put(88,82.49){$\cdot$}
\put(88.5,81.9){$\cdot$}
\put(89,81.29){$\cdot$}
\put(89.5,80.65){$\cdot$}
\put(90,80){$\cdot$}
\put(90.5,79.32){$\cdot$}
\put(91,78.62){$\cdot$}
\put(91.5,77.89){$\cdot$}
\put(92,77.13){$\cdot$}
\put(92.5,76.34){$\cdot$}
\put(93,75.51){$\cdot$}
\put(93.5,74.65){$\cdot$}
\put(94,73.74){$\cdot$}
\put(94.5,72.79){$\cdot$}
\put(96.5,73.74){\vector(1,-2){1}}
\put(80,20){3}
\put(97,31){$\BH$}\put(105,38){$p-q\equiv 6\!\!\!\!\pmod{8}$}
\put(112,30){$(-,+,-)$}
\put(112,23){$(-,-,+)$}
\put(112,16){$(-,-,-)$}
\put(112,9){$(-,+,+)$}
\put(75,6.7){$\cdot$}
\put(75.5,7){$\cdot$}
\put(76,7.29){$\cdot$}
\put(76.5,7.6){$\cdot$}
\put(77,7.91){$\cdot$}
\put(77.5,8.24){$\cdot$}
\put(78,8.57){$\cdot$}
\put(78.5,8.91){$\cdot$}
\put(79,9.27){$\cdot$}
\put(79.5,9.63){$\cdot$}
\put(80,10){$\cdot$}
\put(80.5,10.38){$\cdot$}
\put(81,10.77){$\cdot$}
\put(81.5,11.17){$\cdot$}
\put(82,11.58){$\cdot$}
\put(82.5,12.00){$\cdot$}
\put(83,12.44){$\cdot$}
\put(83.5,12.88){$\cdot$}
\put(84,13.34){$\cdot$}
\put(84.5,13.8){$\cdot$}
\put(85,14.29){$\cdot$}
\put(85.5,14.79){$\cdot$}
\put(86,15.3){$\cdot$}
\put(86.5,15.82){$\cdot$}
\put(87,16.37){$\cdot$}
\put(87.5,16.92){$\cdot$}
\put(88,17.5){$\cdot$}
\put(88.5,18.09){$\cdot$}
\put(89,18.71){$\cdot$}
\put(89.5,19.34){$\cdot$}
\put(90,20){$\cdot$}
\put(90.5,20.68){$\cdot$}
\put(91,21.38){$\cdot$}
\put(91.5,22.11){$\cdot$}
\put(92,22.87){$\cdot$}
\put(92.5,23.66){$\cdot$}
\put(93,24.48){$\cdot$}
\put(93.5,25.34){$\cdot$}
\put(94,26.25){$\cdot$}
\put(94.5,27.20){$\cdot$}
\put(95,28.20){$\cdot$}
\put(20,80){7}
\put(25,93.3){$\cdot$}
\put(24.5,93){$\cdot$}
\put(24,92.7){$\cdot$}
\put(23.5,92.49){$\cdot$}
\put(23,92.08){$\cdot$}
\put(22.5,91.75){$\cdot$}
\put(22,91.42){$\cdot$}
\put(21.5,91.08){$\cdot$}
\put(21,90.73){$\cdot$}
\put(20.5,90.37){$\cdot$}
\put(20,90){$\cdot$}
\put(19.5,89.62){$\cdot$}
\put(19,89.23){$\cdot$}
\put(18.5,88.83){$\cdot$}
\put(18,88.42){$\cdot$}
\put(17.5,87.99){$\cdot$}
\put(17,87.56){$\cdot$}
\put(16.5,87.12){$\cdot$}
\put(16,86.66){$\cdot$}
\put(15.5,86.19){$\cdot$}
\put(15,85.70){$\cdot$}
\put(14.5,85.21){$\cdot$}
\put(14,84.69){$\cdot$}
\put(13.5,84.17){$\cdot$}
\put(13,83.63){$\cdot$}
\put(12.5,83.07){$\cdot$}
\put(12,82.49){$\cdot$}
\put(11.5,81.9){$\cdot$}
\put(11,81.29){$\cdot$}
\put(10.5,80.65){$\cdot$}
\put(10,80){$\cdot$}
\put(9.5,79.32){$\cdot$}
\put(9,78.62){$\cdot$}
\put(8.5,77.89){$\cdot$}
\put(8,77.13){$\cdot$}
\put(7.5,76.34){$\cdot$}
\put(7,75.51){$\cdot$}
\put(6.5,74.65){$\cdot$}
\put(6,73.79){$\cdot$}
\put(5.5,72.79){$\cdot$}
\put(5,71.79){$\cdot$}
\put(20,20){5}
\put(25,6.7){$\cdot$}
\put(24.5,7){$\cdot$}
\put(24,7.29){$\cdot$}
\put(23.5,7.6){$\cdot$}
\put(23,7.91){$\cdot$}
\put(22.5,8.24){$\cdot$}
\put(22,8.57){$\cdot$}
\put(21.5,8.91){$\cdot$}
\put(21,9.27){$\cdot$}
\put(20.5,9.63){$\cdot$}
\put(20,10){$\cdot$}
\put(19.5,10.38){$\cdot$}
\put(19,10.77){$\cdot$}
\put(18.5,11.17){$\cdot$}
\put(18,11.58){$\cdot$}
\put(17.5,12){$\cdot$}
\put(17,12.44){$\cdot$}
\put(16.5,12.88){$\cdot$}
\put(16,13.34){$\cdot$}
\put(15.5,13.8){$\cdot$}
\put(15,14.29){$\cdot$}
\put(14.5,14.79){$\cdot$}
\put(14,15.3){$\cdot$}
\put(13.5,15.82){$\cdot$}
\put(13,16.37){$\cdot$}
\put(12.5,16.92){$\cdot$}
\put(12,17.5){$\cdot$}
\put(11.5,18.09){$\cdot$}
\put(11,18.71){$\cdot$}
\put(10.5,19.34){$\cdot$}
\put(10,20){$\cdot$}
\put(9.5,20.68){$\cdot$}
\put(9,21.38){$\cdot$}
\put(8.5,22.11){$\cdot$}
\put(8,22.87){$\cdot$}
\put(7.5,23.66){$\cdot$}
\put(7,24.48){$\cdot$}
\put(6.5,25.34){$\cdot$}
\put(6,26.25){$\cdot$}
\put(5.5,27.20){$\cdot$}
\put(5,28.20){$\cdot$}
\put(13,97){$\R\oplus\R$}\put(-55,87){$p-q\equiv 1\!\!\!\!\pmod{8}$}
\put(-23,95){$(+,+,-)$}
\put(-23,102){$(+,-,+)$}
\put(-23,109){$(+,-,-)$}
\put(-23,116){$(+,+,+)$}
\put(-23,123){$(-,+,+)$}
\put(-23,130){$(-,-,-)$}
\put(-23,137){$(-,+,-)$}
\put(-23,144){$(-,-,+)$}
\put(50,93){0}
\put(50,100){$\cdot$}
\put(49.5,99.99){$\cdot$}
\put(49,99.98){$\cdot$}
\put(48.5,99.97){$\cdot$}
\put(48,99.96){$\cdot$}
\put(47.5,99.94){$\cdot$}
\put(47,99.91){$\cdot$}
\put(46.5,99.86){$\cdot$}
\put(46,99.84){$\cdot$}
\put(45.5,99.8){$\cdot$}
\put(45,99.75){$\cdot$}
\put(44.5,99.7){$\cdot$}
\put(44,99.64){$\cdot$}
\put(43.5,99.57){$\cdot$}
\put(43,99.51){$\cdot$}
\put(42.5,99.43){$\cdot$}
\put(42,99.35){$\cdot$}
\put(41.5,99.27){$\cdot$}
\put(41,99.18){$\cdot$}
\put(40.5,99.09){$\cdot$}
\put(40,98.99){$\cdot$}
\put(39.5,98.88){$\cdot$}
\put(39,98.77){$\cdot$}
\put(38.5,98.66){$\cdot$}
\put(38,98.54){$\cdot$}
\put(37.5,98.41){$\cdot$}
\put(37,98.28){$\cdot$}
\put(50.5,99.99){$\cdot$}
\put(51,99.98){$\cdot$}
\put(51.5,99.97){$\cdot$}
\put(52,99.96){$\cdot$}
\put(52.5,99.94){$\cdot$}
\put(53,99.91){$\cdot$}
\put(53.5,99.86){$\cdot$}
\put(54,99.84){$\cdot$}
\put(54.5,99.8){$\cdot$}
\put(55,99.75){$\cdot$}
\put(55.5,99.7){$\cdot$}
\put(56,99.64){$\cdot$}
\put(56.5,99.57){$\cdot$}
\put(57,99.51){$\cdot$}
\put(57.5,99.43){$\cdot$}
\put(58,99.35){$\cdot$}
\put(58.5,99.27){$\cdot$}
\put(59,99.18){$\cdot$}
\put(59.5,99.09){$\cdot$}
\put(60,98.99){$\cdot$}
\put(60.5,98.88){$\cdot$}
\put(61,98.77){$\cdot$}
\put(61.5,98.66){$\cdot$}
\put(62,98.54){$\cdot$}
\put(62.5,98.41){$\cdot$}
\put(63,98.28){$\cdot$}
\put(68,97){$\R$}\put(75,100){$p-q\equiv 0\!\!\!\!\pmod{8}$}
\put(85,109){$(+,+,-)$}
\put(85,116){$(+,-,+)$}
\put(85,123){$(+,-,-)$}
\put(85,130){$(+,+,+)$}
\put(50,7){4}
\put(68,2){$\BH\oplus\BH$}\put(90,-4){$p-q\equiv 5\!\!\!\!\pmod{8}$}
\put(95,-13){$(-,+,-)$}
\put(95,-20){$(-,-,+)$}
\put(95,-27){$(-,-,-)$}
\put(95,-34){$(-,+,+)$}
\put(95,-41){$(+,+,+)$}
\put(95,-48){$(+,-,-)$}
\put(95,-55){$(+,-,+)$}
\put(95,-62){$(+,+,-)$}
\put(50,0){$\cdot$}
\put(50.5,0){$\cdot$}
\put(51,0.01){$\cdot$}
\put(51.5,0.02){$\cdot$}
\put(52,0.04){$\cdot$}
\put(52.5,0.06){$\cdot$}
\put(53,0.09){$\cdot$}
\put(53.5,0.12){$\cdot$}
\put(54,0.16){$\cdot$}
\put(54.5,0.2){$\cdot$}
\put(55,0.25){$\cdot$}
\put(55.5,0.3){$\cdot$}
\put(56,0.36){$\cdot$}
\put(56.5,0.42){$\cdot$}
\put(57,0.49){$\cdot$}
\put(57.5,0.56){$\cdot$}
\put(58,0.64){$\cdot$}
\put(58.5,0.73){$\cdot$}
\put(59,0.82){$\cdot$}
\put(59.5,0.91){$\cdot$}
\put(60,1.01){$\cdot$}
\put(60.5,1.11){$\cdot$}
\put(61,1.22){$\cdot$}
\put(61.5,1.34){$\cdot$}
\put(62,1.46){$\cdot$}
\put(62.5,1.59){$\cdot$}
\put(63,1.72){$\cdot$}
\put(49.5,0){$\cdot$}
\put(49,0.01){$\cdot$}
\put(48.5,0.02){$\cdot$}
\put(48,0.04){$\cdot$}
\put(47.5,0.06){$\cdot$}
\put(47,0.09){$\cdot$}
\put(46.5,0.12){$\cdot$}
\put(46,0.16){$\cdot$}
\put(45.5,0.2){$\cdot$}
\put(45,0.25){$\cdot$}
\put(44.5,0.3){$\cdot$}
\put(44,0.36){$\cdot$}
\put(43.5,0.42){$\cdot$}
\put(43,0.49){$\cdot$}
\put(42.5,0.56){$\cdot$}
\put(42,0.64){$\cdot$}
\put(41.5,0.73){$\cdot$}
\put(41,0.82){$\cdot$}
\put(40.5,0.91){$\cdot$}
\put(40,1.01){$\cdot$}
\put(39.5,1.11){$\cdot$}
\put(39,1.22){$\cdot$}
\put(38.5,1.34){$\cdot$}
\put(38,1.46){$\cdot$}
\put(37.5,1.59){$\cdot$}
\put(37,1.72){$\cdot$}
\put(28,3){$\BH$}\put(-40,-4){$p-q\equiv 4\!\!\!\!\pmod{8}$}
\put(-20,-13){$(+,+,+)$}
\put(-20,-20){$(+,-,-)$}
\put(-20,-27){$(+,-,+)$}
\put(-20,-34){$(+,+,-)$}
\put(93,50){2}
\put(98.28,63){$\cdot$}
\put(98.41,62.5){$\cdot$}
\put(98.54,62){$\cdot$}
\put(98.66,61.5){$\cdot$}
\put(98.77,61){$\cdot$}
\put(98.88,60.5){$\cdot$}
\put(98.99,60){$\cdot$}
\put(99.09,59.5){$\cdot$}
\put(99.18,59){$\cdot$}
\put(99.27,58.5){$\cdot$}
\put(99.35,58){$\cdot$}
\put(99.43,57.5){$\cdot$}
\put(99.51,57){$\cdot$}
\put(99.57,56.5){$\cdot$}
\put(99.64,56){$\cdot$}
\put(99.7,55.5){$\cdot$}
\put(99.75,55){$\cdot$}
\put(99.8,54.5){$\cdot$}
\put(99.84,54){$\cdot$}
\put(99.86,53.5){$\cdot$}
\put(99.91,53){$\cdot$}
\put(99.94,52.5){$\cdot$}
\put(99.96,52){$\cdot$}
\put(99.97,51.5){$\cdot$}
\put(99.98,51){$\cdot$}
\put(99.99,50.5){$\cdot$}
\put(100,50){$\cdot$}
\put(98.28,37){$\cdot$}
\put(98.41,37.5){$\cdot$}
\put(98.54,38){$\cdot$}
\put(98.66,38.5){$\cdot$}
\put(98.77,39){$\cdot$}
\put(98.88,39.5){$\cdot$}
\put(98.99,40){$\cdot$}
\put(99.09,40.5){$\cdot$}
\put(99.18,41){$\cdot$}
\put(99.27,41.5){$\cdot$}
\put(99.35,42){$\cdot$}
\put(99.43,42.5){$\cdot$}
\put(99.51,43){$\cdot$}
\put(99.57,43.5){$\cdot$}
\put(99.64,44){$\cdot$}
\put(99.7,44.5){$\cdot$}
\put(99.75,45){$\cdot$}
\put(99.8,45.5){$\cdot$}
\put(99.84,46){$\cdot$}
\put(99.86,46.5){$\cdot$}
\put(99.91,47){$\cdot$}
\put(99.94,47.5){$\cdot$}
\put(99.96,48){$\cdot$}
\put(99.97,48.5){$\cdot$}
\put(99.98,49){$\cdot$}
\put(99.99,49.5){$\cdot$}
\put(7,50){6}
\put(1,32){$\C$}\put(-65,29){$p-q\equiv 3\!\!\!\!\pmod{8}$}
\put(-35,20){$(-,-,-)$}
\put(-35,13){$(+,+,+)$}
\put(1.72,63){$\cdot$}
\put(1.59,62.5){$\cdot$}
\put(1.46,62){$\cdot$}
\put(1.34,61.5){$\cdot$}
\put(1.22,61){$\cdot$}
\put(1.11,60.5){$\cdot$}
\put(1.01,60){$\cdot$}
\put(0.99,59.5){$\cdot$}
\put(0.82,59){$\cdot$}
\put(0.73,58.5){$\cdot$}
\put(0.64,58){$\cdot$}
\put(0.56,57.5){$\cdot$}
\put(0.49,57){$\cdot$}
\put(0.42,56.5){$\cdot$}
\put(0.36,56){$\cdot$}
\put(0.3,55.5){$\cdot$}
\put(0.25,55){$\cdot$}
\put(0.2,54.5){$\cdot$}
\put(0.16,54){$\cdot$}
\put(0.12,53.5){$\cdot$}
\put(0.09,53){$\cdot$}
\put(0.06,52.5){$\cdot$}
\put(0.04,52){$\cdot$}
\put(0.02,51.5){$\cdot$}
\put(0.01,51){$\cdot$}
\put(0,50.5){$\cdot$}
\put(0,50){$\cdot$}
\put(1.72,37){$\cdot$}
\put(1.59,37.5){$\cdot$}
\put(1.46,38){$\cdot$}
\put(1.34,38.5){$\cdot$}
\put(1.22,39){$\cdot$}
\put(1.11,39.5){$\cdot$}
\put(1.01,40){$\cdot$}
\put(0.99,40.5){$\cdot$}
\put(0.82,41){$\cdot$}
\put(0.73,41.5){$\cdot$}
\put(0.64,42){$\cdot$}
\put(0.56,42.5){$\cdot$}
\put(0.49,43){$\cdot$}
\put(0.42,43.5){$\cdot$}
\put(0.36,44){$\cdot$}
\put(0.3,44.5){$\cdot$}
\put(0.25,45){$\cdot$}
\put(0.2,45.5){$\cdot$}
\put(0.16,46){$\cdot$}
\put(0.12,46.5){$\cdot$}
\put(0.09,47){$\cdot$}
\put(0.06,47.5){$\cdot$}
\put(0.04,48){$\cdot$}
\put(0.02,48.5){$\cdot$}
\put(0.01,49){$\cdot$}
\put(0,49.5){$\cdot$}
\put(0.5,67){$\R$}\put(-65,75){$p-q\equiv 2\!\!\!\!\pmod{8}$}
\put(-45,65){$(-,-,-)$}
\put(-45,58){$(-,+,+)$}
\put(-45,51){$(-,-,+)$}
\put(-45,44){$(-,+,-)$}
\put(50,50){\line(5,2){60}}
\put(50,50){\line(2,5){30}}
\put(50,50){\line(-2,-5){30}}
\put(50,50){\line(-2,5){30}}
\put(50,50){\line(2,-5){30}}
\put(50,50){\line(-5,-2){60}}
\put(50,50){\line(5,-2){60}}
\put(50,50){\line(-5,2){60}}
\end{picture}
\]

\vspace{2ex}
\begin{center}
\begin{minipage}{25pc}{\small
{\bf Fig.2} The cyclic structure of the generalized group 
$BW^{a,b,c}_{\R}$.}
\end{minipage}
\end{center}
\medskip
Further, over the field $\F=\C$, there exist two types of the complex
Clifford algebras: $\C_n$ and $\C_{n+1}\simeq\C_n\oplus\C_n$. Therefore,
a Brauer--Wall group $BW_{\C}$ acting on the set of 
these two types is isomorphic
to the cyclic group $\dZ_2$. The cyclic structure of the group
$BW_{\C}\simeq\dZ_2$ may be represented on the following Trautman diagram
(Fig. 3) by means of a transition $\C^+_n\stackrel{h}{\longrightarrow}\C_n$
(the round on the diagram is realized by an hour--hand). At this point, the
type of the algebra on the diagram is defined by an equality
$n=h+2r$, where $h\in\{0,1\}$, $r\in\dZ$.

\[
\unitlength=0.5mm
\begin{picture}(50.00,60.00)
\put(5,25){1}
\put(42,25){0}
\put(22,-4){$\C_{{\rm even}}$}
\put(22,55){$\C_{{\rm odd}}$}
\put(3,-13){$n\equiv 0\!\!\!\!\pmod{2}$}
\put(3,64){$n\equiv 1\!\!\!\!\pmod{2}$}
\put(20,49.49){$\cdot$}
\put(19.5,49.39){$\cdot$}
\put(19,49.27){$\cdot$}
\put(18.5,49.14){$\cdot$}
\put(18,49){$\cdot$}
\put(17.5,48.85){$\cdot$}
\put(17,48.68){$\cdot$}
\put(16.5,48.51){$\cdot$}
\put(16,48.32){$\cdot$}
\put(15.5,48.12){$\cdot$}
\put(15,47.91){$\cdot$}
\put(14.5,47.69){$\cdot$}
\put(14,47.45){$\cdot$}
\put(13.5,47.2){$\cdot$}
\put(13,46.93){$\cdot$}
\put(12.5,46.65){$\cdot$}
\put(12,46.35){$\cdot$}
\put(11.5,46.04){$\cdot$}
\put(11,45.71){$\cdot$}
\put(10.5,45.36){$\cdot$}
\put(10,45){$\cdot$}
\put(9.5,44.61){$\cdot$}
\put(9,44.21){$\cdot$}
\put(8.5,43.78){$\cdot$}
\put(8,43.33){$\cdot$}
\put(7.5,42.85){$\cdot$}
\put(7,42.35){$\cdot$}
\put(6.5,41.81){$\cdot$}
\put(6,41.25){$\cdot$}
\put(5.5,40.64){$\cdot$}
\put(5,40){$\cdot$}
\put(4.5,39.3){$\cdot$}
\put(4,38.56){$\cdot$}
\put(3.5,37.76){$\cdot$}
\put(3,36.87){$\cdot$}
\put(2.5,35.89){$\cdot$}
\put(2,34.79){$\cdot$}
\put(1.5,33.53){$\cdot$}
\put(1,32){$\cdot$}
\put(0.5,29.97){$\cdot$}
\put(30,49.49){$\cdot$}
\put(30.5,49.39){$\cdot$}
\put(31,49.27){$\cdot$}
\put(31.5,49.14){$\cdot$}
\put(32,49){$\cdot$}
\put(32.5,48.85){$\cdot$}
\put(33,48.68){$\cdot$}
\put(33.5,48.51){$\cdot$}
\put(34,48.32){$\cdot$}
\put(34.5,48.12){$\cdot$}
\put(35,47.91){$\cdot$}
\put(35.5,47.69){$\cdot$}
\put(36,47.45){$\cdot$}
\put(36.5,47.2){$\cdot$}
\put(37,46.93){$\cdot$}
\put(37.5,46.65){$\cdot$}
\put(38,46.35){$\cdot$}
\put(38.5,46.04){$\cdot$}
\put(39,45.71){$\cdot$}
\put(39.5,45.36){$\cdot$}
\put(40,45){$\cdot$}
\put(40.5,44.61){$\cdot$}
\put(41,44.21){$\cdot$}
\put(41.5,43.78){$\cdot$}
\put(42,43.33){$\cdot$}
\put(42.5,42.85){$\cdot$}
\put(43,42.35){$\cdot$}
\put(43.5,41.81){$\cdot$}
\put(44,41.25){$\cdot$}
\put(44.5,40.64){$\cdot$}
\put(45,40){$\cdot$}
\put(45.5,39.3){$\cdot$}
\put(46,38.56){$\cdot$}
\put(46.5,37.76){$\cdot$}
\put(47,36.87){$\cdot$}
\put(47.5,35.89){$\cdot$}
\put(48,34.79){$\cdot$}
\put(48.5,33.53){$\cdot$}
\put(49,32){$\cdot$}
\put(49.5,29.97){$\cdot$}
\put(0,25){$\cdot$}
\put(0,24.5){$\cdot$}
\put(0.02,24){$\cdot$}
\put(0.04,23.5){$\cdot$}
\put(0.08,23){$\cdot$}
\put(0.12,22.5){$\cdot$}
\put(0.18,22){$\cdot$}
\put(0.25,21.5){$\cdot$}
\put(0.32,21){$\cdot$}
\put(0.4,20.5){$\cdot$}
\put(0.5,20){$\cdot$}
\put(0.61,19.5){$\cdot$}
\put(0.73,19){$\cdot$}
\put(0.85,18.5){$\cdot$}
\put(1,18){$\cdot$}
\put(1.15,17.5){$\cdot$}
\put(1.31,17){$\cdot$}
\put(1.49,16.5){$\cdot$}
\put(1.68,16){$\cdot$}
\put(1.88,15.5){$\cdot$}
\put(2.09,15){$\cdot$}
\put(2.31,14.5){$\cdot$}
\put(2.55,14){$\cdot$}
\put(2.8,13.5){$\cdot$}
\put(3.06,13){$\cdot$}
\put(0,25.5){$\cdot$}
\put(0.02,26){$\cdot$}
\put(0.04,26.5){$\cdot$}
\put(0.08,27){$\cdot$}
\put(0.12,27.5){$\cdot$}
\put(0.18,28){$\cdot$}
\put(0.25,28.5){$\cdot$}
\put(0.32,29){$\cdot$}
\put(0.4,29.5){$\cdot$}
\put(0.5,30){$\cdot$}
\put(0.61,30.5){$\cdot$}
\put(0.73,31){$\cdot$}
\put(0.85,31.5){$\cdot$}
\put(1,32){$\cdot$}
\put(1.15,32.5){$\cdot$}
\put(1.31,33){$\cdot$}
\put(1.49,33.5){$\cdot$}
\put(1.68,34){$\cdot$}
\put(1.88,34.5){$\cdot$}
\put(2.09,35){$\cdot$}
\put(2.31,35.5){$\cdot$}
\put(2.55,36){$\cdot$}
\put(2.8,36.5){$\cdot$}
\put(3.06,37){$\cdot$}
\put(50,25){$\cdot$}
\put(49.99,24.5){$\cdot$}
\put(49.98,24){$\cdot$}
\put(49.95,23.5){$\cdot$}
\put(49.92,23){$\cdot$}
\put(49.87,22.5){$\cdot$}
\put(49.82,22){$\cdot$}
\put(49.75,21.5){$\cdot$}
\put(49.68,21){$\cdot$}
\put(49.51,20.5){$\cdot$}
\put(49.49,20){$\cdot$}
\put(49.39,19.5){$\cdot$}
\put(49.27,19){$\cdot$}
\put(49.14,18.5){$\cdot$}
\put(49,18){$\cdot$}
\put(48.85,17.5){$\cdot$}
\put(48.69,17){$\cdot$}
\put(48.51,16.5){$\cdot$}
\put(48.32,16){$\cdot$}
\put(48.12,15.5){$\cdot$}
\put(47.91,15){$\cdot$}
\put(47.69,14.5){$\cdot$}
\put(47.45,14){$\cdot$}
\put(47.2,13.5){$\cdot$}
\put(46.93,13){$\cdot$}
\put(50,25){$\cdot$}
\put(49.99,25.5){$\cdot$}
\put(49.98,26){$\cdot$}
\put(49.95,26.5){$\cdot$}
\put(49.92,27){$\cdot$}
\put(49.87,27.5){$\cdot$}
\put(49.82,28){$\cdot$}
\put(49.75,28.5){$\cdot$}
\put(49.68,29){$\cdot$}
\put(49.51,29.5){$\cdot$}
\put(49.49,30){$\cdot$}
\put(49.39,30.5){$\cdot$}
\put(49.27,31){$\cdot$}
\put(49.14,31.5){$\cdot$}
\put(49,32){$\cdot$}
\put(48.85,32.5){$\cdot$}
\put(48.69,33){$\cdot$}
\put(48.51,33.5){$\cdot$}
\put(48.32,34){$\cdot$}
\put(48.12,34.5){$\cdot$}
\put(47.91,35){$\cdot$}
\put(47.69,35.5){$\cdot$}
\put(47.45,36){$\cdot$}
\put(47.2,36.5){$\cdot$}
\put(46.93,37){$\cdot$}
\put(20,0.5){$\cdot$}
\put(19.5,0.61){$\cdot$}
\put(19,0.73){$\cdot$}
\put(18.5,0.86){$\cdot$}
\put(18,1){$\cdot$}
\put(17.5,1.15){$\cdot$}
\put(17,1.31){$\cdot$}
\put(16.5,1.49){$\cdot$}
\put(16,1.68){$\cdot$}
\put(15.5,1.87){$\cdot$}
\put(15,2.09){$\cdot$}
\put(14.5,2.31){$\cdot$}
\put(14,2.55){$\cdot$}
\put(13.5,2.8){$\cdot$}
\put(13,3.06){$\cdot$}
\put(12.5,3.35){$\cdot$}
\put(12,3.64){$\cdot$}
\put(11.5,3.96){$\cdot$}
\put(11,4.29){$\cdot$}
\put(10.5,4.63){$\cdot$}
\put(10,5){$\cdot$}
\put(9.5,5.38){$\cdot$}
\put(9,5.79){$\cdot$}
\put(8.5,6.22){$\cdot$}
\put(8,6.67){$\cdot$}
\put(7.5,7.15){$\cdot$}
\put(7,7.65){$\cdot$}
\put(6.5,8.18){$\cdot$}
\put(6,8.75){$\cdot$}
\put(5.5,9.35){$\cdot$}
\put(5,10){$\cdot$}
\put(4.5,10.69){$\cdot$}
\put(4,11.43){$\cdot$}
\put(3.5,12.24){$\cdot$}
\put(3,13.12){$\cdot$}
\put(2.5,14.10){$\cdot$}
\put(2,15.20){$\cdot$}
\put(1.5,16.47){$\cdot$}
\put(1,18){$\cdot$}
\put(0.5,20.02){$\cdot$}
\put(30,0.5){$\cdot$}
\put(30.5,0.61){$\cdot$}
\put(31,0.73){$\cdot$}
\put(31.5,0.86){$\cdot$}
\put(32,1){$\cdot$}
\put(32.5,1.15){$\cdot$}
\put(33,1.31){$\cdot$}
\put(33.5,1.49){$\cdot$}
\put(34,1.68){$\cdot$}
\put(34.5,1.87){$\cdot$}
\put(35,2.09){$\cdot$}
\put(35.5,2.31){$\cdot$}
\put(36,2.55){$\cdot$}
\put(36.5,2.8){$\cdot$}
\put(37,3.06){$\cdot$}
\put(37.5,3.35){$\cdot$}
\put(38,3.64){$\cdot$}
\put(38.5,3.96){$\cdot$}
\put(39,4.29){$\cdot$}
\put(39.5,4.63){$\cdot$}
\put(40,5){$\cdot$}
\put(40.5,5.38){$\cdot$}
\put(41,5.79){$\cdot$}
\put(41.5,6.22){$\cdot$}
\put(42,6.67){$\cdot$}
\put(42.5,7.15){$\cdot$}
\put(43,7.65){$\cdot$}
\put(43.5,8.18){$\cdot$}
\put(44,8.75){$\cdot$}
\put(44.5,9.35){$\cdot$}
\put(45,10){$\cdot$}
\put(45.5,10.69){$\cdot$}
\put(46,11.43){$\cdot$}
\put(46.5,12.24){$\cdot$}
\put(47,13.12){$\cdot$}
\put(47.5,14.10){$\cdot$}
\put(48,15.20){$\cdot$}
\put(48.5,16.47){$\cdot$}
\put(49,18){$\cdot$}
\put(49.5,20.02){$\cdot$}

\end{picture}
\]
\vspace{2ex}
\begin{center}
\begin{minipage}{25pc}{\small
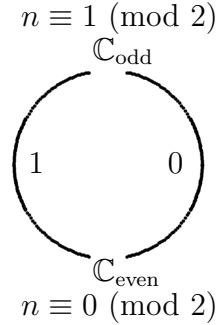
{\bf Fig.3} The Trautman diagram for the Brauer--Wall group
$BW_{\C}\simeq\dZ_2$.}
\end{minipage}
\end{center}
\medskip 
It is obvious that a group structure over $\C_n$, defined by the group
$BW_{\C}\simeq\dZ_2$, immediately relates with a modulo 2 periodicity of the
complex Clifford algebras \cite{AtBSh,Kar79}: $\C_{n+2}\simeq\C_n\otimes\C_2$.

From Theorem \ref{taut}, it follows that the algebra $\C_{2m}\simeq
\M_{2^m}(\C)$ admits the automorphism group $\sAut_-(\C_{2m})\simeq
\dZ_2\otimes\dZ_2$ with the signature $(+,+,+)$ if $m$ is even, and
respectively the group $\sAut_+(\C_{2m})\simeq Q_4/\dZ_2$ with the signature
$(-,-,-)$ if $m$ is odd. In connection with this, the second complex type
$\C_{2m+1}\simeq\C_{2m}\oplus\C_{2m}$ also admits both the previously
mentioned automorphism groups. Therefore, if we take into account this
relation, the cyclic structure of a generalized group $BW^{a,b,c}_{\C}$
would look as follows (Fig. 4). Both complex types $n\equiv 0\pmod{2}$
and $n\equiv 1\pmod{2}$ form an axis of the second order, therefore,
$BW^{a,b,c}_{\C}\simeq\dZ_2$.

\[
\unitlength=0.5mm
\begin{picture}(50.00,80.00)(0,-5)
\put(5,25){1}
\put(42,25){0}
\put(22,-4){$\C_{{\rm even}}$}
\put(22,55){$\C_{{\rm odd}}$}
\put(28,-30){$n\equiv 0\!\!\!\!\pmod{2}$}
\put(28,-13){$(+,+,+)$}
\put(28,-20){$(-,-,-)$}
\put(-31,79){$n\equiv 1\!\!\!\!\pmod{2}$}
\put(-11,70){$(-,-,-)$}
\put(-11,63){$(+,+,+)$}
\put(20,49.49){$\cdot$}
\put(19.5,49.39){$\cdot$}
\put(19,49.27){$\cdot$}
\put(18.5,49.14){$\cdot$}
\put(18,49){$\cdot$}
\put(17.5,48.85){$\cdot$}
\put(17,48.68){$\cdot$}
\put(16.5,48.51){$\cdot$}
\put(16,48.32){$\cdot$}
\put(15.5,48.12){$\cdot$}
\put(15,47.91){$\cdot$}
\put(14.5,47.69){$\cdot$}
\put(14,47.45){$\cdot$}
\put(13.5,47.2){$\cdot$}
\put(13,46.93){$\cdot$}
\put(12.5,46.65){$\cdot$}
\put(12,46.35){$\cdot$}
\put(11.5,46.04){$\cdot$}
\put(11,45.71){$\cdot$}
\put(10.5,45.36){$\cdot$}
\put(10,45){$\cdot$}
\put(9.5,44.61){$\cdot$}
\put(9,44.21){$\cdot$}
\put(8.5,43.78){$\cdot$}
\put(8,43.33){$\cdot$}
\put(7.5,42.85){$\cdot$}
\put(7,42.35){$\cdot$}
\put(6.5,41.81){$\cdot$}
\put(6,41.25){$\cdot$}
\put(5.5,40.64){$\cdot$}
\put(5,40){$\cdot$}
\put(4.5,39.3){$\cdot$}
\put(4,38.56){$\cdot$}
\put(3.5,37.76){$\cdot$}
\put(3,36.87){$\cdot$}
\put(2.5,35.89){$\cdot$}
\put(2,34.79){$\cdot$}
\put(1.5,33.53){$\cdot$}
\put(1,32){$\cdot$}
\put(0.5,29.97){$\cdot$}
\put(30,49.49){$\cdot$}
\put(30.5,49.39){$\cdot$}
\put(31,49.27){$\cdot$}
\put(31.5,49.14){$\cdot$}
\put(32,49){$\cdot$}
\put(32.5,48.85){$\cdot$}
\put(33,48.68){$\cdot$}
\put(33.5,48.51){$\cdot$}
\put(34,48.32){$\cdot$}
\put(34.5,48.12){$\cdot$}
\put(35,47.91){$\cdot$}
\put(35.5,47.69){$\cdot$}
\put(36,47.45){$\cdot$}
\put(36.5,47.2){$\cdot$}
\put(37,46.93){$\cdot$}
\put(37.5,46.65){$\cdot$}
\put(38,46.35){$\cdot$}
\put(38.5,46.04){$\cdot$}
\put(39,45.71){$\cdot$}
\put(39.5,45.36){$\cdot$}
\put(40,45){$\cdot$}
\put(40.5,44.61){$\cdot$}
\put(41,44.21){$\cdot$}
\put(41.5,43.78){$\cdot$}
\put(42,43.33){$\cdot$}
\put(42.5,42.85){$\cdot$}
\put(43,42.35){$\cdot$}
\put(43.5,41.81){$\cdot$}
\put(44,41.25){$\cdot$}
\put(44.5,40.64){$\cdot$}
\put(45,40){$\cdot$}
\put(45.5,39.3){$\cdot$}
\put(46,38.56){$\cdot$}
\put(46.5,37.76){$\cdot$}
\put(47,36.87){$\cdot$}
\put(47.5,35.89){$\cdot$}
\put(48,34.79){$\cdot$}
\put(48.5,33.53){$\cdot$}
\put(49,32){$\cdot$}
\put(49.5,29.97){$\cdot$}
\put(0,25){$\cdot$}
\put(0,24.5){$\cdot$}
\put(0.02,24){$\cdot$}
\put(0.04,23.5){$\cdot$}
\put(0.08,23){$\cdot$}
\put(0.12,22.5){$\cdot$}
\put(0.18,22){$\cdot$}
\put(0.25,21.5){$\cdot$}
\put(0.32,21){$\cdot$}
\put(0.4,20.5){$\cdot$}
\put(0.5,20){$\cdot$}
\put(0.61,19.5){$\cdot$}
\put(0.73,19){$\cdot$}
\put(0.85,18.5){$\cdot$}
\put(1,18){$\cdot$}
\put(1.15,17.5){$\cdot$}
\put(1.31,17){$\cdot$}
\put(1.49,16.5){$\cdot$}
\put(1.68,16){$\cdot$}
\put(1.88,15.5){$\cdot$}
\put(2.09,15){$\cdot$}
\put(2.31,14.5){$\cdot$}
\put(2.55,14){$\cdot$}
\put(2.8,13.5){$\cdot$}
\put(3.06,13){$\cdot$}
\put(0,25.5){$\cdot$}
\put(0.02,26){$\cdot$}
\put(0.04,26.5){$\cdot$}
\put(0.08,27){$\cdot$}
\put(0.12,27.5){$\cdot$}
\put(0.18,28){$\cdot$}
\put(0.25,28.5){$\cdot$}
\put(0.32,29){$\cdot$}
\put(0.4,29.5){$\cdot$}
\put(0.5,30){$\cdot$}
\put(0.61,30.5){$\cdot$}
\put(0.73,31){$\cdot$}
\put(0.85,31.5){$\cdot$}
\put(1,32){$\cdot$}
\put(1.15,32.5){$\cdot$}
\put(1.31,33){$\cdot$}
\put(1.49,33.5){$\cdot$}
\put(1.68,34){$\cdot$}
\put(1.88,34.5){$\cdot$}
\put(2.09,35){$\cdot$}
\put(2.31,35.5){$\cdot$}
\put(2.55,36){$\cdot$}
\put(2.8,36.5){$\cdot$}
\put(3.06,37){$\cdot$}
\put(50,25){$\cdot$}
\put(49.99,24.5){$\cdot$}
\put(49.98,24){$\cdot$}
\put(49.95,23.5){$\cdot$}
\put(49.92,23){$\cdot$}
\put(49.87,22.5){$\cdot$}
\put(49.82,22){$\cdot$}
\put(49.75,21.5){$\cdot$}
\put(49.68,21){$\cdot$}
\put(49.51,20.5){$\cdot$}
\put(49.49,20){$\cdot$}
\put(49.39,19.5){$\cdot$}
\put(49.27,19){$\cdot$}
\put(49.14,18.5){$\cdot$}
\put(49,18){$\cdot$}
\put(48.85,17.5){$\cdot$}
\put(48.69,17){$\cdot$}
\put(48.51,16.5){$\cdot$}
\put(48.32,16){$\cdot$}
\put(48.12,15.5){$\cdot$}
\put(47.91,15){$\cdot$}
\put(47.69,14.5){$\cdot$}
\put(47.45,14){$\cdot$}
\put(47.2,13.5){$\cdot$}
\put(46.93,13){$\cdot$}
\put(50,25){$\cdot$}
\put(49.99,25.5){$\cdot$}
\put(49.98,26){$\cdot$}
\put(49.95,26.5){$\cdot$}
\put(49.92,27){$\cdot$}
\put(49.87,27.5){$\cdot$}
\put(49.82,28){$\cdot$}
\put(49.75,28.5){$\cdot$}
\put(49.68,29){$\cdot$}
\put(49.51,29.5){$\cdot$}
\put(49.49,30){$\cdot$}
\put(49.39,30.5){$\cdot$}
\put(49.27,31){$\cdot$}
\put(49.14,31.5){$\cdot$}
\put(49,32){$\cdot$}
\put(48.85,32.5){$\cdot$}
\put(48.69,33){$\cdot$}
\put(48.51,33.5){$\cdot$}
\put(48.32,34){$\cdot$}
\put(48.12,34.5){$\cdot$}
\put(47.91,35){$\cdot$}
\put(47.69,35.5){$\cdot$}
\put(47.45,36){$\cdot$}
\put(47.2,36.5){$\cdot$}
\put(46.93,37){$\cdot$}
\put(20,0.5){$\cdot$}
\put(19.5,0.61){$\cdot$}
\put(19,0.73){$\cdot$}
\put(18.5,0.86){$\cdot$}
\put(18,1){$\cdot$}
\put(17.5,1.15){$\cdot$}
\put(17,1.31){$\cdot$}
\put(16.5,1.49){$\cdot$}
\put(16,1.68){$\cdot$}
\put(15.5,1.87){$\cdot$}
\put(15,2.09){$\cdot$}
\put(14.5,2.31){$\cdot$}
\put(14,2.55){$\cdot$}
\put(13.5,2.8){$\cdot$}
\put(13,3.06){$\cdot$}
\put(12.5,3.35){$\cdot$}
\put(12,3.64){$\cdot$}
\put(11.5,3.96){$\cdot$}
\put(11,4.29){$\cdot$}
\put(10.5,4.63){$\cdot$}
\put(10,5){$\cdot$}
\put(9.5,5.38){$\cdot$}
\put(9,5.79){$\cdot$}
\put(8.5,6.22){$\cdot$}
\put(8,6.67){$\cdot$}
\put(7.5,7.15){$\cdot$}
\put(7,7.65){$\cdot$}
\put(6.5,8.18){$\cdot$}
\put(6,8.75){$\cdot$}
\put(5.5,9.35){$\cdot$}
\put(5,10){$\cdot$}
\put(4.5,10.69){$\cdot$}
\put(4,11.43){$\cdot$}
\put(3.5,12.24){$\cdot$}
\put(3,13.12){$\cdot$}
\put(2.5,14.10){$\cdot$}
\put(2,15.20){$\cdot$}
\put(1.5,16.47){$\cdot$}
\put(1,18){$\cdot$}
\put(0.5,20.02){$\cdot$}
\put(30,0.5){$\cdot$}
\put(30.5,0.61){$\cdot$}
\put(31,0.73){$\cdot$}
\put(31.5,0.86){$\cdot$}
\put(32,1){$\cdot$}
\put(32.5,1.15){$\cdot$}
\put(33,1.31){$\cdot$}
\put(33.5,1.49){$\cdot$}
\put(34,1.68){$\cdot$}
\put(34.5,1.87){$\cdot$}
\put(35,2.09){$\cdot$}
\put(35.5,2.31){$\cdot$}
\put(36,2.55){$\cdot$}
\put(36.5,2.8){$\cdot$}
\put(37,3.06){$\cdot$}
\put(37.5,3.35){$\cdot$}
\put(38,3.64){$\cdot$}
\put(38.5,3.96){$\cdot$}
\put(39,4.29){$\cdot$}
\put(39.5,4.63){$\cdot$}
\put(40,5){$\cdot$}
\put(40.5,5.38){$\cdot$}
\put(41,5.79){$\cdot$}
\put(41.5,6.22){$\cdot$}
\put(42,6.67){$\cdot$}
\put(42.5,7.15){$\cdot$}
\put(43,7.65){$\cdot$}
\put(43.5,8.18){$\cdot$}
\put(44,8.75){$\cdot$}
\put(44.5,9.35){$\cdot$}
\put(45,10){$\cdot$}
\put(45.5,10.69){$\cdot$}
\put(46,11.43){$\cdot$}
\put(46.5,12.24){$\cdot$}
\put(47,13.12){$\cdot$}
\put(47.5,14.10){$\cdot$}
\put(48,15.20){$\cdot$}
\put(48.5,16.47){$\cdot$}
\put(49,18){$\cdot$}
\put(49.5,20.02){$\cdot$}
\put(25,25){\line(0,1){50}}
\put(25,25){\line(0,-1){50}}
\end{picture}
\]
\vspace{2ex}
\begin{center}
\begin{minipage}{25pc}{\small
{\bf Fig.4} The cyclic structure of the generalized group
$BW^{a,b,c}_{\C}$.}
\end{minipage}
\end{center}
\medskip
\section*{Acknowledgements}
I am grateful to Prof. C.T.C. Wall and to Prof. A. Trautman for sending me
their interesting papers, and also to Prof. R. Ab\l amowicz for helpful
discussions about {\sc CLIFFORD}.

\end{document}